\documentclass[aps,pra,epsf,superscriptaddress,amsmath,amssymb,amsfonts,twocolumn,showpacs,nofootinbib]{revtex4-1}

\usepackage[T1]{fontenc}
\usepackage{graphicx}
\usepackage{dcolumn}
\usepackage{bm}
\usepackage{braket}
\usepackage{amsmath}
\usepackage{mathtools}
\usepackage{graphicx,color,xcolor}
\usepackage{hyperref}
\usepackage[capitalize]{cleveref}
\newcommand{\abs}[1]{\left| #1 \right|} 
\usepackage{multirow}

\usepackage[normalem]{ulem} 

\usepackage[sectionbib]{bibunits}
\usepackage{bibunits}

\begin{document}

\begin{bibunit}[apsrev4-1]

\title{Signatures of rigidity and second sound in dipolar supersolids}

\author{G. A. Bougas}
\affiliation{Department of Physics and LAMOR, Missouri University of Science and Technology, Rolla, MO 65409, USA}

\author{T. Bland}
\affiliation{Division of Mathematical Physics and NanoLund, Lund University, SE-221 00 Lund, Sweden}

\author{H.~R.~Sadeghpour}
\affiliation{ITAMP, Center for Astrophysics $|$ Harvard \& Smithsonian Cambridge, Massachusetts 02138, USA}

\author{S. I. Mistakidis}
\affiliation{Department of Physics and LAMOR, Missouri University of Science and Technology, Rolla, MO 65409, USA}

\date{\today}

\begin{abstract} 
We propose a dynamical protocol to probe the rigidity and phase coherence of dipolar supersolids by merging initially separated fragments in quasi-one-dimensional (1D) double-well potentials. Simulations based on the extended Gross-Pitaevskii equation reveal distinct dynamical signatures across phases. Supersolids exhibit damped crystal oscillations following barrier removal, with the damping rate reflecting superfluid connectivity. A phase-imprinted jump additionally triggers metastable dark solitary waves, which excites second sound,  
as revealed by an out-of-phase drift between the droplet lattice and the superfluid background. Our results show a realizable path to dynamically detect the second sound and rigidity of supersolids, as well as to realize and probe soliton formation. 
\end{abstract}

\maketitle

\paragraph*{\textit { Introduction.}}    \label{sec:intro}
Supersolidity, 
originally proposed to explain the low-temperature behavior of solid He~\cite{leggett_can_1970,Chester1970,chan_overview_2013,balibar_enigma_2010}, refers to the spontaneous formation of a crystalline structure atop a superfluid background. Conclusive experimental evidence for this phase has recently emerged with dipolar quantum gases~\cite{bottcher2019transient,chomaz2019long,Tanzi_observation_2019,norcia_two-dimensional_2021,recati_supersolidity_2023}, composed of highly magnetic atoms--typically lanthanides~\cite{lu2011strongly,aikawa2012bose,chomaz_dipolar_2022}--that feature a combination of short-range and anisotropic long-range  interactions~\cite{chomaz_dipolar_2022,lahaye_physics_2009}. 
Their interplay with confinement leads to roton softening~\cite{petter_probing_2019,Hertkorn_Density_fluc_2021}, stabilized by repulsive quantum fluctuations often described by the Lee-Huang-Yang (LHY)  correction~\cite{kadau_observing_2016,lee1957eigenvalues,Lima_quantum_2011}.

The relative strength of dipolar and contact interactions can be tuned using external magnetic fields~\cite{Maier_broad_2015,Tang_anisotropic_2016}. Increasing the dipolar interaction ratio enables the emergence of the supersolid phase, while deeper in the dipolar regime, arrays of isolated, incoherent droplets~\cite{ferrier-barbut_observation_2016,schmitt_self-bound_2016,chomaz2016quantum} occur. 
Confining geometry also plays a key role, influencing both crystal structure~\cite{Hertkorn_pattern_2021,poli_maintaining_2021} and collective excitations~\cite{natale2019,hertkorn_supersolidity_2021,Schmidt_Roton_excitation2021} in these long-range interacting systems. Additional morphologies such as honeycomb or labyrinthine (stripe) states~\cite{zhang2019supersolidity,zhang2021phases,Hertkorn_pattern_2021,poli_maintaining_2021,ripley_two-dimensional_2023}, though yet to be observed, may arise at larger atom numbers and stronger interactions.

Beyond observing density modulation associated with translational symmetry breaking~\cite{recati_supersolidity_2023},  establishing rigidity and phase coherence through dynamical excitations is crucial. 
Phase coherence has been demonstrated via time-of-flight experiments~\cite{bottcher2019transient,chomaz2019long,Tanzi_observation_2019} and after quenches from the droplet phase~\cite{Ilzhofer2021}. Recently, phase modulation has been used to induce oscillations of the superfluid background between crystal peaks~\cite{Biagioni_measurement_2024,donelli_self_2025}, enabling measurement of the superfluid fraction~\cite{Biagioni_measurement_2024}. Superfluidity has been further probed through excitation of the scissors mode \cite{tanzi2021evidence,norcia_can_2022} and observation of quantum vortices~\cite{casotti_observation_2024,poli2024synchronization}. The solid rigidity in a supersolid has been theoretically explored via the emergence of shear waves~\cite{Yapa_supersonic_2024} and through connections between elastic properties and sound velocities~\cite{Platt_sounds_2024,sindik2024sound,poli_excitations_2024,rakic2024elastic,zawislak2024anomalous}.

A dynamical protocol for probing superfluidity, successfully applied in non-dipolar systems, involves observing the interference pattern of two condensates after merging~\cite{andrews_observation_1997,Shin_atom_2004,kohstall_observation_2011}.
Time-of-fligh interference reveals their relative phase,
while in-trap merging instead generates solitons and shock waves~\cite{Engels_DS_shock,Engels_DSW}, decaying into vortices. 
Similar excitations arise from phase imprinting, including dark solitons~\cite{Burger_dark_1999} and vortices~\cite{Matthews_vortices_1999}, whose dynamics act as “quantum canaries” probing the superfluid background~\cite{anglin2008quantum}. An interesting question is: how do these canaries fly through a supersolid?
Dipolar dark solitons have been predicted to exhibit strong sensitivity to interactions~\cite{pawlowski2015dipolar,bland2015controllable,edmonds2016exploring,Bland_interaction_2017,kopycinski2023ultrawide}, yet their properties--and potential as probes of 1D supersolids--remain largely unexplored.

\begin{figure}
\centering
\includegraphics[width=1\columnwidth]{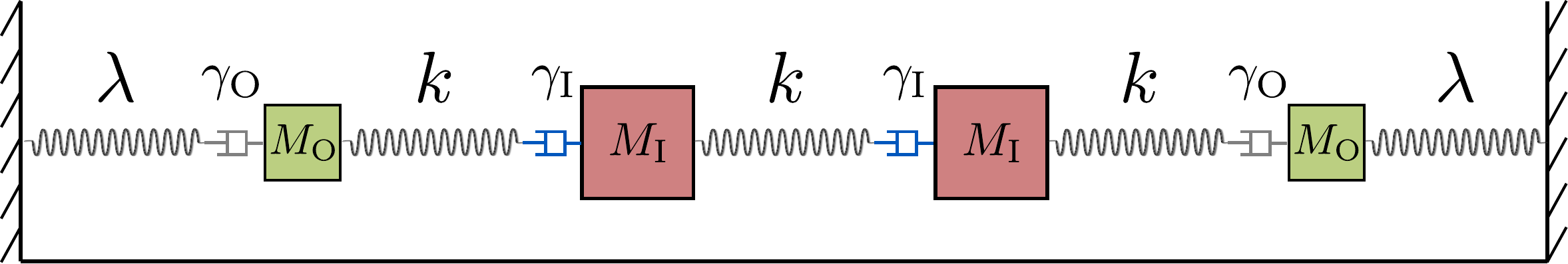}
\caption{Schematic illustration of the coupled damped springs model capturing the crystal oscillations in quasi-1D droplet arrays and supersolids. The spring constants are $k$, and $\lambda$, and the inner and outer masses are $M_\text{I}$ and $M_\text{O}$ respectively, with $M_\text{I}>M_\text{O}$. The damper elements denote the damping coefficients with strength $\gamma_{\rm{O}}$ and $\gamma_{\rm{I}}$ for the outer and inner droplets.}
\label{Fig:Springs_sketch}
\end{figure}

Here, we propose a protocol to dynamically probe supersolidity by merging two initially separated structures confined in a double-well potential, modeled using the 3D extended Gross-Pitaevskii equation (eGPE)~\cite{Santos2016filemanets,bisset_ground-state_2016,chomaz_dipolar_2022}. 
We find that the resulting crystal dynamics are accurately described by a damped coupled-oscillator model [see also the schematic in Fig.~\ref{Fig:Springs_sketch}], with the fitted damping parameter serving as a direct measure of the superfluid connectivity between droplets. 

Combining phase imprinting with barrier removal, the crystal drifts collectively while the superfluid flows in the opposite direction, signaling the excitation of the second sound mode~\cite{hertkorn_decoupled_2024,Guo2019}. The drift speed is tuned by adjusting the phase jump, and the crystalline structure remains intact throughout the evolution. Notably, for an initial phase jump of $\pi$, a 
solitary wave
forms, which decays via sound emission into the crystal before exciting the second sound mode after a significant delay. Measuring this delay time would offer a direct signature of dark solitary waves 
in supersolids, 
providing a unique probe of their superfluid properties.

\paragraph*{\textit { Setup}.}    \label{sec:theory}
We consider the zero temperature dynamics of $N = 8 \times 10^4$ $^{164}$Dy atoms of mass $m$, whose dipole moments are polarized along the $z$ axis. 
For $^{164}$Dy, the dipolar length is  fixed to $a_\text{dd} = 131~a_0$ with $a_0$  the Bohr radius. 
The dipolar quantum gas is initially held in a double well potential, elongated across one spatial direction,
\begin{gather}
V(\boldsymbol{r}) = \frac{m}{2} (\omega_x^2 x^2 + \omega_y^2 y^2 + \omega_z^2 z^2)\,+ V_0 e^{-x^2/2 W_0^2},
\label{Eq:Potential}
\end{gather}
where $\boldsymbol{r}=(x,y,z)$ and $(\omega_x,\omega_y,\omega_z) = 2\pi \times (19,53,81) ~ \rm{Hz}$. 
The double-well barrier height and width are denoted by $V_0$ and $W_0$ respectively.
The harmonic oscillator frequencies $\omega_i$, with $i=x,y,z$, 
are chosen to be similar to recent quasi-1D dipolar gas experiments~\cite{tanzi2019supersolid,bottcher2019transient}.
The corresponding harmonic oscillator lengths read $l_i = \sqrt{\hbar/m \omega_i}$. The quasi-1D geometry here refers to the arangement of droplet crystals along a single spatial direction~\cite{hertkorn_supersolidity_2021}.

\begin{figure}[t!]
\centering
\includegraphics[width=1\columnwidth]{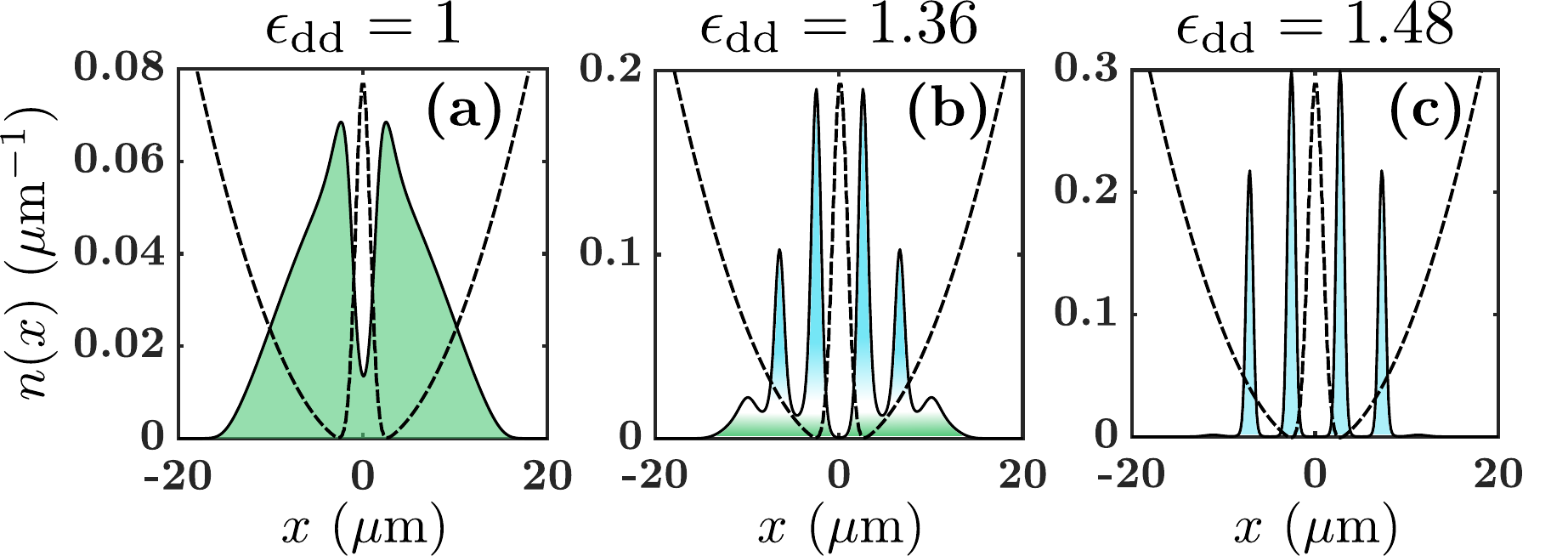}
\caption{Integrated density profiles $n(x) = \int \text{d}z \text{d}y ~ \abs{\Psi(\boldsymbol{r})}^2$ of $N=8 \times 10^4$ ${}^{164}$Dy atoms in the presence of a barrier. As $\epsilon_\text{dd}$ increases (see titles) the dipolar gas transitions from (a) a superfluid to (b) a supersolid and (c) to isolated droplets. Green (light-blue) shaded areas mark the superfluid (crystal) fractions. The rescaled 1D double well potential $V/(\hbar \omega_x l_x)$ is depicted by the dashed line in all panels, characterized by $V_0 = 10~ \hbar \omega_x$ and $W_0 = 0.5~l_x$.}
\label{Fig:Chemical_pot_scan}
\end{figure}

The phases of the quasi-1D dipolar gas trapped in the above potential are adequately captured by the 3D eGPE~\cite{Santos2016filemanets,bisset_ground-state_2016,chomaz2016quantum,ferrier-barbut_observation_2016,chomaz_dipolar_2022}, see also the Supplemental Material (SM)~\cite{supp}, dictating the evolution of the wavefunction $\Psi(\boldsymbol{r},t)$.
Depending on the ratio of dipolar over contact scattering lengths, $\epsilon_{\rm{dd}} \equiv a_{\rm{dd}}/a$, tunable via Fano-Feshbach resonances~\cite{Chin_Feshbach_2010,Maier_broad_2015}, three distinct phases occur; the superfluid [Fig.~\ref{Fig:Chemical_pot_scan}(a)], the supersolid [Fig.~\ref{Fig:Chemical_pot_scan}(b)], and isolated droplets [Fig.~\ref{Fig:Chemical_pot_scan}(c)].

\paragraph*{\textit {Damped oscillators model.}}   \label{sec:Barrier}

To capture the rigid crystal dynamics in the droplet and supersolid phases, we introduce a model of damped coupled oscillators, extending the framework of Ref.~\cite{Mukherjee_classical_2023}. 
Such a scheme provides an analytically tractable phenomenological treatment of supersolids and droplets, whose theoretical description is challenging.
Here, droplet peaks are treated as massive particles, with the inner and outer droplet pairs assigned effective masses $M_\text{I}$ and $M_\text{O}$, respectively. Each crystal pair is connected by a spring with constant $k$, as illustrated in Fig.~\ref{Fig:Springs_sketch}, while the outer droplets are additionally anchored to fixed external points by springs of constant $\lambda$. The damping  coefficients are $\gamma_{\rm{I}}$ and $\gamma_{\rm{O}}$, associated with the motion of the inner and outer droplets, respectively. The dynamics of the droplet positions, $X_i$, are then governed by the Newtonian system
\begin{gather}
\begin{pmatrix}
M_\text{O}\ddot X_1 \\ M_\text{I} \ddot X_2 \\  M_\text{I} \ddot X_3 \\ M_\text{O} \ddot X_4 
\end{pmatrix} =
\begin{pmatrix}
-\lambda -k & k & 0 & 0 \\
k & -2k & k & 0 \\
0 & k & -2k & k \\
0 & 0 & k &  -\lambda -k 
\end{pmatrix}
\begin{pmatrix}
\Delta X_1 \\ \Delta X_2 \\ \Delta X_3 \\ \Delta X_4
\end{pmatrix} \nonumber \\
-\begin{pmatrix}
\gamma_{\rm{O}} & 0 & 0 & 0 \\
0 & \gamma_{\rm{I}} & 0 & 0 \\
0 & 0 & \gamma_{\rm{I}} & 0 \\
0 & 0 & 0 & \gamma_{\rm{O}}
\end{pmatrix}
\begin{pmatrix}
\dot X_1 \\ \dot X_2 \\ \dot X_3 \\ \dot X_4
\end{pmatrix}\,.
\label{Eq:Springs_decay}
\end{gather}
Here, the displacements $\Delta X_i = X_i -X_{i0}$, with  $i=1,\ldots,4$, and $X_{i0}$ represents the equilibrium position of the $i$-th crystal prior to the barrier removal [see Fig.~\ref{Fig:Chemical_pot_scan}(b), (c)]. 
The fitting  procedure of the parameters of 
the above model to the eGPE results will be described below.

In the following,  the central barrier of the double-well exhibits a height $V_0=10 ~ \hbar \omega_x$ and width  $W_0 = 0.5 ~ l_x$. 
In this geometry, the dynamics occur predominantly in the elongated $x$-direction, with only small amplitude collective motion in the transverse $yz$-plane and no pattern formation (see also SM~\cite{supp}).

\begin{figure}[t!]
\centering
\includegraphics[width=1\columnwidth]{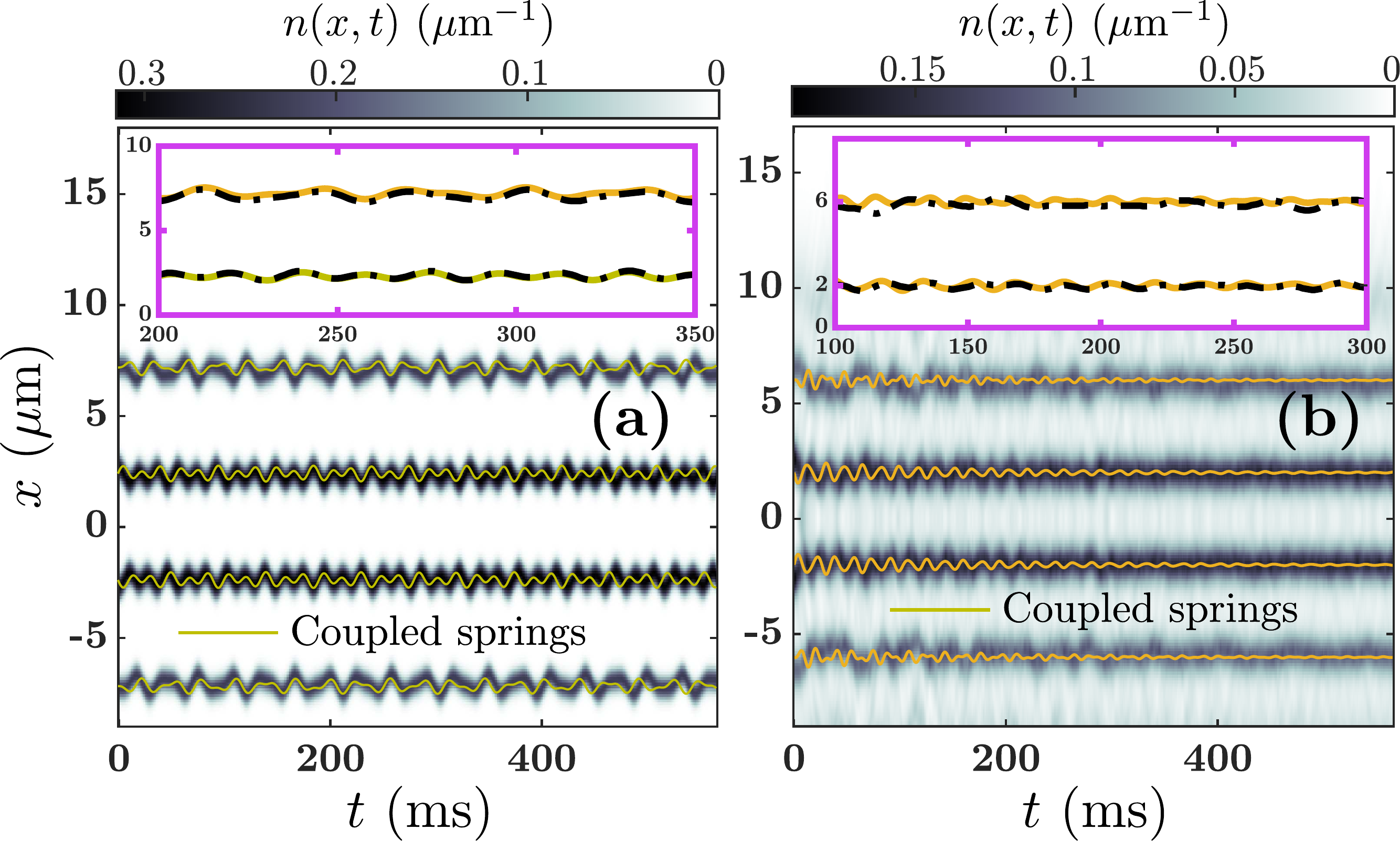}
\caption{Merging dipolar gases through barrier removal. Density response of quasi-1D (a) droplet lattices ($\epsilon_{\rm{dd}} = 1.48$), and (b) supersolids ($\epsilon_{\rm{dd}} = 1.36$) following barrier quench. In both phases, the yellow solid lines model the motion of four droplets using a damped coupled oscillator system with zero [panel (a)] and finite [panel (b)] damping. 
Insets show a magnified view of the dynamics of the upper inner and outer droplets. The dash-dotted lines correspond to a fit of the droplets position. Barrier potential characteristics are the same as in Fig.~\ref{Fig:Chemical_pot_scan}.}
\label{Fig:1D_dyn_without_phase}
\end{figure}

\paragraph*{\textit {Normal mode dynamics of rigid droplet crystals}.}  \label{sec:drops1D}
To assess how the density-modulated phases respond to the quench induced by barrier removal at $t = 0$, we follow their subsequent evolution in Fig.~\ref{Fig:1D_dyn_without_phase}, by tracking the 1D densities $n(x,t)=\int dydz~ \abs{\Psi(\boldsymbol{r},t)}^2$. In the isolated droplet 
regime ($\epsilon_\text{dd} = 1.48$), 
the inner ($x\approx\pm 2\,\mu$m) and outer ($x\approx\pm7\,\mu$m) droplet pairs oscillate out-of-phase following the barrier removal, while retaining their shapes--highlighting their rigidity [Fig.~\ref{Fig:1D_dyn_without_phase}(a)]. This motion, driven by the sudden barrier release and weak background-mediated forces, resembles the dynamics of linearly coupled oscillators~\cite{Mukherjee_classical_2023}.

To describe this, we apply a coupled springs model [Eq.~\eqref{Eq:Springs_decay}] with $k = \lambda$ and no damping ($\gamma_{\rm{I}} = \gamma_{\rm{O}} = 0$), assuming heavier inner droplets ($M_\text{I} > M_\text{O}$), consistent with their peak amplitudes [see Fig.~\ref{Fig:1D_dyn_without_phase}(a)]. 
Eqs.~\eqref{Eq:Springs_decay} are solved numerically, treating the natural frequencies $\sqrt{k/M_{\rm{O}}}$, $\sqrt{k/M_{\rm{I}}}$ and the initial velocities, $\dot{X}_1(0)$, $\dot{X}_2(0)$, as fitting parameters, obtained through a nonlinear least-squares analysis [see also SM~\cite{supp} for details].
The initial positions, $X_i(0)$ match the initial droplet peak positions, while $\dot{X}_3(0)=-\dot{X}_2(0)$, $\dot{X}_4(0) = -\dot{X}_1(0)$ due to symmetry of the crystals. Moreover, the initial estimates of the model's natural frequencies stem from the frequency spectra of the droplet peak positions.

The model qualitatively reproduces the droplet dynamics, as shown by the yellow lines in Fig.~\ref{Fig:1D_dyn_without_phase}(a) and its inset,
and the agreement improves for narrower barriers (not shown), where the perturbation is weaker~\cite{Mukherjee_classical_2023}.

\paragraph*{\textit {Probing the rigidity of supersolids}.}  \label{sec:SS1D}
We now examine the barrier-release dynamics of the dipolar supersolid at $\epsilon_\text{dd} = 1.36$, where both superfluid and solid-like responses are expected. The resulting density evolution is shown in Fig.~\ref{Fig:1D_dyn_without_phase}(b). A complex behavior arises from the coexistence of a droplet crystal and a non-negligible superfluid background, which enables particle tunneling between droplets~\cite{Turmanov_oscillations_2021}.

\begin{figure}[t!]
\centering
\includegraphics[width=1\columnwidth]{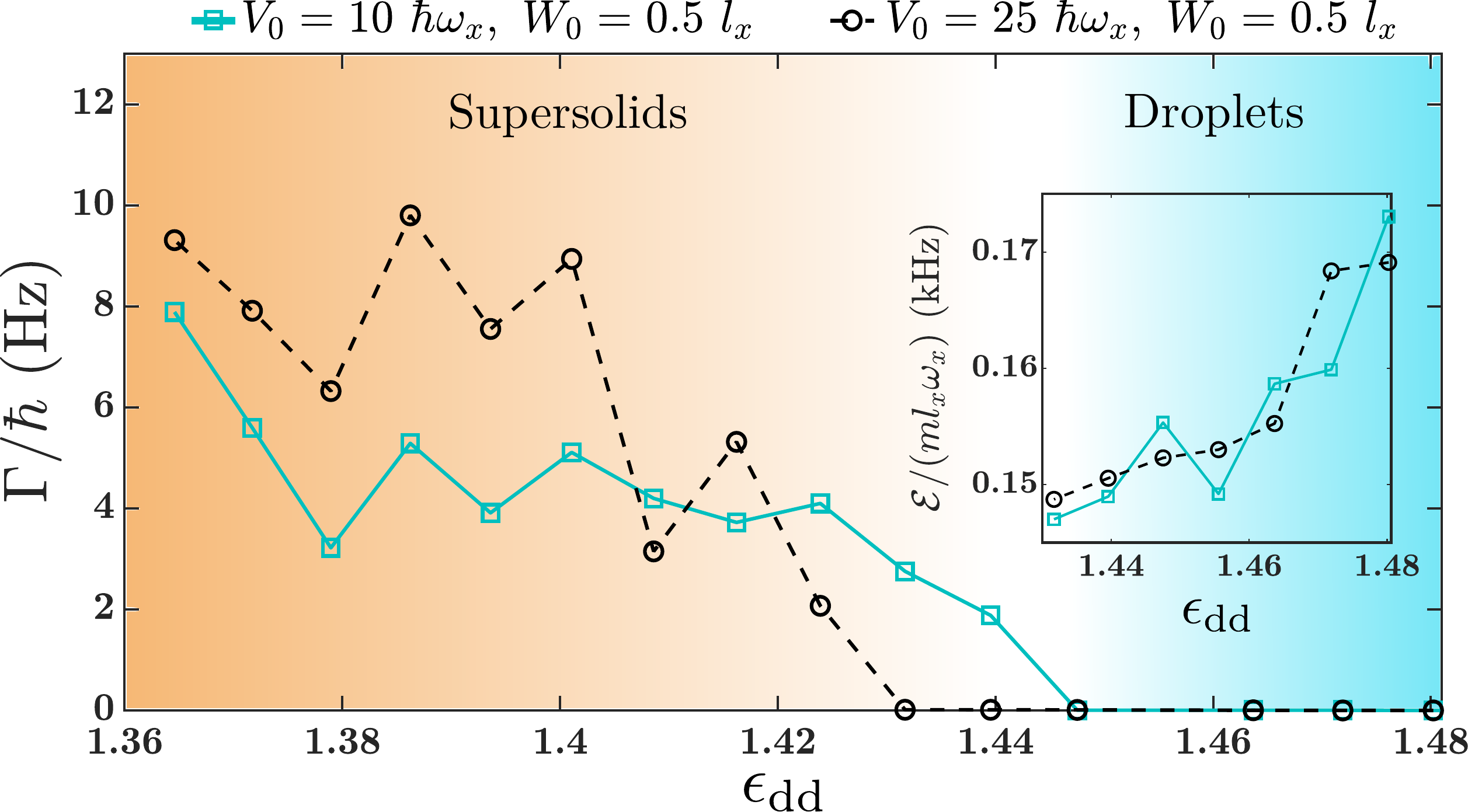}
\caption{
Decay rate $\Gamma$ of the damped crystal motion in the quasi-1D system with respect to $\epsilon_\text{dd}$, demonstrated for two initial barrier amplitudes (see legend). $\Gamma$ is extracted by fitting the droplet motion to a damped exponential function (see text). 
The corresponding standard deviations are of the order of $1~\rm{Hz}$.
Background color indicates the transition from supersolids to isolated droplets. 
In the isolated droplet case, the dipolar gas exhibits enhanced stiffness, quantified via the Young modulus (inset).}
\label{Fig:Normal_supersolid}
\end{figure}

Following the quench, the droplets are displaced and undergo small-amplitude oscillations with relative phase differences, resembling the motion seen in the isolated droplet phase [Fig.~\ref{Fig:1D_dyn_without_phase}(a)]. However, unlike the isolated droplet regime, these oscillations are damped over time due to coupling with the superfluid background. We quantify this damping by tracking the droplet positions and fitting the envelope of their oscillations with an exponential function of the form $A + B e^{-\Gamma t / \hbar}$. The extracted decay rates, $\Gamma$, as a function of the interaction parameter $\epsilon_\text{dd}$ are shown in Fig.~\ref{Fig:Normal_supersolid}.

As $\epsilon_\text{dd}$ decreases, the superfluid component becomes stronger, resulting in increased damping. 
This indicates that the superfluid background acts as a ``viscous''  medium, with the damping rate providing a direct measure of superfluid fraction. This may help interpreting recent observations~\cite{liebster2025observation} of rapidly decaying transverse phonons in the density-modulated regime.
Our results suggest that longer-lived excitations could persist if similar dynamics were explored at greater modulation depths, approaching the equivalent isolated droplet regime.

This damping trend persists across different barrier parameters: larger values of $V_0$ and $W_0$ generally lead to higher decay rates for fixed $\epsilon_\text{dd}$, as stronger initial forces are imparted to the droplets [Fig.~\ref{Fig:Normal_supersolid}]. Simultaneously, the superfluid background becomes excited and emits sound waves, which remain confined between the droplets. While solitary wave formation in this regime cannot be ruled out, any such structures are extremely shallow due to the weak disturbance of the background and are indistinguishable from sound waves.
Note that counter-propagating gray solitary wave pairs are generated in superfluids upon the barrier release [see SM~\cite{supp}], similar to their non-dipolar counterparts.

Building on the observed crystal dynamics, we extend our coupled oscillator model to include damping, setting $\gamma_{\rm{O}}/(2M_\text{O}) = \gamma_{\rm{I}}/(2M_\text{I}) = \Gamma/\hbar$. The spring constants are now differentiated such that $\lambda$ connects the outer droplets to the wall, while $k$ couples the oscillators. This distinction accounts for the weaker superfluid background at the edges of the supersolid compared to its center after the barrier release [Fig.~\ref{Fig:1D_dyn_without_phase}(b)]. 
Eqs.~\eqref{Eq:Springs_decay} are treated as in the droplets case, with $\sqrt{\lambda/M_{\rm{O}}}$ being an additional fitting parameter.
Due to the abrupt initial redistribution of the droplet peaks, the coupled oscillators model is constructed so that it captures droplet peak oscillations around their positions at later time instants.
The classical model qualitatively reproduces the motion of the inner droplets, as shown by the yellow solid lines in Fig.~\ref{Fig:1D_dyn_without_phase}(b) and its inset, offering a simplified yet insightful picture of damped crystal dynamics in a supersolid.
Due to the connecting superfluid background, a generalized model with next-to-nearest neighbor couplings between the droplet peaks has also  been considered, which does not offer however significant improvement  
over the current modeling, see also SM~\cite{supp}.

Based on the classical oscillator model, further insights into the elastic properties of supersolids can be obtained. We consider an ideal, infinite one-dimensional lattice of supersolid droplets, with a unit cell defined by the two inner sites~\cite{Biagioni_measurement_2024} and lattice spacing $a$ set by their separation. The position of the $n$-th site (droplet), $X_n$, satisfies the differential equation $\ddot{X}_n = -\frac{k}{M_\text{I}} [2\Delta X_n - \Delta X_{n-1} - \Delta X_{n+1}] - 2\frac{\Gamma}{\hbar} \dot{X}_n $
where $\Delta X_i$ is the displacement of the droplet from equilibrium. To describe crystal excitations, we seek traveling wave solutions of the form $X_n = X_{n0} + A e^{i(qna - \omega t)}$, where $q \in [-\pi/a, \pi/a]$ is the wavevector, and $A$ is a small amplitude. 
Substituting into the equation yields the dispersion relation $\omega(q) = - i \Gamma/ \hbar + \sqrt{(4k/M_\text{I})\sin^2(qa/2) - \left(\Gamma/\hbar\right)^2}$.

In the droplet regime where $\Gamma \to 0$ [Fig.~\ref{Fig:Normal_supersolid}], the long-wavelength limit $qa \to 0$ yields a phonon speed $c = a\sqrt{k/M_\text{I}}$~\cite{Platt_sounds_2024}. Comparing this to the elastic relation $c = \sqrt{\mathcal{E}/\rho}$, with $\rho = M_\text{I}/a$, allows us to extract the Young’s modulus as $\mathcal{E} = ka$~\cite{dove_introduction_1993}. As shown in the inset of Fig.~\ref{Fig:Normal_supersolid}, $\mathcal{E}$ increases with $\epsilon_\text{dd}$, indicating greater stiffness deeper in the droplet regime. In contrast, lowering $\epsilon_\text{dd}$ toward the supersolid region leads to a softer, more elastic response~\cite{poli_excitations_2024}.

In supersolids,  
crystal vibrations become damped due to the finite superfluid background. Moreover, long-wavelength excitations become overdamped, since the radicand in $\omega(q)$ turns negative.
A stronger superfluid background increases this damping, effectively filtering out a broader band of long-wavelength excitations. In contrast, shorter-wavelength modes can still propagate, giving the crystal chain viscoelastic behavior--exhibiting both elastic and dissipative characteristics akin to materials with complex elastic moduli~\cite{bland_linear_1960}.

\begin{figure}[t!]
\centering
\includegraphics[width=1\columnwidth]{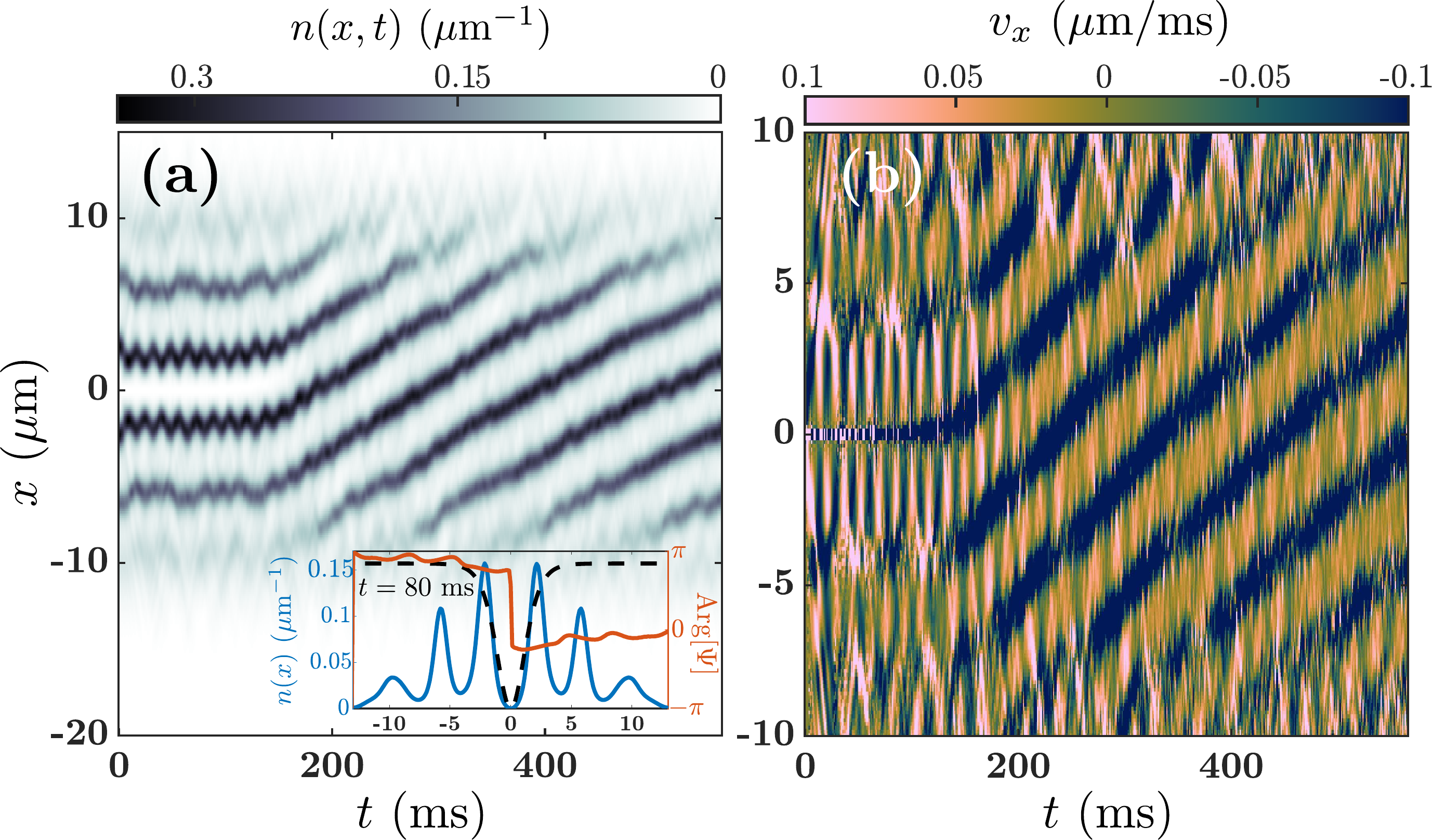}
\caption{Merging dipolar  supersolids ($\epsilon_{\rm{dd}} = 1.36$) featuring a $\pi$-phase discontinuity via barrier removal.
(a) Initially, a dark solitary wave forms at the trap center, visible in the density and phase profiles (inset). Once the solitary wave departs, the supersolid crystal undergoes second sound motion.  
(b) Time evolution of the superfluid velocity, $v_x = (\hbar/m) \partial_x \theta$, showing out-of-phase motion between the supersolid crystal and its superfluid background. Blue regions correspond to superfluid flow toward the $-x$ direction. All other parameters are the same as in  Fig.~\ref{Fig:1D_dyn_without_phase}.
}
\label{Fig:1D_dyn_with_phase}
\end{figure}

\paragraph*{\textit {Controlled excitation of the second sound}.}   \label{sec:phase}
A distinguishing feature of supersolids is their phase coherence, maintained through a connecting superfluid background. This property has been demonstrated in experiments using time-of-flight and in-situ imaging techniques~\cite{bottcher2019transient,chomaz2019long,Tanzi_observation_2019,Ilzhofer2021,norcia_two-dimensional_2021,bland_two-dimensional_2022}. In contrast, while a pure superfluid exhibits global phase coherence~\cite{andrews_observation_1997,Hadzibabic_interference_2004}, the droplet array is incoherent between isolated superfluid droplets~\cite{Ilzhofer2021}. To harness these coherence properties, we imprint a phase jump of $\Delta \varphi = \pi$ across the quasi-1D double well, assigning $\varphi = 0$ to the left ($x < 0$) and $\varphi = \pi$ to the right ($x > 0$) side of the dipolar gas. Simultaneously, the central barrier is removed at $t = 0$. Phase imprinting on the bulk has been widely used in Bose gases to generate solitonic excitations~\cite{Burger_dark_1999,denschlag_generating_2000,meyer2017observation} and vortices~\cite{Matthews_vortices_1999,Leanhardt_imprinting_2002}. Note that, since the independent droplet state lacks global phase coherence, phase imprinting has no effect on their dynamics, resulting in identical behavior to that shown in Fig.~\ref{Fig:1D_dyn_without_phase}(a).

In sharp contrast to the independent droplet regime, phase imprinting in the supersolid phase leads to a much richer dynamical response [Fig.~\ref{Fig:1D_dyn_with_phase}]. At early times, barrier removal through the superfluid background excites vibrations of the crystal peaks and generates sound waves through interference. During this phase, the $\Delta \varphi = \pi$ imprint is preserved at the trap center, where a pronounced density dip forms atop the background--indicative of a dark solitary wave [inset of Fig.~\ref{Fig:1D_dyn_with_phase}(a)]. The soliton core is notably broad and resides in the interstitial region between droplets, resembling the wide vortex cores predicted in dipolar supersolids~\cite{Gallemi_quantized_2020,Ancilotto_vortex_2021}.

Remarkably, after $t \approx 180$\,ms, interactions between the solitary wave, sound waves, and crystal lattice cause the soliton to drift. Unlike in superfluids, the solitary wave does not oscillate, see also SM~\cite{supp}, but instead transfers its momentum to the crystal, setting the droplets into motion. 
A similar effect has been predicted in rotating toroidal supersolids, where a soliton nucleated in the wake of a moving barrier induces a crystal drift. However, this drift is suppressed by the continued presence of the barrier, leading instead to the nucleation of a persistent current~\cite{tengstrand2023toroidal}. This momentum transfer is the inverse of vortex nucleation in supersolids, where momentum flows from the rigid structure to the superfluid component~\cite{poli2024synchronization}; here, momentum is transferred from the solitary wave to the rigid part of the system.

Importantly, this drifting motion is driven by the imprinted phase jump, not by the barrier removal. The barrier primarily serves to reduce background excitations by enforcing a node at the trap center during imprinting. We confirm this by observing similar drift without an initial central barrier, where the crystal moves in sync but the superfluid background exhibits significantly stronger excitations.

Despite the collective in-sync motion of the droplets, the center-of-mass of the gas remains stationary throughout the evolution, as verified by the static mean position $\braket{x(t)} = \int \text{d}^3\boldsymbol{r} ~ x \abs{\Psi(\boldsymbol{r},t)}^2$ (not shown). The droplet motion is compensated by an opposite drift of the superfluid background in the $-x$ direction, a clear manifestation of second sound, characterized by out-of-phase motion between the superfluid and crystalline components~\cite{Guo2019,hertkorn_decoupled_2024}. This behavior is evident in the superfluid velocity field, $v_x = \frac{\hbar}{m} \partial_x \theta(x,0,0)$, where $\theta$ is the local phase at $y = z = 0$ [Fig.~\ref{Fig:1D_dyn_with_phase}(b)]. Note that the direction of droplets motion is determined by the direction in which the solitary wave departs. In the presence of noise in the initial state, emulating experimental conditions, this direction becomes random. The slanted blue stripes, corresponding to negative velocities, indicate flow through the superfluid substrate connecting the density crystal sites. This background motion arises from the phase difference imprinted across the initially separated fragments, effectively acting as a momentum kick~\cite{Mukherjee_classical_2023}. These results highlight a distinctive feature of solitary waves in dipolar supersolids: rather than oscillating, they serve as a controlled mechanism for exciting the second sound mode.

\begin{figure}[t!]
\centering
\includegraphics[width=1\columnwidth]{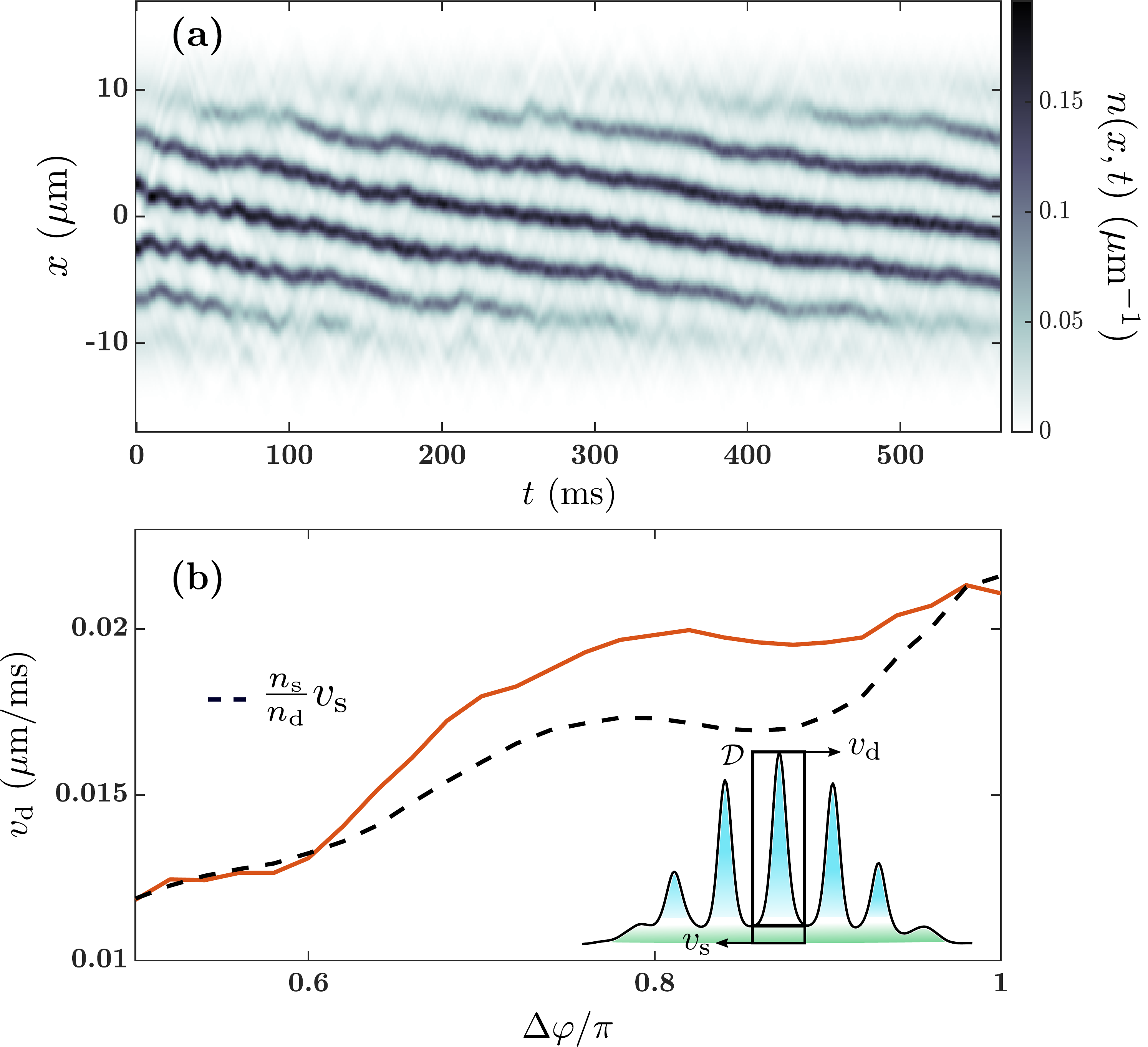}
\caption{Controllable second sound mode excitation.  
(a) Density evolution of a supersolid ($\epsilon_\text{dd} = 1.36$) following a $\Delta \varphi = \pi/2$ phase imprint.  
(b) Crystal drift velocity $v_\text{d}$ as a function of the imprinted phase jump $\Delta \varphi/\pi$. The dashed line presents an analytical estimate based on superfluid velocity and momentum conservation. The inset illustrates the opposite flow between a crystal site within the spatial region $\mathcal{D}$ and its underlying superfluid segment. All other system parameters match those in Fig.~\ref{Fig:1D_dyn_with_phase}.}
\label{Fig:Drift_speed}
\end{figure}

The drift velocity of the supersolid crystal can be controlled by varying the amplitude of the initial phase jump, $\pi/2 \leq \Delta \varphi \leq \pi$. A representative case is depicted in Fig.~\ref{Fig:Drift_speed}(a) for $\Delta \varphi = \pi/2$, where the crystal moves more slowly compared to the $\Delta \varphi = \pi$ case [Fig.~\ref{Fig:1D_dyn_with_phase}(a)]. Unlike the $\pi$-imprint, which traps a black solitary wave and induces a random second sound direction, a gray solitary wave forms and escapes immediately when $\Delta \varphi < \pi$, fixing the direction of second sound via the initial phase gradient.

To quantify the crystal drift, we assume approximately linear motion, $\braket{x_\text{c}(t)} = \braket{x_\text{c}(0)} + v_\text{d} t$, and extract the velocity as $v_\text{d} = -\braket{x_\text{c}(0)}/t$, taking the average over all the time instants at which 
the different crystals reach $\braket{x_\text{c}(t)} = 0$.
Here, $\braket{x_c(t)}$ corresponds to the mean position of individual droplet peaks.
As shown in Fig.~\ref{Fig:Drift_speed}(b), $v_\text{d}$ increases nearly linearly with $\Delta \varphi$. This scaling reflects the fact that a larger phase jump corresponds to a stronger phase gradient and thus a higher superfluid velocity, $v_x = \hbar/m\, \partial_x \theta(x,0,0)$. Due to the counterflow between the crystal sites and the superfluid, this enhances the crystal drift.
For $\Delta \varphi/\pi \gtrsim 0.8$, the solitary wave resulting from the phase imprint persists for a long time before eventually drifting away from the trap center [cf. Fig.~\ref{Fig:1D_dyn_with_phase}(a) and Fig.~\ref{Fig:Drift_speed}(a)]. This persistence hinders the nearly linear increase of $v_{\text{d}}$ for $\Delta \varphi/\pi \gtrsim 0.8$ [Fig.~\ref{Fig:Drift_speed}(b)], leading to a plateau at $v_{\rm{d}} \simeq 0.02 ~ \rm{\mu m}/\rm{ms}$, since the crystal remains nearly intact for an extended time period, affecting the averaging process described above.
For completeness, we note that phase jumps smaller than $\pi/2$ still induce crystal motion, but it becomes nonlinear and periodically reverses direction over time, 
complicating the definition of a drift velocity.

The increase of the drift velocity with larger $\Delta \varphi$ can also be understood through momentum conservation. Since the center-of-mass remains nearly stationary during the evolution, total momentum is conserved across the supersolid. Focusing on a spatial region $\mathcal{D}$ around a given crystal site, the momentum of the crystal must be balanced by that of the underlying superfluid segment outside of $\mathcal{D}$, leading to the relation $n_\text{s} v_\text{s} = n_\text{d} v_\text{d}$, see also the sketch in the inset of Fig.~\ref{Fig:Drift_speed}(b). Here,  $n_\text{s}$ and $n_\text{d}$ are the approximate densities of the superfluid and the crystal outside of and within $\mathcal{D}$, respectively, and $v_\text{s}$, $v_\text{d}$ their respective velocities.

The superfluid velocity $v_\text{s}$ is estimated by averaging over regions with positive $v_x$, corresponding to the substrate.
This gives an approximate drift velocity of $(n_\text{s}/n_\text{d})v_\text{s}$, represented by the dashed line in Fig.~\ref{Fig:Drift_speed}(b), which closely follows the eGPE results for $v_\text{d}$ (solid line). Deviations arise due to time-dependent variations in $n_\text{s}$ and $n_\text{d}$ during the evolution [see Fig.~\ref{Fig:Drift_speed}(a)].

\paragraph*{\textit {Summary \& Outlook}.}  \label{sec:Summary}
A dynamical protocol was proposed to explore the rigidity and coherence properties of quasi-1D supersolid configurations in dipolar quantum gases, including the controlled excitation of the second sound mode. Specifically, we considered $^{164}$Dy atoms 
in quasi-1D double-well potentials. 
The central barrier splits the gas into two fragments whose interference 
is analyzed following barrier removal.

Supersolids undergo damped oscillatory motion, 
with the damping rate reflecting the degree of superfluid connectivity. This response is captured by a damped coupled oscillator model, extending earlier work on classical droplet dynamics~\cite{Mukherjee_classical_2023}. The damping rate, shown to increase with the superfluid fraction, provides a direct measure of phase coherence. In contrast, droplet lattices exhibit undamped, rigid-body motion. 
We further demonstrated that phase imprinting across the barrier 
in supersolids leads to a long-lived stationary dark solitary wave. 
The persistent phase jump offers an additional measure of coherence and should be observable in time-of-flight expansion. After a delay, the solitary wave transfers momentum to the crystals, controllably exciting the second sound mode~\cite{Guo2019,hertkorn_decoupled_2024}.

Further study is needed to connect
the decay rate of supersolid crystal oscillations and the underlying superfluid properties~\cite{Biagioni_measurement_2024}, e.g. by developing alternative effective models, capturing more accurately the crystals motion and their rigidity~\cite{poli_excitations_2024}.
In this context, understanding the role of temperature effects--e.g., by employing suitably modified versions of the eGPE~\cite{sanchez_heating_2023}--may shed light on how thermal fluctuations influence the decay rate and collective drift velocity of the crystal. 
Finally, higher-dimensional generalizations to nucleate vortices beyond magnetostirring~\cite{prasad_instability_2019,Klaus2022,bland_vortices_2023,casotti_observation_2024,poli2024synchronization}, e.g. by exploiting the snake instability for vortex dipole formation~\cite{prep}, or ring dark solitons decaying into vortex necklaces as demonstrated in non-dipolar superfluids~\cite{Theocharis_ring} are desirable.

\paragraph*{\textit {Acknowledgments}.}
S.I.M. and G.A.B. would like to thank J. E. Medvedeva and A. V. Chernatynskiy for fruitful discussions on viscoelastic materials.
S.I.M acknowledges support by the Army Research Office under Award number: W911NF-26-1-A043.
H.R.S. acknowledges support for ITAMP by the NSF.
T.B. thanks the Knut and Alice Wallenberg Foundation (GrantNo. KAW 2018.0217) and the Swedish Research Council (Grant No. 2022-03654 vr).

\putbib[Dipolar_Gases]

\end{bibunit}

\clearpage

\begin{bibunit}[apsrev4-1]


\onecolumngrid
\setcounter{equation}{0}
\setcounter{figure}{0}
\setcounter{section}{0}
\makeatletter
\renewcommand{\theequation}{S\arabic{equation}}
\renewcommand{\thefigure}{S\arabic{figure}}
\renewcommand{\bibnumfmt}[1]{[S#1]}
\renewcommand{\citenumfont}[1]{S#1}
\renewcommand{\thesection}{\arabic{section}}
\setcounter{page}1
\def\thepage{S\arabic{page}}

\begin{center}
	{\Large\bfseries Supplementary Material: Signatures of rigidity and second sound in dipolar supersolids \\ 
 }
\end{center}

\section{Extended Gross-Pitaevskii equation}

To describe the stationary and dynamical properties of dipolar gases, the following extended Gross-Pitaevskii equation~\cite{Santos2016filemanets,bisset_ground-state_2016,chomaz2016quantum,ferrier-barbut_observation_2016} is employed,
\begin{equation}
i \hbar \partial_t \Psi(\boldsymbol{r},t) = \Bigg [ - \frac{\hbar^2}{2m} \nabla^2 +V(\boldsymbol{r})  + \frac{4\pi \hbar^2 a}{m} \abs{\Psi(\boldsymbol{r},t)}^2  
+ \int \text{d}^3\boldsymbol{r'}~ U_\text{dd}(\boldsymbol{r} -\boldsymbol{r'})\abs{\Psi(\boldsymbol{r'},t)}^2  
+  \gamma (\epsilon_\text{dd}) \abs{\Psi(\boldsymbol{r},t)}^3 \Bigg ] \Psi(\boldsymbol{r},t), 
\label{Eq:MF_equation}
\end{equation}
where $\Psi(\boldsymbol{r},t)$ corresponds to the 3D wavefunction. 
In addition to the kinetic energy (first term on the right-hand side of Eq.~(\ref{Eq:MF_equation})) and the external trapping potential (second term), the behavior of the dipolar gas is governed by the interplay between short-range (third term) and long-range (fourth term) interactions. The short-range interaction is characterized by the $s$-wave scattering length $a$.
The dipolar long-range interaction potential,
\begin{equation}
U_\text{dd}(\boldsymbol{r}) = \frac{3\hbar^2 a_\text{dd}}{m}\left[\frac{1-3\cos^2\theta}{\boldsymbol{r}^3}\right],
\label{Eq:Dipolar_interactions}
\end{equation}
is anisotropic, as evidenced by the angle $\theta$ between the line connecting two dipoles and the polarization axis.
Moreover, in our study, the dipolar length is fixed at $a_\text{dd} = 131~a_0$ for $^{164}$Dy, where $a_0$ denotes the Bohr radius.

Importantly, $U_\text{dd}(\boldsymbol{r})$ can turn attractive as well, leading to a predicted collapse of the dipolar gas upon increasing the interaction ratio $\epsilon_\text{dd} = a_\text{dd}/a$~\cite{Parker_structure_2009,Bohn_collapse_2009}, a behavior that is in contrast to experimental observations~\cite{kadau_observing_2016,chomaz_dipolar_2022}. This can be theoretically circumvented by incorporating  the first order LHY correction (fifth term in Eq.~(\ref{Eq:MF_equation})) to the mean-field energy functional of the dipolar gas within the local density approximation. 
The contribution of this quantum correction is repulsive and arrests the wave collapse for large $\epsilon_\text{dd}$, leading to a more accurate description of the dipolar gas in this regime~\cite{ferrier-barbut_observation_2016}.
The associated coefficient to this term is~\cite{Lima_quantum_2011,schutzhold_mean_2006} $\gamma(\epsilon_\text{dd})=\frac{128 \hbar^2\sqrt{\pi}a^{5/2}}{3m}  \left(1+\frac{3}{2} \epsilon^2_\text{dd}\right)$.
The inclusion of this term in conjunction with the competition among (repulsive) short-range and (attractive or repulsive) long-range interactions facilitates the gradual formation of the supersolid and droplet lattice arrangements. 
This occurs for decreasing short-range scattering lengths where the relative strength of the long-range dipolar 
interaction becomes dominant.

\section{Solitary wave emission and phase  imprinting in the superfluid}

\begin{figure}[h!]
\centering
\includegraphics[width=1\columnwidth]{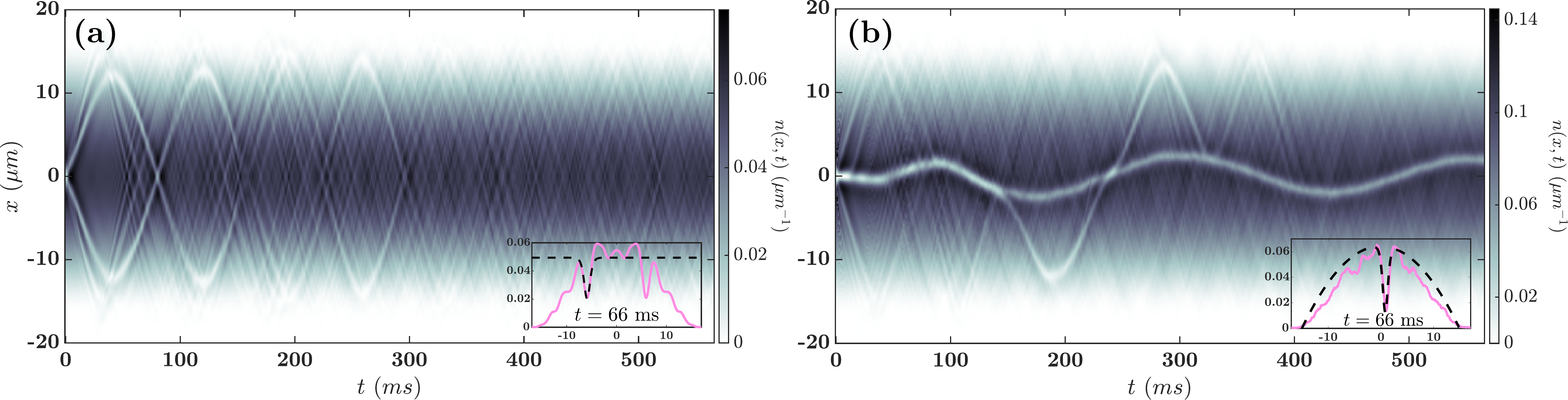}
\caption{(a) The barrier removal in the dipolar superfluid ($\epsilon_{\rm{dd}} = 1$) generates gray solitary waves; the inset presents a fit (dashed line) of the gray soliton waveform [Eq.~(\ref{Eq:Gray_soliton})] to the eGPE density at $t = 66~{\rm ms}$. (b) Superfluid dynamics following a phase ($\pi$-discontinuity) quench and barrier release, resulting in the generation of an oscillating gray solitary wave within the superfluid background. Inset: density snapshot with a fit (dashed line) to the gray soliton waveform [Eq.~(\ref{Eq:Gray_soliton})] atop a Thomas-Fermi profile.}
    \label{Fig:Superfluid}
\end{figure}

In the superfluid phase ($\epsilon_\text{dd}=1$), a sudden ramp-down of the central barrier at $t=0$ induces destructive interference, as seen in the integrated density evolution $n(x,t) = \int \text{d}y\,\text{d}z\,|\Psi(\boldsymbol{r},t)|^2$ [Fig.~\ref{Fig:Superfluid}(a)]. This process spontaneously generates a pair of counter-propagating gray solitary waves and sound waves at the trap center. To confirm the solitonic nature of the resulting density notches, we fit the profiles at selected times using the standard gray soliton waveform~\cite{Kivshar_dark_1998,Frantzeskakis_dark_2010} ,
 \begin{equation}
 \Phi(x) = B \tanh \left[ D(x - x_0) \right] + iA,
 \label{Eq:Gray_soliton}
 \end{equation}
where $B$ is the background density, $D$ the inverse soliton width, $x_0$ the soliton center, and $A$ determines the velocity. Dark solitons are commonly called black when they are stationary, and gray otherwise. As shown in the inset of Fig.~\ref{Fig:Superfluid}(a) for $t=66$~ms, the fit agrees well, and a small phase jump of approximately $0.2\pi$ across the solitary wave cores confirms their motion. Previous studies have predicted that dipolar solitons should exhibit density oscillations around their cores~\cite{pawlowski2015dipolar,bland2015controllable,Bland_interaction_2017}, although such features are difficult to resolve here amidst background noise.

Each solitary wave travels toward the trap edges, reflects, and undergoes inelastic collisions with its counterpart over several cycles. As evolution progresses, they gradually become shallower due to accumulated radiation and eventually dissipate into the background. In the absence of the LHY correction, collisions between dipolar solitons have been shown to be only weakly inelastic~\cite{pawlowski2015dipolar,bland2015controllable,edmonds2016exploring}. However, in the present setting, where the LHY term is included and fast-moving solitary waves are generated via barrier removal, the solitons are short-lived.

The clearest solitary wave formation occurs for a barrier width $W_0 = 0.5~l_x$, matching the soliton width ($\sim 0.4~l_x$). Wider or narrower barriers suppress this effect. Increasing the barrier height $V_0$ enhances soliton contrast~\cite{Ros_demand_2021}, producing darker solitary waves with larger phase jumps approaching $\pi$, consistent with observations in non-dipolar superfluids~\cite{Weller_experimental_2008}.

The response of the superfluid to a $\Delta \varphi = \pi$ phase imprinting [Fig.~\ref{Fig:Superfluid}(b)] closely resembles that of non-dipolar systems, initially generating a single black solitary wave at the trap center~\cite{denschlag_generating_2000,Becker_oscillations_2008}, that quickly decays into a gray solitary wave through interaction with sound modes and begins oscillating. This solitary wave, fitted using Eq.~\eqref{Eq:Gray_soliton} atop a Thomas-Fermi background, exhibits a phase jump less than $\pi$, consistent with its motion at later times. It oscillates around the trap center with a frequency $\sim 0.45\, \omega_x/\sqrt{2}$, lower than the non-dipolar prediction~\cite{Busch_motion_2000}, as reported for dipolar superfluids in the absence of LHY corrections~\cite{Bland_interaction_2017}.

\begin{figure}[h!]
\centering
\includegraphics[width=1\columnwidth]{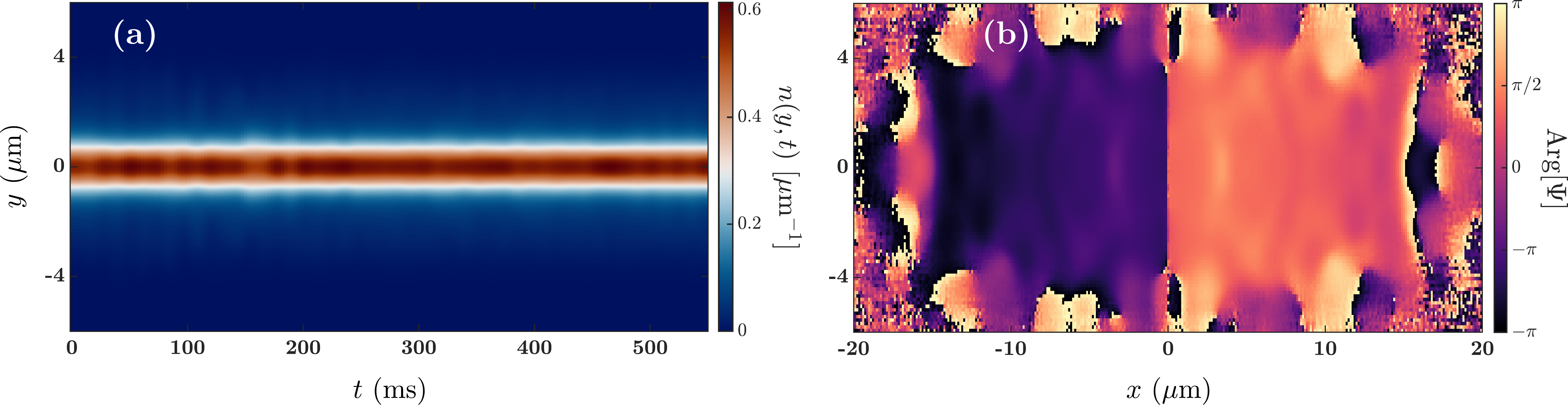}
\caption{(a) Dynamics of the integrated density profile, $n(y,t)$, along the transverse $y$ spatial direction following the barrier release and phase imprinting. An overall small amplitude collective motion occurs. (b) The phase imprinting leads to the nucleation of a dark solitary wave at the trap center, as identified by the $ \approx \pi$ jump of the phase profile of the gas across the $x-y$ plane ($z=0$) at $t=100 ~ \rm{ms}$. In all cases, the relevant system  parameters refer to $\rm{\epsilon_{dd}}=1.36$, $V_0 = 10 ~ \hbar \omega_x$, and $W_0 = 0.5 ~ l_x$.}
\label{Fig:Transverse_dynamics}
\end{figure}

\section{Excitations in the transverse directions} \label{Sec:Transverse_dynamics}

As mentioned in the main text, the dynamical response of the dipolar gas in the directions transverse to its elongated axis majorly corresponds to a collective motion.
To illustrate this phenomenon, we focus on the dynamics of a supersolid, $\rm{\epsilon_{dd}} = 1.36$, following the barrier release along with a phase imprinting [Fig.~\ref{Fig:Transverse_dynamics}(a)].
In the $y$ spatial direction, the cloud undergoes a  small-amplitude collective motion, as demonstrated by the integrated density profile, $n(y,t) = \int dx dz ~ \abs{\Psi(\boldsymbol{r},t)}^2$ depicted in Fig.~\ref{Fig:Transverse_dynamics}(a).
A similar response occurs also for the other transverse direction, not shown for brevity.
Moreover, in the absence of phase imprinting the transverse dynamics is similar to the one presented in Fig.~\ref{Fig:Transverse_dynamics}(a), i.e. only collective excitations occur.

Due to the introduced $\pi$ phase jump at $t=0$, a dark solitary wave forms at the trap center [Fig.~5(a) in the main text], persisting up to long evolution times, before eventually drifitng and exciting the second sound.
In the $x-y$ plane, the solitary wave features a $\simeq \pi$ phase jump stripe at $x=0$, see Fig.~\ref{Fig:Transverse_dynamics}
(b) for a time snapshot of the phase at $z=0$ and $t = 100 ~ \rm{ms}$.
The stripe persists until the solitary wave eventually drifts away without any sign of vortex creation along $x=0$.
Only a small number of vortices are nucleated [identified by the $2\pi$ circulation in Fig.~\ref{Fig:Transverse_dynamics}(b)], which appear however at the outskirts of the dipolar supersolid.
The vortices always remain at the periphery of the cloud without interfering with the supersolid crystals, even at longer evolution times.

\section{Rigid dynamics for large particle numbers} 
\label{Sec:Large_particles}

\begin{figure}[t!]
\centering
\includegraphics[width=1\columnwidth]{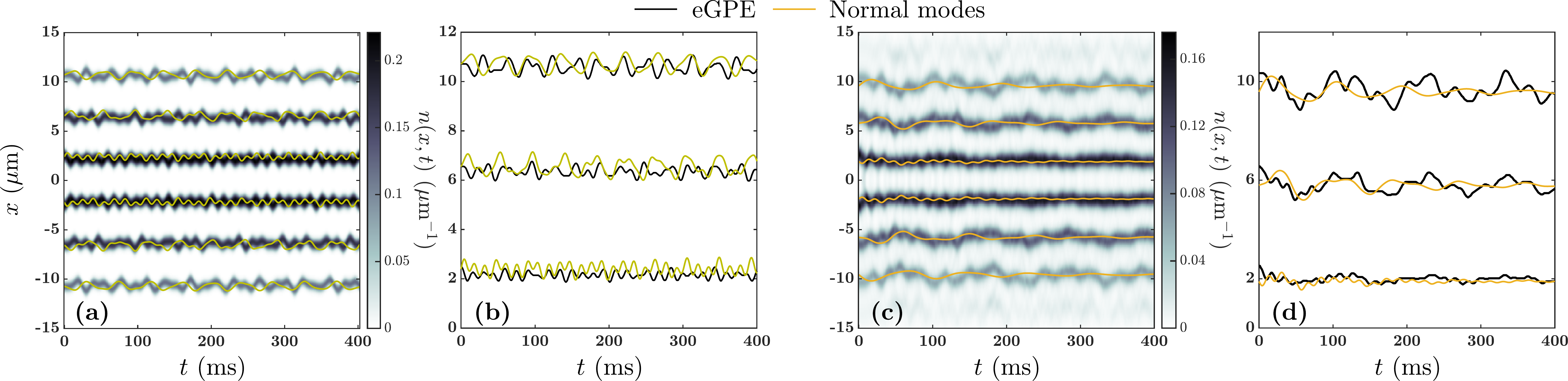}
\caption{Comparison of the damped coupled oscillators model with the dynamical response of (a), (b) isolated droplets featuring $\rm{\epsilon_{dd}}=1.48$ and (c), (d) supersolids characterized by $\rm{\epsilon_{dd}}=1.36$ for large particle number, $N= 16 \times 10^4$. In panels (b), (d) the positions of the droplet peaks (solid black lines) are directly compared with the results of the linear chain model (yellow solid lines). A better qualitative agreement occurs in the case of isolated droplets. All other parameters are the same as in Fig.~3 in the main text.}
\label{Fig:Multiple_peaks}
\end{figure}

To elaborate on the general applicability of the coupled oscillator model used in the main text, we next tackle the barrier release dynamics of droplets and supersolids with a large particle number, $N=16\times 10^4$.
For such configurations, six droplet peaks can be accommodated in these phases [e.g. Fig.~\ref{Fig:Multiple_peaks}(a)], in contrast to the four presented in the main text.
Overall, the same dynamical behavior is observed as for smaller particle number, as inferred by the integrated density $n(x,t)$ evolution shown in Fig.~\ref{Fig:Multiple_peaks}(a), (c).
Namely, small-amplitude oscillations occur in the case of the isolated droplets [Fig.~\ref{Fig:Multiple_peaks}(a)], while the droplet peaks in supersolids undergo damped oscillations, retaining their rigid character [Fig.~\ref{Fig:Multiple_peaks}(c)].
In order to reproduce such a dynamical response, the coupled oscillators model [Eq.~(2) in the main text] is extended so that it takes into account six coupled oscillators. 
Explicitly,
\begin{equation}
\begin{pmatrix}
M^{(1)}_\text{O}\ddot X_1 \\ M^{(2)}_\text{O} \ddot X_2 \\  M_\text{I} \ddot X_3 \\ M_\text{I} \ddot X_4  \\
M^{(2)}_\text{O} \ddot X_5 \\
M^{(1)}_\text{O} \ddot X_6
\end{pmatrix} =
\begin{pmatrix}
-\lambda -k & k & 0 & 0 & 0 & 0 \\
k & -2k & k & 0 & 0 & 0 \\
0 & k & -2k & k & 0 & 0 \\
0 & 0 & k &  -2k & k  & 0  \\
0 & 0 & 0  & k & -2k & k \\
0 & 0 & 0 & 0 & k & -\lambda -k
\end{pmatrix}
\begin{pmatrix}
\Delta X_1 \\ \Delta X_2 \\ \Delta X_3 \\ \Delta X_4
 \\ \Delta X_5 \\ \Delta X_6
\end{pmatrix} 
-\begin{pmatrix}
\gamma^{(1)}_{\rm{O}} & 0 & 0 & 0 & 0 & 0 \\
0 & \gamma^{(2)}_{\rm{O}} & 0 & 0 & 0 & 0 \\
0 & 0 & \gamma_{\rm{I}} & 0 & 0 & 0 \\
0 & 0 & 0 & \gamma_{\rm{I}} & 0 & 0 \\
0 & 0 & 0 & 0 &  \gamma^{(1)}_{\rm{O}} & 0 \\
0 & 0 & 0 & 0 & 0 & \gamma^{(1)}_{\rm{O}}
\end{pmatrix}
\begin{pmatrix}
\dot X_1 \\ \dot X_2 \\ \dot X_3 \\ \dot X_4 \\ \dot X_5 \\ \dot X_6
\end{pmatrix}\,.
\label{Eq:Springs_double_particles}
\end{equation}
Similarly to the analysis in the main text, $k$, $\lambda$ are the spring constants, while $M^{(1)}_{\rm{O}}$, $M^{(2)}_{\rm{O}}$, $M_{\rm{I}}$ correspond to the masses of the two outer and inner droplet peaks respectively.
In this extended model five parameter are introduced, i.e. four natural frequencies, $\sqrt{\lambda/M^{(1)}_{\rm{O}}}$, $\sqrt{k/M^{(1)}_{\rm{O}}}$, $\sqrt{k/M^{(2)}_{\rm{O}}}$, $\sqrt{k/M_{\rm{I}}}$, and one damping rate $\Gamma/\hbar = \gamma^{(1)}_{\rm{O}}/(2M^{(1)}_{\rm{O}}) = \gamma^{(2)}_{\rm{O}}/(2M^{(2)}_{\rm{O}}) = \gamma_{\rm{I}}/(2M_{\rm{I}}) $.

In spite of the large number of model parameters, the droplet peak positions resulting from Eq.~\eqref{Eq:Springs_double_particles} are on a good qualitative agreement with the dynamics stemming from the eGPE [Fig.~\ref{Fig:Multiple_peaks}(b), (d)].
Similarly to the case of smaller particle number, the linear-chain model is particularly suited for capturing the dynamics of isolated droplets [Fig.~\ref{Fig:Multiple_peaks}(b)].
For the supersolid phase, the superfluid substrate becomes significantly perturbed due to the interference stemming from the initially segregated density segments.
This behavior leads to an abrupt rearrangement of the droplet peaks during the initial time instants, as well as to the decay of their oscillations at late times.
A crucial difference with the dynamics of four droplet peaks ($N=8\times 10^4$) [Fig.~3(b) in the main text] is that the decay of the oscillations amplitude is smaller in Fig.~\ref{Fig:Multiple_peaks}(c).
For large particle numbers, the superfluid background is highly suppressed due to the more pronounced crystal component.
This observation further corroborates the connection between the decay and the superfluid substrate presented in Fig.~4 in the main text.

\section{Oscillators model with next-to-nearest neighbor couplings} \label{Sec:NNN}

\begin{figure}[t!]
\centering
\includegraphics[width=0.8\columnwidth]{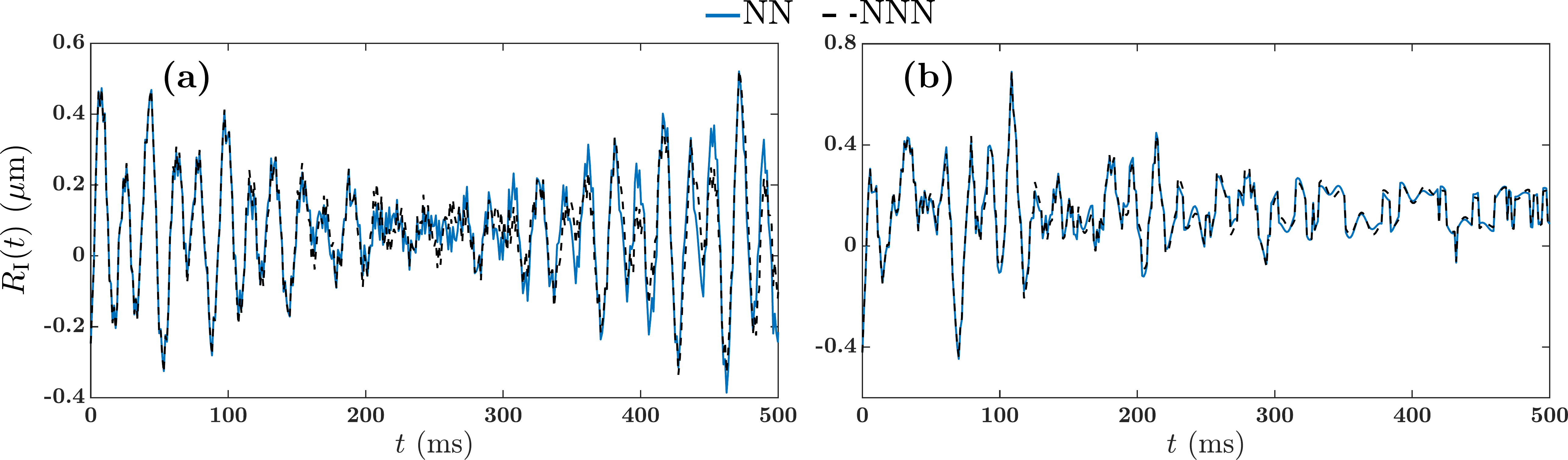}
\caption{Residuals between the inner droplet peak positions stemming from the coupled oscillators model ($X_{\rm{I}}(t)$), and the eGPE simulations ($\bar{X}_{\rm{I}}(t)$), $R_{\rm{I}}(t) = X_{\rm{I}}(t) - \bar{X}_{\rm{I}}(t)$ in the (a) isolated droplet with $\rm{\epsilon_{dd}=1.48}$ and (b) supersolid featuring $\rm{\epsilon_{dd}}=1.36$ phase. The accuracy of the model remains the same regardless of whether we consider nearest neighbor (NN) or next-to-nearest neighbor (NNN) couplings.}
\label{Fig:Residuals}
\end{figure}

We have shown that the dynamical behavior of the individual droplet peaks after the barrier removal can be qualitatively captured by a damped coupled oscillators model [Eq.~(2) in the main text].
In this approach, every droplet peak is regarded as a massive particle that is directly linked to its nearest neighbors by a spring, i.e. it is a nearest neighbor (NN) model.
However, dipolar gases are characterized by long-range interactions that might affect the dynamical behavior of the individual droplet peaks after the barrier removal beyond the nearest neighbor approximation.
To explore this possibility, we extend our damped coupled oscillators model by including next-to-nearest neighbor (NNN) harmonic couplings.
Explicitly, the NNN model takes the following form,
\begin{gather}
\begin{pmatrix}
M_\text{O}\ddot X_1 \\ M_\text{I} \ddot X_2 \\  M_\text{I} \ddot X_3 \\ M_\text{O} \ddot X_4 
\end{pmatrix} =
\begin{pmatrix}
-\lambda -k & k & \nu & 0 \\
k & -2k & k & \nu \\
\nu & k & -2k & k \\
0 & \nu & k &  -\lambda -k 
\end{pmatrix}
\begin{pmatrix}
\Delta X_1 \\ \Delta X_2 \\ \Delta X_3 \\ \Delta X_4
\end{pmatrix}
-\begin{pmatrix}
\gamma_{\rm{O}} & 0 & 0 & 0 \\
0 & \gamma_{\rm{I}} & 0 & 0 \\
0 & 0 & \gamma_{\rm{I}} & 0 \\
0 & 0 & 0 & \gamma_{\rm{O}}
\end{pmatrix}
\begin{pmatrix}
\dot X_1 \\ \dot X_2 \\ \dot X_3 \\ \dot X_4
\end{pmatrix}\,.
\label{Eq:NNN_model}
\end{gather}
The crucial difference with the Newtonian system presented in Eq.~(2) in the main text is that an additional spring constant, $\nu$, is introduced.
The linear couplings matrix [first matrix in Eq.~\eqref{Eq:NNN_model}] becomes now pentadiagonal, reflecting the beyond nearest neighbor couplings.

To assess the accuracy of the NNN model with respect to the NN model, we employ the residual measure $R_i(t) = X_i(t)- \bar{X}_i(t), ~ i=1,\ldots,4$.
The latter quantity directly compares the droplet peak positions stemming from the eGPE calculations [$\bar{X}_i(t)$] and the coupled oscillators model [$X_i(t)$].
We find that both the NN and the NNN models behave similarly, regardless of the phase of the dipolar gas, see Fig.~\ref{Fig:Residuals} for the residuals of the inner droplets in the droplet [Fig.~\ref{Fig:Residuals}(a)] and the supersolid [Fig.~\ref{Fig:Residuals}(b)] states. 
The same also applies to the residuals pertaining to the outer droplet peaks (not shown for brevity).
By virtue of the Occam's razor, we deem that the NN damped coupled oscillators model yields an adequate description of the crystals dynamics induced by the  barrier removal. 
Nevertheless, a more sophisticated model is necessary for quantitatively capturing the dynamics of the droplet peaks in the supersolid states.

Table~\ref{Tab:Fit_parameters} presents the fitted velocities and natural frequencies for both the NN and the NNN models, exemplarily, in the isolated droplets phase, $\epsilon_{\rm{dd}}=1.48$. These parameters are estimated through a nonlinear least-squares analysis, minimizing the residual measures. Due to the large number of fitting parameters, such an optimization procedure crucially depends on the initial guesses for the natural frequencies and velocities.
The latter are carefully chosen, relying on the normal mode analysis of the coupled oscillators model~\cite{Mukherjee_classical_2023}.

\begin{table}[t!]
\centering
\setlength{\tabcolsep}{12pt}
\renewcommand{\arraystretch}{1.5}
\begin{tabular}{|c|c|c|} \hline
\rm{Parameters} & \rm{NN model [Eq.~(2)]} & \rm{NNN model [Eq.~(\ref{Eq:NNN_model})]}   \\  \hline \hline
$\dot{X}_1(0)~(l_x \omega_x)$ &  $-0.3457 \pm 0.0164$  &  $-0.3009 \pm 0.0191$  \\ \hline
$\dot{X}_2(0)~(l_x\omega_x)$ &  $0.1124 \pm 0.0189$ &  $0.0191 \pm 0.0242$  \\  \hline
$\sqrt{k/M_{\rm{O}}}/\omega_x$ & $1.5195 \pm 0.0025$ & $1.3882 \pm 0.0306$ \\ \hline
$\sqrt{k/M_{\rm{I}}}/\omega_x$ & $1.4711\pm 0.0021$ &  $1.5554 \pm 0.01082$  \\ \hline
$\sqrt{\nu/M_{\rm{O}}}/\omega_x$ & - & $0.6561\pm 0.1081$  \\ \hline
\end{tabular}
\caption{
Fitted velocities and natural frequencies (first column) for the coupled oscillators model with either NN (second column) or both NN and NNN (third column) couplings in the case of isolated droplets ($\epsilon_{\rm{dd}}=1.48$). In all models the parameters are estimated by employing nonlinear least-squares, while the standard deviation of each fitting parameter is also provided. Note that $\sqrt{\nu/M_{\rm{I}}}$ in the NNN model is similar to the other natural frequency, $\sqrt{\nu/M_{\rm{O}}}$ (not shown).}
\label{Tab:Fit_parameters}
\end{table}

\section{Tunneling rates}  \label{Sec:Tunneling}

\begin{figure}[t!]
\centering
\includegraphics[width=0.6\columnwidth]{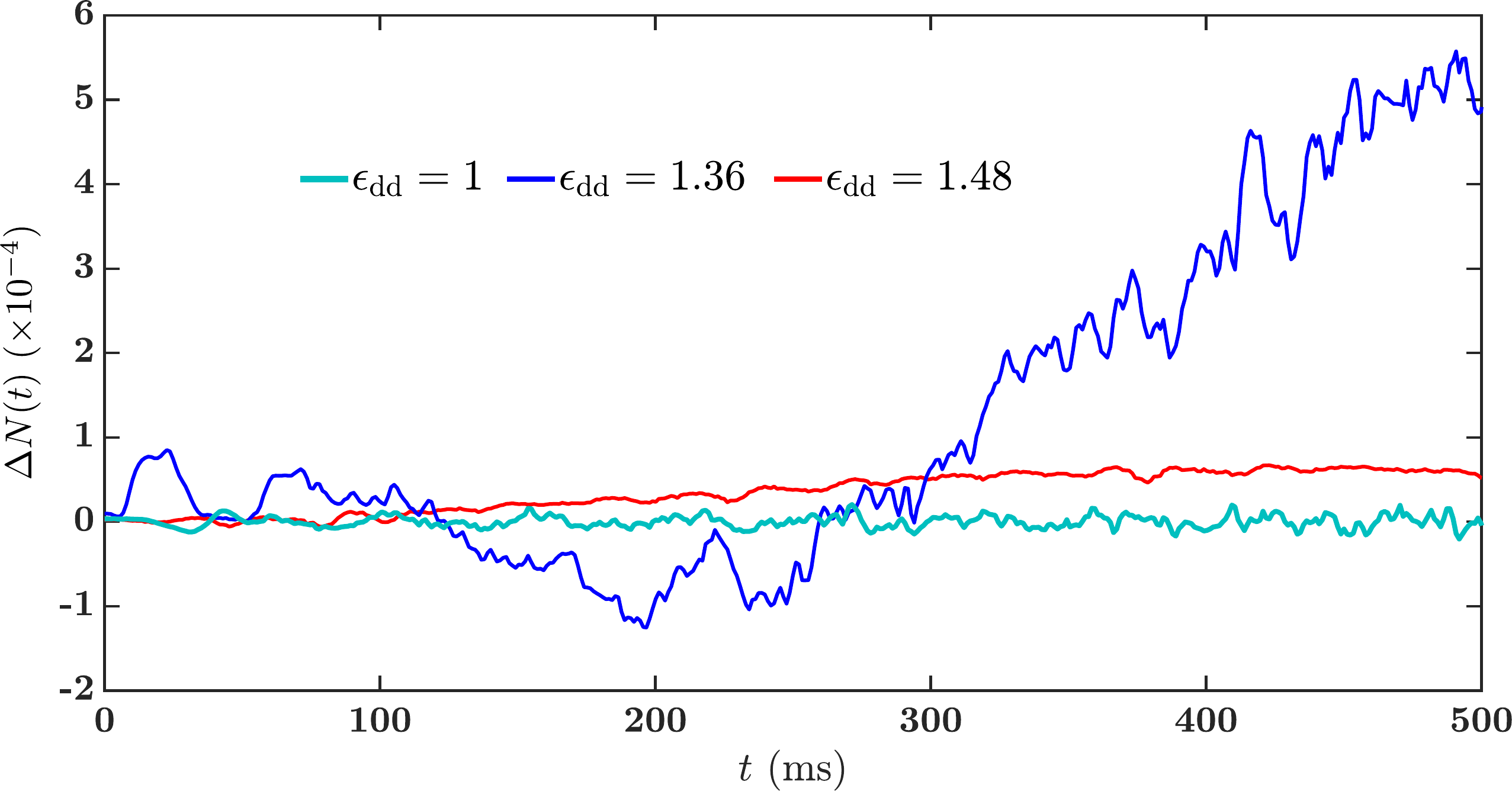}
\caption{Relative population imbalance $\Delta N$ (see also the text) between the two wells in the case of superfluids, supesolids, and droplets (see legend). Due to the parity symmetry of the potential barrier, tunneling between the two wells is suppressed.}
\label{Fig:Tunneling}
\end{figure}

The dipolar gas is initialized in a double-well potential along its elongated axis for each of its three phases — superfluid, supersolid, and isolated droplets [see also Fig. 2 in the main text].
Such an external potential is parity symmetric, i.e. no tilt occurs between the left and right potential wells [see also Eq.~(1) in the main text].
As such, no tunneling is expected to occur between the left and right segments of the dipolar gas after the barrier removal.
To demonstrate this behavior, we investigate the population imbalance between the left and right halves of the potential, i.e., $\Delta N(t) = \frac{N_L(t)-N_R(t)}{N}$~\cite{Mistakidis2024}.
Here,
$N_L(t) = \int_{-\infty}^0 dx ~ n(x,t)$, $N_R(t) = \int_0^{\infty} dx ~ n(x,t)$, and $n(x,t) = \int dy dz ~ |\Psi(\boldsymbol{r},t)|^2$ is the 1D integrated density, while $N$ is the total atom number.
For all dipolar gas phases considered herein, $\Delta N(t)$ remains small during the entire evolution time [Fig.~\ref{Fig:Tunneling}], suggesting the absence of any appreciable tunneling among the left and right sides of the original double-well potential.

\section{Simulation details}\label{Ap:Num}

To numerically address the ground state and nonequilibrium dynamics of the 3D dipolar gas, we recast the eGPE given in Eq.~\eqref{Eq:MF_equation} into a dimensionless form. 
This is achieved by expressing spatial and temporal scales in units of the harmonic oscillator length $l_x$ and the inverse trap frequency $\omega_x^{-1}$, respectively.

A uniform cubic spatial grid of $1024 \times 128 \times 128$ 
points is used to discretize the quasi-1D 
domain, with discretization steps $\delta x = \delta y = 0.07~l_x$, $\delta z = 0.2~l_x$. 
The time step is set to $\delta t = 10^{-4} \omega_x^{-1}$, ensuring that $(\omega_x \delta t)^2 < \delta x \delta y / l_x^2$. 
This condition guarantees conservation of both particle number and energy throughout the real-time evolution. 
Time propagation, both in imaginary time (for finding the ground state) and real time (for quench dynamics), is performed using the Crank–Nicolson method~\cite{crank_practical_1947,antoine_computational_2013}.

A key numerical challenge involves handling the dipolar interaction term in the eGPE. 
This term diverges as $r^{-3}$ at small interparticle distances, requiring regularization. 
We overcome this using the convolution theorem~\cite{arfken_mathematical_1972,Goral_ground_2002}, which allows the dipolar interaction to be computed as
\begin{align}
\int \text{d}^3\boldsymbol{r}' ~ U_\text{dd}(\boldsymbol{r} - \boldsymbol{r}') \abs{\Psi(\boldsymbol{r}',t)}^2 
= \mathcal{F}^{-1} \Big[ \mathcal{F}[U_\text{dd}] \cdot \mathcal{F}[\abs{\Psi}^2] \Big],
\end{align}
where $\mathcal{F}$ ($\mathcal{F}^{-1}$) denotes the (inverse) Fourier transform. 
The advantage is that the Fourier transform of $U_\text{dd}$ is regular~\cite{Goral_ground_2002}, making this term numerically well-defined and stable.

To compute the ground state, a suitable initial wavefunction ansatz is essential for identifying the lowest-energy configuration based on symmetry considerations. 
In the considered quasi-1D geometry, we use
\begin{align}
\Psi(x,y,z) = \mathcal{A} \, e^{-\left( \frac{x^2}{2l_x^2} + \frac{y^2}{2l_y^2} + \frac{z^2}{2l_z^2} \right)} \sin^2\left(\frac{kx}{l_x}\right),
\end{align}
where $\mathcal{A}$ is a normalization constant, and the parameter $k$ is varied to identify the (lowest in energy) ground state configuration. 
For the ground state, numerical convergence of the order of $10^{-4}$ and $10^{-8}$ is ensured at the wavefunction and energy levels respectively. 
Turning to the quench dynamics, the total particle number and energy are numerically conserved within the order of $10^{-6}$ throughout the real time  evolution.

\putbib[Dipolar_Gases]

\end{bibunit}


\begin{thebibliography}{97}%
\makeatletter
\providecommand \@ifxundefined [1]{%
 \@ifx{#1\undefined}
}%
\providecommand \@ifnum [1]{%
 \ifnum #1\expandafter \@firstoftwo
 \else \expandafter \@secondoftwo
 \fi
}%
\providecommand \@ifx [1]{%
 \ifx #1\expandafter \@firstoftwo
 \else \expandafter \@secondoftwo
 \fi
}%
\providecommand \natexlab [1]{#1}%
\providecommand \enquote  [1]{``#1''}%
\providecommand \bibnamefont  [1]{#1}%
\providecommand \bibfnamefont [1]{#1}%
\providecommand \citenamefont [1]{#1}%
\providecommand \href@noop [0]{\@secondoftwo}%
\providecommand \href [0]{\begingroup \@sanitize@url \@href}%
\providecommand \@href[1]{\@@startlink{#1}\@@href}%
\providecommand \@@href[1]{\endgroup#1\@@endlink}%
\providecommand \@sanitize@url [0]{\catcode `\\12\catcode `\$12\catcode `\&12\catcode `\#12\catcode `\^12\catcode `\_12\catcode `\%12\relax}%
\providecommand \@@startlink[1]{}%
\providecommand \@@endlink[0]{}%
\providecommand \url  [0]{\begingroup\@sanitize@url \@url }%
\providecommand \@url [1]{\endgroup\@href {#1}{\urlprefix }}%
\providecommand \urlprefix  [0]{URL }%
\providecommand \Eprint [0]{\href }%
\providecommand \doibase [0]{http://dx.doi.org/}%
\providecommand \selectlanguage [0]{\@gobble}%
\providecommand \bibinfo  [0]{\@secondoftwo}%
\providecommand \bibfield  [0]{\@secondoftwo}%
\providecommand \translation [1]{[#1]}%
\providecommand \BibitemOpen [0]{}%
\providecommand \bibitemStop [0]{}%
\providecommand \bibitemNoStop [0]{.\EOS\space}%
\providecommand \EOS [0]{\spacefactor3000\relax}%
\providecommand \BibitemShut  [1]{\csname bibitem#1\endcsname}%
\let\auto@bib@innerbib\@empty
\bibitem [{\citenamefont {Leggett}(1970)}]{leggett_can_1970}%
  \BibitemOpen
  \bibfield  {author} {\bibinfo {author} {\bibfnamefont {A.~J.}\ \bibnamefont {Leggett}},\ }\href {\doibase 10.1103/PhysRevLett.25.1543} {\bibfield  {journal} {\bibinfo  {journal} {Phys. Rev. Lett.}\ }\textbf {\bibinfo {volume} {25}},\ \bibinfo {pages} {1543} (\bibinfo {year} {1970})}\BibitemShut {NoStop}%
\bibitem [{\citenamefont {Chester}(1970)}]{Chester1970}%
  \BibitemOpen
  \bibfield  {author} {\bibinfo {author} {\bibfnamefont {G.~V.}\ \bibnamefont {Chester}},\ }\href {\doibase 10.1103/PhysRevA.2.256} {\bibfield  {journal} {\bibinfo  {journal} {Phys. Rev. A}\ }\textbf {\bibinfo {volume} {2}},\ \bibinfo {pages} {256} (\bibinfo {year} {1970})}\BibitemShut {NoStop}%
\bibitem [{\citenamefont {Chan}\ \emph {et~al.}(2013)\citenamefont {Chan}, \citenamefont {Hallock},\ and\ \citenamefont {Reatto}}]{chan_overview_2013}%
  \BibitemOpen
  \bibfield  {author} {\bibinfo {author} {\bibfnamefont {M.~H.-W.}\ \bibnamefont {Chan}}, \bibinfo {author} {\bibfnamefont {R.}~\bibnamefont {Hallock}}, \ and\ \bibinfo {author} {\bibfnamefont {L.}~\bibnamefont {Reatto}},\ }\href {\doibase 10.1007/s10909-013-0882-x} {\bibfield  {journal} {\bibinfo  {journal} {J. Low Temp. Phys.}\ }\textbf {\bibinfo {volume} {172}},\ \bibinfo {pages} {317} (\bibinfo {year} {2013})}\BibitemShut {NoStop}%
\bibitem [{\citenamefont {Balibar}(2010)}]{balibar_enigma_2010}%
  \BibitemOpen
  \bibfield  {author} {\bibinfo {author} {\bibfnamefont {S.}~\bibnamefont {Balibar}},\ }\href {\doibase 10.1038/nature08913} {\bibfield  {journal} {\bibinfo  {journal} {Nature}\ }\textbf {\bibinfo {volume} {464}},\ \bibinfo {pages} {176} (\bibinfo {year} {2010})}\BibitemShut {NoStop}%
\bibitem [{\citenamefont {B\"ottcher}\ \emph {et~al.}(2019)\citenamefont {B\"ottcher}, \citenamefont {Schmidt}, \citenamefont {Wenzel}, \citenamefont {Hertkorn}, \citenamefont {Guo}, \citenamefont {Langen},\ and\ \citenamefont {Pfau}}]{bottcher2019transient}%
  \BibitemOpen
  \bibfield  {author} {\bibinfo {author} {\bibfnamefont {F.}~\bibnamefont {B\"ottcher}}, \bibinfo {author} {\bibfnamefont {J.-N.}\ \bibnamefont {Schmidt}}, \bibinfo {author} {\bibfnamefont {M.}~\bibnamefont {Wenzel}}, \bibinfo {author} {\bibfnamefont {J.}~\bibnamefont {Hertkorn}}, \bibinfo {author} {\bibfnamefont {M.}~\bibnamefont {Guo}}, \bibinfo {author} {\bibfnamefont {T.}~\bibnamefont {Langen}}, \ and\ \bibinfo {author} {\bibfnamefont {T.}~\bibnamefont {Pfau}},\ }\href {\doibase 10.1103/PhysRevX.9.011051} {\bibfield  {journal} {\bibinfo  {journal} {Phys. Rev. X}\ }\textbf {\bibinfo {volume} {9}},\ \bibinfo {pages} {011051} (\bibinfo {year} {2019})}\BibitemShut {NoStop}%
\bibitem [{\citenamefont {Chomaz}\ \emph {et~al.}(2019)\citenamefont {Chomaz}, \citenamefont {Petter}, \citenamefont {Ilzh\"ofer}, \citenamefont {Natale}, \citenamefont {Trautmann}, \citenamefont {Politi}, \citenamefont {Durastante}, \citenamefont {van Bijnen}, \citenamefont {Patscheider}, \citenamefont {Sohmen}, \citenamefont {Mark},\ and\ \citenamefont {Ferlaino}}]{chomaz2019long}%
  \BibitemOpen
  \bibfield  {author} {\bibinfo {author} {\bibfnamefont {L.}~\bibnamefont {Chomaz}}, \bibinfo {author} {\bibfnamefont {D.}~\bibnamefont {Petter}}, \bibinfo {author} {\bibfnamefont {P.}~\bibnamefont {Ilzh\"ofer}}, \bibinfo {author} {\bibfnamefont {G.}~\bibnamefont {Natale}}, \bibinfo {author} {\bibfnamefont {A.}~\bibnamefont {Trautmann}}, \bibinfo {author} {\bibfnamefont {C.}~\bibnamefont {Politi}}, \bibinfo {author} {\bibfnamefont {G.}~\bibnamefont {Durastante}}, \bibinfo {author} {\bibfnamefont {R.~M.~W.}\ \bibnamefont {van Bijnen}}, \bibinfo {author} {\bibfnamefont {A.}~\bibnamefont {Patscheider}}, \bibinfo {author} {\bibfnamefont {M.}~\bibnamefont {Sohmen}}, \bibinfo {author} {\bibfnamefont {M.~J.}\ \bibnamefont {Mark}}, \ and\ \bibinfo {author} {\bibfnamefont {F.}~\bibnamefont {Ferlaino}},\ }\href {\doibase 10.1103/PhysRevX.9.021012} {\bibfield  {journal} {\bibinfo  {journal} {Phys. Rev. X}\ }\textbf {\bibinfo {volume} {9}},\ \bibinfo {pages} {021012} (\bibinfo {year} {2019})}\BibitemShut {NoStop}%
\bibitem [{\citenamefont {Tanzi}\ \emph {et~al.}(2019{\natexlab{a}})\citenamefont {Tanzi}, \citenamefont {Lucioni}, \citenamefont {Fam\`a}, \citenamefont {Catani}, \citenamefont {Fioretti}, \citenamefont {Gabbanini}, \citenamefont {Bisset}, \citenamefont {Santos},\ and\ \citenamefont {Modugno}}]{Tanzi_observation_2019}%
  \BibitemOpen
  \bibfield  {author} {\bibinfo {author} {\bibfnamefont {L.}~\bibnamefont {Tanzi}}, \bibinfo {author} {\bibfnamefont {E.}~\bibnamefont {Lucioni}}, \bibinfo {author} {\bibfnamefont {F.}~\bibnamefont {Fam\`a}}, \bibinfo {author} {\bibfnamefont {J.}~\bibnamefont {Catani}}, \bibinfo {author} {\bibfnamefont {A.}~\bibnamefont {Fioretti}}, \bibinfo {author} {\bibfnamefont {C.}~\bibnamefont {Gabbanini}}, \bibinfo {author} {\bibfnamefont {R.~N.}\ \bibnamefont {Bisset}}, \bibinfo {author} {\bibfnamefont {L.}~\bibnamefont {Santos}}, \ and\ \bibinfo {author} {\bibfnamefont {G.}~\bibnamefont {Modugno}},\ }\href {\doibase 10.1103/PhysRevLett.122.130405} {\bibfield  {journal} {\bibinfo  {journal} {Phys. Rev. Lett.}\ }\textbf {\bibinfo {volume} {122}},\ \bibinfo {pages} {130405} (\bibinfo {year} {2019}{\natexlab{a}})}\BibitemShut {NoStop}%
\bibitem [{\citenamefont {Norcia}\ \emph {et~al.}(2021)\citenamefont {Norcia}, \citenamefont {Politi}, \citenamefont {Klaus}, \citenamefont {Poli}, \citenamefont {Sohmen}, \citenamefont {Mark}, \citenamefont {Bisset}, \citenamefont {Santos},\ and\ \citenamefont {Ferlaino}}]{norcia_two-dimensional_2021}%
  \BibitemOpen
  \bibfield  {author} {\bibinfo {author} {\bibfnamefont {M.~A.}\ \bibnamefont {Norcia}}, \bibinfo {author} {\bibfnamefont {C.}~\bibnamefont {Politi}}, \bibinfo {author} {\bibfnamefont {L.}~\bibnamefont {Klaus}}, \bibinfo {author} {\bibfnamefont {E.}~\bibnamefont {Poli}}, \bibinfo {author} {\bibfnamefont {M.}~\bibnamefont {Sohmen}}, \bibinfo {author} {\bibfnamefont {M.~J.}\ \bibnamefont {Mark}}, \bibinfo {author} {\bibfnamefont {R.~N.}\ \bibnamefont {Bisset}}, \bibinfo {author} {\bibfnamefont {L.}~\bibnamefont {Santos}}, \ and\ \bibinfo {author} {\bibfnamefont {F.}~\bibnamefont {Ferlaino}},\ }\href {\doibase 10.1038/s41586-021-03725-7} {\bibfield  {journal} {\bibinfo  {journal} {Nature}\ }\textbf {\bibinfo {volume} {596}},\ \bibinfo {pages} {357} (\bibinfo {year} {2021})}\BibitemShut {NoStop}%
\bibitem [{\citenamefont {Recati}\ and\ \citenamefont {Stringari}(2023)}]{recati_supersolidity_2023}%
  \BibitemOpen
  \bibfield  {author} {\bibinfo {author} {\bibfnamefont {A.}~\bibnamefont {Recati}}\ and\ \bibinfo {author} {\bibfnamefont {S.}~\bibnamefont {Stringari}},\ }\href {\doibase 10.1038/s42254-023-00648-2} {\bibfield  {journal} {\bibinfo  {journal} {Nat. Rev. Phys.}\ }\textbf {\bibinfo {volume} {5}},\ \bibinfo {pages} {735} (\bibinfo {year} {2023})}\BibitemShut {NoStop}%
\bibitem [{\citenamefont {Lu}\ \emph {et~al.}(2011)\citenamefont {Lu}, \citenamefont {Burdick}, \citenamefont {Youn},\ and\ \citenamefont {Lev}}]{lu2011strongly}%
  \BibitemOpen
  \bibfield  {author} {\bibinfo {author} {\bibfnamefont {M.}~\bibnamefont {Lu}}, \bibinfo {author} {\bibfnamefont {N.~Q.}\ \bibnamefont {Burdick}}, \bibinfo {author} {\bibfnamefont {S.~H.}\ \bibnamefont {Youn}}, \ and\ \bibinfo {author} {\bibfnamefont {B.~L.}\ \bibnamefont {Lev}},\ }\href {\doibase 10.1103/PhysRevLett.107.190401} {\bibfield  {journal} {\bibinfo  {journal} {Phys. Rev. Lett.}\ }\textbf {\bibinfo {volume} {107}},\ \bibinfo {pages} {190401} (\bibinfo {year} {2011})}\BibitemShut {NoStop}%
\bibitem [{\citenamefont {Aikawa}\ \emph {et~al.}(2012)\citenamefont {Aikawa}, \citenamefont {Frisch}, \citenamefont {Mark}, \citenamefont {Baier}, \citenamefont {Rietzler}, \citenamefont {Grimm},\ and\ \citenamefont {Ferlaino}}]{aikawa2012bose}%
  \BibitemOpen
  \bibfield  {author} {\bibinfo {author} {\bibfnamefont {K.}~\bibnamefont {Aikawa}}, \bibinfo {author} {\bibfnamefont {A.}~\bibnamefont {Frisch}}, \bibinfo {author} {\bibfnamefont {M.}~\bibnamefont {Mark}}, \bibinfo {author} {\bibfnamefont {S.}~\bibnamefont {Baier}}, \bibinfo {author} {\bibfnamefont {A.}~\bibnamefont {Rietzler}}, \bibinfo {author} {\bibfnamefont {R.}~\bibnamefont {Grimm}}, \ and\ \bibinfo {author} {\bibfnamefont {F.}~\bibnamefont {Ferlaino}},\ }\href {\doibase 10.1103/PhysRevLett.108.210401} {\bibfield  {journal} {\bibinfo  {journal} {Phys. Rev. Lett.}\ }\textbf {\bibinfo {volume} {108}},\ \bibinfo {pages} {210401} (\bibinfo {year} {2012})}\BibitemShut {NoStop}%
\bibitem [{\citenamefont {Chomaz}\ \emph {et~al.}(2022)\citenamefont {Chomaz}, \citenamefont {Ferrier-Barbut}, \citenamefont {Ferlaino}, \citenamefont {Laburthe-Tolra}, \citenamefont {Lev},\ and\ \citenamefont {Pfau}}]{chomaz_dipolar_2022}%
  \BibitemOpen
  \bibfield  {author} {\bibinfo {author} {\bibfnamefont {L.}~\bibnamefont {Chomaz}}, \bibinfo {author} {\bibfnamefont {I.}~\bibnamefont {Ferrier-Barbut}}, \bibinfo {author} {\bibfnamefont {F.}~\bibnamefont {Ferlaino}}, \bibinfo {author} {\bibfnamefont {B.}~\bibnamefont {Laburthe-Tolra}}, \bibinfo {author} {\bibfnamefont {B.~L.}\ \bibnamefont {Lev}}, \ and\ \bibinfo {author} {\bibfnamefont {T.}~\bibnamefont {Pfau}},\ }\href {\doibase 10.1088/1361-6633/aca814} {\bibfield  {journal} {\bibinfo  {journal} {Rep. Prog. Phys.}\ }\textbf {\bibinfo {volume} {86}},\ \bibinfo {pages} {026401} (\bibinfo {year} {2022})}\BibitemShut {NoStop}%
\bibitem [{\citenamefont {Lahaye}\ \emph {et~al.}(2009)\citenamefont {Lahaye}, \citenamefont {Menotti}, \citenamefont {Santos}, \citenamefont {Lewenstein},\ and\ \citenamefont {Pfau}}]{lahaye_physics_2009}%
  \BibitemOpen
  \bibfield  {author} {\bibinfo {author} {\bibfnamefont {T.}~\bibnamefont {Lahaye}}, \bibinfo {author} {\bibfnamefont {C.}~\bibnamefont {Menotti}}, \bibinfo {author} {\bibfnamefont {L.}~\bibnamefont {Santos}}, \bibinfo {author} {\bibfnamefont {M.}~\bibnamefont {Lewenstein}}, \ and\ \bibinfo {author} {\bibfnamefont {T.}~\bibnamefont {Pfau}},\ }\href {\doibase 10.1088/0034-4885/72/12/126401} {\bibfield  {journal} {\bibinfo  {journal} {Rep. Prog. Phys.}\ }\textbf {\bibinfo {volume} {72}},\ \bibinfo {pages} {126401} (\bibinfo {year} {2009})}\BibitemShut {NoStop}%
\bibitem [{\citenamefont {Petter}\ \emph {et~al.}(2019)\citenamefont {Petter}, \citenamefont {Natale}, \citenamefont {van Bijnen}, \citenamefont {Patscheider}, \citenamefont {Mark}, \citenamefont {Chomaz},\ and\ \citenamefont {Ferlaino}}]{petter_probing_2019}%
  \BibitemOpen
  \bibfield  {author} {\bibinfo {author} {\bibfnamefont {D.}~\bibnamefont {Petter}}, \bibinfo {author} {\bibfnamefont {G.}~\bibnamefont {Natale}}, \bibinfo {author} {\bibfnamefont {R.}~\bibnamefont {van Bijnen}}, \bibinfo {author} {\bibfnamefont {A.}~\bibnamefont {Patscheider}}, \bibinfo {author} {\bibfnamefont {M.}~\bibnamefont {Mark}}, \bibinfo {author} {\bibfnamefont {L.}~\bibnamefont {Chomaz}}, \ and\ \bibinfo {author} {\bibfnamefont {F.}~\bibnamefont {Ferlaino}},\ }\href {\doibase 10.1103/PhysRevLett.122.183401} {\bibfield  {journal} {\bibinfo  {journal} {Phys. Rev. Lett.}\ }\textbf {\bibinfo {volume} {122}},\ \bibinfo {pages} {183401} (\bibinfo {year} {2019})}\BibitemShut {NoStop}%
\bibitem [{\citenamefont {Hertkorn}\ \emph {et~al.}(2021{\natexlab{a}})\citenamefont {Hertkorn}, \citenamefont {Schmidt}, \citenamefont {B\"ottcher}, \citenamefont {Guo}, \citenamefont {Schmidt}, \citenamefont {Ng}, \citenamefont {Graham}, \citenamefont {B\"uchler}, \citenamefont {Langen}, \citenamefont {Zwierlein},\ and\ \citenamefont {Pfau}}]{Hertkorn_Density_fluc_2021}%
  \BibitemOpen
  \bibfield  {author} {\bibinfo {author} {\bibfnamefont {J.}~\bibnamefont {Hertkorn}}, \bibinfo {author} {\bibfnamefont {J.-N.}\ \bibnamefont {Schmidt}}, \bibinfo {author} {\bibfnamefont {F.}~\bibnamefont {B\"ottcher}}, \bibinfo {author} {\bibfnamefont {M.}~\bibnamefont {Guo}}, \bibinfo {author} {\bibfnamefont {M.}~\bibnamefont {Schmidt}}, \bibinfo {author} {\bibfnamefont {K.~S.~H.}\ \bibnamefont {Ng}}, \bibinfo {author} {\bibfnamefont {S.~D.}\ \bibnamefont {Graham}}, \bibinfo {author} {\bibfnamefont {H.~P.}\ \bibnamefont {B\"uchler}}, \bibinfo {author} {\bibfnamefont {T.}~\bibnamefont {Langen}}, \bibinfo {author} {\bibfnamefont {M.}~\bibnamefont {Zwierlein}}, \ and\ \bibinfo {author} {\bibfnamefont {T.}~\bibnamefont {Pfau}},\ }\href {\doibase 10.1103/PhysRevX.11.011037} {\bibfield  {journal} {\bibinfo  {journal} {Phys. Rev. X}\ }\textbf {\bibinfo {volume} {11}},\ \bibinfo {pages} {011037} (\bibinfo {year} {2021}{\natexlab{a}})}\BibitemShut {NoStop}%
\bibitem [{\citenamefont {Kadau}\ \emph {et~al.}(2016)\citenamefont {Kadau}, \citenamefont {Schmitt}, \citenamefont {Wenzel}, \citenamefont {Wink}, \citenamefont {Maier}, \citenamefont {Ferrier-Barbut},\ and\ \citenamefont {Pfau}}]{kadau_observing_2016}%
  \BibitemOpen
  \bibfield  {author} {\bibinfo {author} {\bibfnamefont {H.}~\bibnamefont {Kadau}}, \bibinfo {author} {\bibfnamefont {M.}~\bibnamefont {Schmitt}}, \bibinfo {author} {\bibfnamefont {M.}~\bibnamefont {Wenzel}}, \bibinfo {author} {\bibfnamefont {C.}~\bibnamefont {Wink}}, \bibinfo {author} {\bibfnamefont {T.}~\bibnamefont {Maier}}, \bibinfo {author} {\bibfnamefont {I.}~\bibnamefont {Ferrier-Barbut}}, \ and\ \bibinfo {author} {\bibfnamefont {T.}~\bibnamefont {Pfau}},\ }\href {\doibase 10.1038/nature16485} {\bibfield  {journal} {\bibinfo  {journal} {Nature}\ }\textbf {\bibinfo {volume} {530}},\ \bibinfo {pages} {194} (\bibinfo {year} {2016})}\BibitemShut {NoStop}%
\bibitem [{\citenamefont {Lee}\ \emph {et~al.}(1957)\citenamefont {Lee}, \citenamefont {Huang},\ and\ \citenamefont {Yang}}]{lee1957eigenvalues}%
  \BibitemOpen
  \bibfield  {author} {\bibinfo {author} {\bibfnamefont {T.~D.}\ \bibnamefont {Lee}}, \bibinfo {author} {\bibfnamefont {K.}~\bibnamefont {Huang}}, \ and\ \bibinfo {author} {\bibfnamefont {C.~N.}\ \bibnamefont {Yang}},\ }\href {\doibase 10.1103/PhysRev.106.1135} {\bibfield  {journal} {\bibinfo  {journal} {Phys. Rev.}\ }\textbf {\bibinfo {volume} {106}},\ \bibinfo {pages} {1135} (\bibinfo {year} {1957})}\BibitemShut {NoStop}%
\bibitem [{\citenamefont {Lima}\ and\ \citenamefont {Pelster}(2011)}]{Lima_quantum_2011}%
  \BibitemOpen
  \bibfield  {author} {\bibinfo {author} {\bibfnamefont {A.~R.~P.}\ \bibnamefont {Lima}}\ and\ \bibinfo {author} {\bibfnamefont {A.}~\bibnamefont {Pelster}},\ }\href {\doibase 10.1103/PhysRevA.84.041604} {\bibfield  {journal} {\bibinfo  {journal} {Phys. Rev. A}\ }\textbf {\bibinfo {volume} {84}},\ \bibinfo {pages} {041604} (\bibinfo {year} {2011})}\BibitemShut {NoStop}%
\bibitem [{\citenamefont {Maier}\ \emph {et~al.}(2015)\citenamefont {Maier}, \citenamefont {Ferrier-Barbut}, \citenamefont {Kadau}, \citenamefont {Schmitt}, \citenamefont {Wenzel}, \citenamefont {Wink}, \citenamefont {Pfau}, \citenamefont {Jachymski},\ and\ \citenamefont {Julienne}}]{Maier_broad_2015}%
  \BibitemOpen
  \bibfield  {author} {\bibinfo {author} {\bibfnamefont {T.}~\bibnamefont {Maier}}, \bibinfo {author} {\bibfnamefont {I.}~\bibnamefont {Ferrier-Barbut}}, \bibinfo {author} {\bibfnamefont {H.}~\bibnamefont {Kadau}}, \bibinfo {author} {\bibfnamefont {M.}~\bibnamefont {Schmitt}}, \bibinfo {author} {\bibfnamefont {M.}~\bibnamefont {Wenzel}}, \bibinfo {author} {\bibfnamefont {C.}~\bibnamefont {Wink}}, \bibinfo {author} {\bibfnamefont {T.}~\bibnamefont {Pfau}}, \bibinfo {author} {\bibfnamefont {K.}~\bibnamefont {Jachymski}}, \ and\ \bibinfo {author} {\bibfnamefont {P.~S.}\ \bibnamefont {Julienne}},\ }\href {\doibase 10.1103/PhysRevA.92.060702} {\bibfield  {journal} {\bibinfo  {journal} {Phys. Rev. A}\ }\textbf {\bibinfo {volume} {92}},\ \bibinfo {pages} {060702} (\bibinfo {year} {2015})}\BibitemShut {NoStop}%
\bibitem [{\citenamefont {Tang}\ \emph {et~al.}(2016)\citenamefont {Tang}, \citenamefont {Sykes}, \citenamefont {Burdick}, \citenamefont {DiSciacca}, \citenamefont {Petrov},\ and\ \citenamefont {Lev}}]{Tang_anisotropic_2016}%
  \BibitemOpen
  \bibfield  {author} {\bibinfo {author} {\bibfnamefont {Y.}~\bibnamefont {Tang}}, \bibinfo {author} {\bibfnamefont {A.~G.}\ \bibnamefont {Sykes}}, \bibinfo {author} {\bibfnamefont {N.~Q.}\ \bibnamefont {Burdick}}, \bibinfo {author} {\bibfnamefont {J.~M.}\ \bibnamefont {DiSciacca}}, \bibinfo {author} {\bibfnamefont {D.~S.}\ \bibnamefont {Petrov}}, \ and\ \bibinfo {author} {\bibfnamefont {B.~L.}\ \bibnamefont {Lev}},\ }\href {\doibase 10.1103/PhysRevLett.117.155301} {\bibfield  {journal} {\bibinfo  {journal} {Phys. Rev. Lett.}\ }\textbf {\bibinfo {volume} {117}},\ \bibinfo {pages} {155301} (\bibinfo {year} {2016})}\BibitemShut {NoStop}%
\bibitem [{\citenamefont {Ferrier-Barbut}\ \emph {et~al.}(2016)\citenamefont {Ferrier-Barbut}, \citenamefont {Kadau}, \citenamefont {Schmitt}, \citenamefont {Wenzel},\ and\ \citenamefont {Pfau}}]{ferrier-barbut_observation_2016}%
  \BibitemOpen
  \bibfield  {author} {\bibinfo {author} {\bibfnamefont {I.}~\bibnamefont {Ferrier-Barbut}}, \bibinfo {author} {\bibfnamefont {H.}~\bibnamefont {Kadau}}, \bibinfo {author} {\bibfnamefont {M.}~\bibnamefont {Schmitt}}, \bibinfo {author} {\bibfnamefont {M.}~\bibnamefont {Wenzel}}, \ and\ \bibinfo {author} {\bibfnamefont {T.}~\bibnamefont {Pfau}},\ }\href {\doibase 10.1103/PhysRevLett.116.215301} {\bibfield  {journal} {\bibinfo  {journal} {Phys. Rev. Lett.}\ }\textbf {\bibinfo {volume} {116}},\ \bibinfo {pages} {215301} (\bibinfo {year} {2016})}\BibitemShut {NoStop}%
\bibitem [{\citenamefont {Schmitt}\ \emph {et~al.}(2016)\citenamefont {Schmitt}, \citenamefont {Wenzel}, \citenamefont {Böttcher}, \citenamefont {Ferrier-Barbut},\ and\ \citenamefont {Pfau}}]{schmitt_self-bound_2016}%
  \BibitemOpen
  \bibfield  {author} {\bibinfo {author} {\bibfnamefont {M.}~\bibnamefont {Schmitt}}, \bibinfo {author} {\bibfnamefont {M.}~\bibnamefont {Wenzel}}, \bibinfo {author} {\bibfnamefont {F.}~\bibnamefont {Böttcher}}, \bibinfo {author} {\bibfnamefont {I.}~\bibnamefont {Ferrier-Barbut}}, \ and\ \bibinfo {author} {\bibfnamefont {T.}~\bibnamefont {Pfau}},\ }\href {\doibase 10.1038/nature20126} {\bibfield  {journal} {\bibinfo  {journal} {Nature}\ }\textbf {\bibinfo {volume} {539}},\ \bibinfo {pages} {259} (\bibinfo {year} {2016})}\BibitemShut {NoStop}%
\bibitem [{\citenamefont {Chomaz}\ \emph {et~al.}(2016)\citenamefont {Chomaz}, \citenamefont {Baier}, \citenamefont {Petter}, \citenamefont {Mark}, \citenamefont {W\"achtler}, \citenamefont {Santos},\ and\ \citenamefont {Ferlaino}}]{chomaz2016quantum}%
  \BibitemOpen
  \bibfield  {author} {\bibinfo {author} {\bibfnamefont {L.}~\bibnamefont {Chomaz}}, \bibinfo {author} {\bibfnamefont {S.}~\bibnamefont {Baier}}, \bibinfo {author} {\bibfnamefont {D.}~\bibnamefont {Petter}}, \bibinfo {author} {\bibfnamefont {M.~J.}\ \bibnamefont {Mark}}, \bibinfo {author} {\bibfnamefont {F.}~\bibnamefont {W\"achtler}}, \bibinfo {author} {\bibfnamefont {L.}~\bibnamefont {Santos}}, \ and\ \bibinfo {author} {\bibfnamefont {F.}~\bibnamefont {Ferlaino}},\ }\href {\doibase 10.1103/PhysRevX.6.041039} {\bibfield  {journal} {\bibinfo  {journal} {Phys. Rev. X}\ }\textbf {\bibinfo {volume} {6}},\ \bibinfo {pages} {041039} (\bibinfo {year} {2016})}\BibitemShut {NoStop}%
\bibitem [{\citenamefont {Hertkorn}\ \emph {et~al.}(2021{\natexlab{b}})\citenamefont {Hertkorn}, \citenamefont {Schmidt}, \citenamefont {Guo}, \citenamefont {B\"ottcher}, \citenamefont {Ng}, \citenamefont {Graham}, \citenamefont {Uerlings}, \citenamefont {Langen}, \citenamefont {Zwierlein},\ and\ \citenamefont {Pfau}}]{Hertkorn_pattern_2021}%
  \BibitemOpen
  \bibfield  {author} {\bibinfo {author} {\bibfnamefont {J.}~\bibnamefont {Hertkorn}}, \bibinfo {author} {\bibfnamefont {J.-N.}\ \bibnamefont {Schmidt}}, \bibinfo {author} {\bibfnamefont {M.}~\bibnamefont {Guo}}, \bibinfo {author} {\bibfnamefont {F.}~\bibnamefont {B\"ottcher}}, \bibinfo {author} {\bibfnamefont {K.~S.~H.}\ \bibnamefont {Ng}}, \bibinfo {author} {\bibfnamefont {S.~D.}\ \bibnamefont {Graham}}, \bibinfo {author} {\bibfnamefont {P.}~\bibnamefont {Uerlings}}, \bibinfo {author} {\bibfnamefont {T.}~\bibnamefont {Langen}}, \bibinfo {author} {\bibfnamefont {M.}~\bibnamefont {Zwierlein}}, \ and\ \bibinfo {author} {\bibfnamefont {T.}~\bibnamefont {Pfau}},\ }\href {\doibase 10.1103/PhysRevResearch.3.033125} {\bibfield  {journal} {\bibinfo  {journal} {Phys. Rev. Res.}\ }\textbf {\bibinfo {volume} {3}},\ \bibinfo {pages} {033125} (\bibinfo {year} {2021}{\natexlab{b}})}\BibitemShut {NoStop}%
\bibitem [{\citenamefont {Poli}\ \emph {et~al.}(2021)\citenamefont {Poli}, \citenamefont {Bland}, \citenamefont {Politi}, \citenamefont {Klaus}, \citenamefont {Norcia}, \citenamefont {Ferlaino}, \citenamefont {Bisset},\ and\ \citenamefont {Santos}}]{poli_maintaining_2021}%
  \BibitemOpen
  \bibfield  {author} {\bibinfo {author} {\bibfnamefont {E.}~\bibnamefont {Poli}}, \bibinfo {author} {\bibfnamefont {T.}~\bibnamefont {Bland}}, \bibinfo {author} {\bibfnamefont {C.}~\bibnamefont {Politi}}, \bibinfo {author} {\bibfnamefont {L.}~\bibnamefont {Klaus}}, \bibinfo {author} {\bibfnamefont {M.~A.}\ \bibnamefont {Norcia}}, \bibinfo {author} {\bibfnamefont {F.}~\bibnamefont {Ferlaino}}, \bibinfo {author} {\bibfnamefont {R.~N.}\ \bibnamefont {Bisset}}, \ and\ \bibinfo {author} {\bibfnamefont {L.}~\bibnamefont {Santos}},\ }\href {\doibase 10.1103/PhysRevA.104.063307} {\bibfield  {journal} {\bibinfo  {journal} {Phys. Rev. A}\ }\textbf {\bibinfo {volume} {104}},\ \bibinfo {pages} {063307} (\bibinfo {year} {2021})}\BibitemShut {NoStop}%
\bibitem [{\citenamefont {Natale}\ \emph {et~al.}(2019)\citenamefont {Natale}, \citenamefont {van Bijnen}, \citenamefont {Patscheider}, \citenamefont {Petter}, \citenamefont {Mark}, \citenamefont {Chomaz},\ and\ \citenamefont {Ferlaino}}]{natale2019}%
  \BibitemOpen
  \bibfield  {author} {\bibinfo {author} {\bibfnamefont {G.}~\bibnamefont {Natale}}, \bibinfo {author} {\bibfnamefont {R.~M.~W.}\ \bibnamefont {van Bijnen}}, \bibinfo {author} {\bibfnamefont {A.}~\bibnamefont {Patscheider}}, \bibinfo {author} {\bibfnamefont {D.}~\bibnamefont {Petter}}, \bibinfo {author} {\bibfnamefont {M.~J.}\ \bibnamefont {Mark}}, \bibinfo {author} {\bibfnamefont {L.}~\bibnamefont {Chomaz}}, \ and\ \bibinfo {author} {\bibfnamefont {F.}~\bibnamefont {Ferlaino}},\ }\href {\doibase 10.1103/PhysRevLett.123.050402} {\bibfield  {journal} {\bibinfo  {journal} {Phys. Rev. Lett.}\ }\textbf {\bibinfo {volume} {123}},\ \bibinfo {pages} {050402} (\bibinfo {year} {2019})}\BibitemShut {NoStop}%
\bibitem [{\citenamefont {Hertkorn}\ \emph {et~al.}(2021{\natexlab{c}})\citenamefont {Hertkorn}, \citenamefont {Schmidt}, \citenamefont {Guo}, \citenamefont {Böttcher}, \citenamefont {Ng}, \citenamefont {Graham}, \citenamefont {Uerlings}, \citenamefont {Büchler}, \citenamefont {Langen}, \citenamefont {Zwierlein},\ and\ \citenamefont {Pfau}}]{hertkorn_supersolidity_2021}%
  \BibitemOpen
  \bibfield  {author} {\bibinfo {author} {\bibfnamefont {J.}~\bibnamefont {Hertkorn}}, \bibinfo {author} {\bibfnamefont {J.-N.}\ \bibnamefont {Schmidt}}, \bibinfo {author} {\bibfnamefont {M.}~\bibnamefont {Guo}}, \bibinfo {author} {\bibfnamefont {F.}~\bibnamefont {Böttcher}}, \bibinfo {author} {\bibfnamefont {K.}~\bibnamefont {Ng}}, \bibinfo {author} {\bibfnamefont {S.}~\bibnamefont {Graham}}, \bibinfo {author} {\bibfnamefont {P.}~\bibnamefont {Uerlings}}, \bibinfo {author} {\bibfnamefont {H.}~\bibnamefont {Büchler}}, \bibinfo {author} {\bibfnamefont {T.}~\bibnamefont {Langen}}, \bibinfo {author} {\bibfnamefont {M.}~\bibnamefont {Zwierlein}}, \ and\ \bibinfo {author} {\bibfnamefont {T.}~\bibnamefont {Pfau}},\ }\href {\doibase 10.1103/PhysRevLett.127.155301} {\bibfield  {journal} {\bibinfo  {journal} {Phys. Rev. Lett.}\ }\textbf {\bibinfo {volume} {127}},\ \bibinfo {pages} {155301} (\bibinfo {year} {2021}{\natexlab{c}})}\BibitemShut {NoStop}%
\bibitem [{\citenamefont {Schmidt}\ \emph {et~al.}(2021)\citenamefont {Schmidt}, \citenamefont {Hertkorn}, \citenamefont {Guo}, \citenamefont {B\"ottcher}, \citenamefont {Schmidt}, \citenamefont {Ng}, \citenamefont {Graham}, \citenamefont {Langen}, \citenamefont {Zwierlein},\ and\ \citenamefont {Pfau}}]{Schmidt_Roton_excitation2021}%
  \BibitemOpen
  \bibfield  {author} {\bibinfo {author} {\bibfnamefont {J.-N.}\ \bibnamefont {Schmidt}}, \bibinfo {author} {\bibfnamefont {J.}~\bibnamefont {Hertkorn}}, \bibinfo {author} {\bibfnamefont {M.}~\bibnamefont {Guo}}, \bibinfo {author} {\bibfnamefont {F.}~\bibnamefont {B\"ottcher}}, \bibinfo {author} {\bibfnamefont {M.}~\bibnamefont {Schmidt}}, \bibinfo {author} {\bibfnamefont {K.~S.~H.}\ \bibnamefont {Ng}}, \bibinfo {author} {\bibfnamefont {S.~D.}\ \bibnamefont {Graham}}, \bibinfo {author} {\bibfnamefont {T.}~\bibnamefont {Langen}}, \bibinfo {author} {\bibfnamefont {M.}~\bibnamefont {Zwierlein}}, \ and\ \bibinfo {author} {\bibfnamefont {T.}~\bibnamefont {Pfau}},\ }\href {\doibase 10.1103/PhysRevLett.126.193002} {\bibfield  {journal} {\bibinfo  {journal} {Phys. Rev. Lett.}\ }\textbf {\bibinfo {volume} {126}},\ \bibinfo {pages} {193002} (\bibinfo {year} {2021})}\BibitemShut {NoStop}%
\bibitem [{\citenamefont {Zhang}\ \emph {et~al.}(2019)\citenamefont {Zhang}, \citenamefont {Maucher},\ and\ \citenamefont {Pohl}}]{zhang2019supersolidity}%
  \BibitemOpen
  \bibfield  {author} {\bibinfo {author} {\bibfnamefont {Y.-C.}\ \bibnamefont {Zhang}}, \bibinfo {author} {\bibfnamefont {F.}~\bibnamefont {Maucher}}, \ and\ \bibinfo {author} {\bibfnamefont {T.}~\bibnamefont {Pohl}},\ }\href {https://doi.org/10.1103/PhysRevLett.123.015301} {\bibfield  {journal} {\bibinfo  {journal} {Phys. Rev. Lett.}\ }\textbf {\bibinfo {volume} {123}},\ \bibinfo {pages} {015301} (\bibinfo {year} {2019})}\BibitemShut {NoStop}%
\bibitem [{\citenamefont {Zhang}\ \emph {et~al.}(2021)\citenamefont {Zhang}, \citenamefont {Pohl},\ and\ \citenamefont {Maucher}}]{zhang2021phases}%
  \BibitemOpen
  \bibfield  {author} {\bibinfo {author} {\bibfnamefont {Y.-C.}\ \bibnamefont {Zhang}}, \bibinfo {author} {\bibfnamefont {T.}~\bibnamefont {Pohl}}, \ and\ \bibinfo {author} {\bibfnamefont {F.}~\bibnamefont {Maucher}},\ }\href {https://doi.org/10.1103/PhysRevA.104.013310} {\bibfield  {journal} {\bibinfo  {journal} {Phys. Rev. A}\ }\textbf {\bibinfo {volume} {104}},\ \bibinfo {pages} {013310} (\bibinfo {year} {2021})}\BibitemShut {NoStop}%
\bibitem [{\citenamefont {Ripley}\ \emph {et~al.}(2023)\citenamefont {Ripley}, \citenamefont {Baillie},\ and\ \citenamefont {Blakie}}]{ripley_two-dimensional_2023}%
  \BibitemOpen
  \bibfield  {author} {\bibinfo {author} {\bibfnamefont {B.~T.~E.}\ \bibnamefont {Ripley}}, \bibinfo {author} {\bibfnamefont {D.}~\bibnamefont {Baillie}}, \ and\ \bibinfo {author} {\bibfnamefont {P.~B.}\ \bibnamefont {Blakie}},\ }\href {\doibase 10.1103/PhysRevA.108.053321} {\bibfield  {journal} {\bibinfo  {journal} {Phys. Rev. A}\ }\textbf {\bibinfo {volume} {108}},\ \bibinfo {pages} {053321} (\bibinfo {year} {2023})}\BibitemShut {NoStop}%
\bibitem [{\citenamefont {Ilzhöfer}\ \emph {et~al.}(2021)\citenamefont {Ilzhöfer}, \citenamefont {Sohmen}, \citenamefont {Durastante}, \citenamefont {Politi}, \citenamefont {Trautmann}, \citenamefont {Natale}, \citenamefont {Morpurgo}, \citenamefont {Giamarchi}, \citenamefont {Chomaz}, \citenamefont {Mark},\ and\ \citenamefont {Ferlaino}}]{Ilzhofer2021}%
  \BibitemOpen
  \bibfield  {author} {\bibinfo {author} {\bibfnamefont {P.}~\bibnamefont {Ilzhöfer}}, \bibinfo {author} {\bibfnamefont {M.}~\bibnamefont {Sohmen}}, \bibinfo {author} {\bibfnamefont {G.}~\bibnamefont {Durastante}}, \bibinfo {author} {\bibfnamefont {C.}~\bibnamefont {Politi}}, \bibinfo {author} {\bibfnamefont {A.}~\bibnamefont {Trautmann}}, \bibinfo {author} {\bibfnamefont {G.}~\bibnamefont {Natale}}, \bibinfo {author} {\bibfnamefont {G.}~\bibnamefont {Morpurgo}}, \bibinfo {author} {\bibfnamefont {T.}~\bibnamefont {Giamarchi}}, \bibinfo {author} {\bibfnamefont {L.}~\bibnamefont {Chomaz}}, \bibinfo {author} {\bibfnamefont {M.~J.}\ \bibnamefont {Mark}}, \ and\ \bibinfo {author} {\bibfnamefont {F.}~\bibnamefont {Ferlaino}},\ }\href {\doibase 10.1038/s41567-020-01100-3} {\bibfield  {journal} {\bibinfo  {journal} {Nature Phys.}\ }\textbf {\bibinfo {volume} {17}},\ \bibinfo {pages} {356–361} (\bibinfo {year} {2021})}\BibitemShut {NoStop}%
\bibitem [{\citenamefont {Biagioni}\ \emph {et~al.}(2024)\citenamefont {Biagioni}, \citenamefont {Antolini}, \citenamefont {Donelli}, \citenamefont {Pezz{\`e}}, \citenamefont {Smerzi}, \citenamefont {Fattori}, \citenamefont {Fioretti}, \citenamefont {Gabbanini}, \citenamefont {Inguscio}, \citenamefont {Tanzi},\ and\ \citenamefont {Modugno}}]{Biagioni_measurement_2024}%
  \BibitemOpen
  \bibfield  {author} {\bibinfo {author} {\bibfnamefont {G.}~\bibnamefont {Biagioni}}, \bibinfo {author} {\bibfnamefont {N.}~\bibnamefont {Antolini}}, \bibinfo {author} {\bibfnamefont {B.}~\bibnamefont {Donelli}}, \bibinfo {author} {\bibfnamefont {L.}~\bibnamefont {Pezz{\`e}}}, \bibinfo {author} {\bibfnamefont {A.}~\bibnamefont {Smerzi}}, \bibinfo {author} {\bibfnamefont {M.}~\bibnamefont {Fattori}}, \bibinfo {author} {\bibfnamefont {A.}~\bibnamefont {Fioretti}}, \bibinfo {author} {\bibfnamefont {C.}~\bibnamefont {Gabbanini}}, \bibinfo {author} {\bibfnamefont {M.}~\bibnamefont {Inguscio}}, \bibinfo {author} {\bibfnamefont {L.}~\bibnamefont {Tanzi}}, \ and\ \bibinfo {author} {\bibfnamefont {G.}~\bibnamefont {Modugno}},\ }\href {\doibase 10.1038/s41586-024-07361-9} {\bibfield  {journal} {\bibinfo  {journal} {Nature}\ }\textbf {\bibinfo {volume} {629}},\ \bibinfo {pages} {773} (\bibinfo {year} {2024})}\BibitemShut {NoStop}%
\bibitem [{\citenamefont {Donelli}\ \emph {et~al.}(2025)\citenamefont {Donelli}, \citenamefont {Antolini}, \citenamefont {Biagioni}, \citenamefont {Fattori}, \citenamefont {Fioretti}, \citenamefont {Gabbanini}, \citenamefont {Inguscio}, \citenamefont {Tanzi}, \citenamefont {Modugno}, \citenamefont {Smerzi},\ and\ \citenamefont {Pezz\`e}}]{donelli_self_2025}%
  \BibitemOpen
  \bibfield  {author} {\bibinfo {author} {\bibfnamefont {B.}~\bibnamefont {Donelli}}, \bibinfo {author} {\bibfnamefont {N.}~\bibnamefont {Antolini}}, \bibinfo {author} {\bibfnamefont {G.}~\bibnamefont {Biagioni}}, \bibinfo {author} {\bibfnamefont {M.}~\bibnamefont {Fattori}}, \bibinfo {author} {\bibfnamefont {A.}~\bibnamefont {Fioretti}}, \bibinfo {author} {\bibfnamefont {C.}~\bibnamefont {Gabbanini}}, \bibinfo {author} {\bibfnamefont {M.}~\bibnamefont {Inguscio}}, \bibinfo {author} {\bibfnamefont {L.}~\bibnamefont {Tanzi}}, \bibinfo {author} {\bibfnamefont {G.}~\bibnamefont {Modugno}}, \bibinfo {author} {\bibfnamefont {A.}~\bibnamefont {Smerzi}}, \ and\ \bibinfo {author} {\bibfnamefont {L.}~\bibnamefont {Pezz\`e}},\ }\href {\doibase 10.1103/hy2k-vqxd} {\bibfield  {journal} {\bibinfo  {journal} {Phys. Rev. A}\ }\textbf {\bibinfo {volume} {112}},\ \bibinfo {pages} {L051302} (\bibinfo {year} {2025})}\BibitemShut {NoStop}%
\bibitem [{\citenamefont {Tanzi}\ \emph {et~al.}(2021)\citenamefont {Tanzi}, \citenamefont {Maloberti}, \citenamefont {Biagioni}, \citenamefont {Fioretti}, \citenamefont {Gabbanini},\ and\ \citenamefont {Modugno}}]{tanzi2021evidence}%
  \BibitemOpen
  \bibfield  {author} {\bibinfo {author} {\bibfnamefont {L.}~\bibnamefont {Tanzi}}, \bibinfo {author} {\bibfnamefont {J.}~\bibnamefont {Maloberti}}, \bibinfo {author} {\bibfnamefont {G.}~\bibnamefont {Biagioni}}, \bibinfo {author} {\bibfnamefont {A.}~\bibnamefont {Fioretti}}, \bibinfo {author} {\bibfnamefont {C.}~\bibnamefont {Gabbanini}}, \ and\ \bibinfo {author} {\bibfnamefont {G.}~\bibnamefont {Modugno}},\ }\href {https://doi.org/10.1126/science.aba4309} {\bibfield  {journal} {\bibinfo  {journal} {Science}\ }\textbf {\bibinfo {volume} {371}},\ \bibinfo {pages} {1162} (\bibinfo {year} {2021})}\BibitemShut {NoStop}%
\bibitem [{\citenamefont {Norcia}\ \emph {et~al.}(2022)\citenamefont {Norcia}, \citenamefont {Poli}, \citenamefont {Politi}, \citenamefont {Klaus}, \citenamefont {Bland}, \citenamefont {Mark}, \citenamefont {Santos}, \citenamefont {Bisset},\ and\ \citenamefont {Ferlaino}}]{norcia_can_2022}%
  \BibitemOpen
  \bibfield  {author} {\bibinfo {author} {\bibfnamefont {M.~A.}\ \bibnamefont {Norcia}}, \bibinfo {author} {\bibfnamefont {E.}~\bibnamefont {Poli}}, \bibinfo {author} {\bibfnamefont {C.}~\bibnamefont {Politi}}, \bibinfo {author} {\bibfnamefont {L.}~\bibnamefont {Klaus}}, \bibinfo {author} {\bibfnamefont {T.}~\bibnamefont {Bland}}, \bibinfo {author} {\bibfnamefont {M.~J.}\ \bibnamefont {Mark}}, \bibinfo {author} {\bibfnamefont {L.}~\bibnamefont {Santos}}, \bibinfo {author} {\bibfnamefont {R.~N.}\ \bibnamefont {Bisset}}, \ and\ \bibinfo {author} {\bibfnamefont {F.}~\bibnamefont {Ferlaino}},\ }\href {\doibase 10.1103/PhysRevLett.129.040403} {\bibfield  {journal} {\bibinfo  {journal} {Phys. Rev. Lett.}\ }\textbf {\bibinfo {volume} {129}},\ \bibinfo {pages} {040403} (\bibinfo {year} {2022})}\BibitemShut {NoStop}%
\bibitem [{\citenamefont {Casotti}\ \emph {et~al.}(2024)\citenamefont {Casotti}, \citenamefont {Poli}, \citenamefont {Klaus}, \citenamefont {Litvinov}, \citenamefont {Ulm}, \citenamefont {Politi}, \citenamefont {Mark}, \citenamefont {Bland},\ and\ \citenamefont {Ferlaino}}]{casotti_observation_2024}%
  \BibitemOpen
  \bibfield  {author} {\bibinfo {author} {\bibfnamefont {E.}~\bibnamefont {Casotti}}, \bibinfo {author} {\bibfnamefont {E.}~\bibnamefont {Poli}}, \bibinfo {author} {\bibfnamefont {L.}~\bibnamefont {Klaus}}, \bibinfo {author} {\bibfnamefont {A.}~\bibnamefont {Litvinov}}, \bibinfo {author} {\bibfnamefont {C.}~\bibnamefont {Ulm}}, \bibinfo {author} {\bibfnamefont {C.}~\bibnamefont {Politi}}, \bibinfo {author} {\bibfnamefont {M.~J.}\ \bibnamefont {Mark}}, \bibinfo {author} {\bibfnamefont {T.}~\bibnamefont {Bland}}, \ and\ \bibinfo {author} {\bibfnamefont {F.}~\bibnamefont {Ferlaino}},\ }\href {\doibase 10.1038/s41586-024-08149-7} {\bibfield  {journal} {\bibinfo  {journal} {Nature}\ }\textbf {\bibinfo {volume} {635}},\ \bibinfo {pages} {327} (\bibinfo {year} {2024})}\BibitemShut {NoStop}%
\bibitem [{\citenamefont {Poli}\ \emph {et~al.}(2025)\citenamefont {Poli}, \citenamefont {Litvinov}, \citenamefont {Casotti}, \citenamefont {Ulm}, \citenamefont {Klaus}, \citenamefont {Mark}, \citenamefont {Lamporesi}, \citenamefont {Bland},\ and\ \citenamefont {Ferlaino}}]{poli2024synchronization}%
  \BibitemOpen
  \bibfield  {author} {\bibinfo {author} {\bibfnamefont {E.}~\bibnamefont {Poli}}, \bibinfo {author} {\bibfnamefont {A.}~\bibnamefont {Litvinov}}, \bibinfo {author} {\bibfnamefont {E.}~\bibnamefont {Casotti}}, \bibinfo {author} {\bibfnamefont {C.}~\bibnamefont {Ulm}}, \bibinfo {author} {\bibfnamefont {L.}~\bibnamefont {Klaus}}, \bibinfo {author} {\bibfnamefont {M.~J.}\ \bibnamefont {Mark}}, \bibinfo {author} {\bibfnamefont {G.}~\bibnamefont {Lamporesi}}, \bibinfo {author} {\bibfnamefont {T.}~\bibnamefont {Bland}}, \ and\ \bibinfo {author} {\bibfnamefont {F.}~\bibnamefont {Ferlaino}},\ }\href {\doibase 10.1038/s41567-025-03065-7} {\bibfield  {journal} {\bibinfo  {journal} {Nat. Phys.}\ } (\bibinfo {year} {2025}),\ 10.1038/s41567-025-03065-7}\BibitemShut {NoStop}%
\bibitem [{\citenamefont {Senarath~Yapa}\ and\ \citenamefont {Bland}(2025)}]{Yapa_supersonic_2024}%
  \BibitemOpen
  \bibfield  {author} {\bibinfo {author} {\bibfnamefont {P.}~\bibnamefont {Senarath~Yapa}}\ and\ \bibinfo {author} {\bibfnamefont {T.}~\bibnamefont {Bland}},\ }\href {\doibase 10.1103/xz57-52ft} {\bibfield  {journal} {\bibinfo  {journal} {Phys. Rev. A}\ }\textbf {\bibinfo {volume} {112}},\ \bibinfo {pages} {L021303} (\bibinfo {year} {2025})}\BibitemShut {NoStop}%
\bibitem [{\citenamefont {Platt}\ \emph {et~al.}(2024)\citenamefont {Platt}, \citenamefont {Baillie},\ and\ \citenamefont {Blakie}}]{Platt_sounds_2024}%
  \BibitemOpen
  \bibfield  {author} {\bibinfo {author} {\bibfnamefont {L.~M.}\ \bibnamefont {Platt}}, \bibinfo {author} {\bibfnamefont {D.}~\bibnamefont {Baillie}}, \ and\ \bibinfo {author} {\bibfnamefont {P.~B.}\ \bibnamefont {Blakie}},\ }\href {\doibase 10.1103/PhysRevA.110.023320} {\bibfield  {journal} {\bibinfo  {journal} {Phys. Rev. A}\ }\textbf {\bibinfo {volume} {110}},\ \bibinfo {pages} {023320} (\bibinfo {year} {2024})}\BibitemShut {NoStop}%
\bibitem [{\citenamefont {{\v{S}}indik}\ \emph {et~al.}(2024)\citenamefont {{\v{S}}indik}, \citenamefont {Zawi{\'s}lak}, \citenamefont {Recati},\ and\ \citenamefont {Stringari}}]{sindik2024sound}%
  \BibitemOpen
  \bibfield  {author} {\bibinfo {author} {\bibfnamefont {M.}~\bibnamefont {{\v{S}}indik}}, \bibinfo {author} {\bibfnamefont {T.}~\bibnamefont {Zawi{\'s}lak}}, \bibinfo {author} {\bibfnamefont {A.}~\bibnamefont {Recati}}, \ and\ \bibinfo {author} {\bibfnamefont {S.}~\bibnamefont {Stringari}},\ }\href {https://doi.org/10.1103/PhysRevLett.132.146001} {\bibfield  {journal} {\bibinfo  {journal} {Phys. Rev. Lett.}\ }\textbf {\bibinfo {volume} {132}},\ \bibinfo {pages} {146001} (\bibinfo {year} {2024})}\BibitemShut {NoStop}%
\bibitem [{\citenamefont {Poli}\ \emph {et~al.}(2024)\citenamefont {Poli}, \citenamefont {Baillie}, \citenamefont {Ferlaino},\ and\ \citenamefont {Blakie}}]{poli_excitations_2024}%
  \BibitemOpen
  \bibfield  {author} {\bibinfo {author} {\bibfnamefont {E.}~\bibnamefont {Poli}}, \bibinfo {author} {\bibfnamefont {D.}~\bibnamefont {Baillie}}, \bibinfo {author} {\bibfnamefont {F.}~\bibnamefont {Ferlaino}}, \ and\ \bibinfo {author} {\bibfnamefont {P.~B.}\ \bibnamefont {Blakie}},\ }\href {\doibase 10.1103/PhysRevA.110.053301} {\bibfield  {journal} {\bibinfo  {journal} {Phys. Rev. A}\ }\textbf {\bibinfo {volume} {110}},\ \bibinfo {pages} {053301} (\bibinfo {year} {2024})}\BibitemShut {NoStop}%
\bibitem [{\citenamefont {Rakic}\ \emph {et~al.}(2024)\citenamefont {Rakic}, \citenamefont {Ho},\ and\ \citenamefont {Lee}}]{rakic2024elastic}%
  \BibitemOpen
  \bibfield  {author} {\bibinfo {author} {\bibfnamefont {M.}~\bibnamefont {Rakic}}, \bibinfo {author} {\bibfnamefont {A.~F.}\ \bibnamefont {Ho}}, \ and\ \bibinfo {author} {\bibfnamefont {D.~K.}\ \bibnamefont {Lee}},\ }\href {https://doi.org/10.1103/PhysRevResearch.6.043040} {\bibfield  {journal} {\bibinfo  {journal} {Phys. Rev. Res.}\ }\textbf {\bibinfo {volume} {6}},\ \bibinfo {pages} {043040} (\bibinfo {year} {2024})}\BibitemShut {NoStop}%
\bibitem [{\citenamefont {Zawi\ifmmode~\acute{s}\else \'{s}\fi{}lak}\ \emph {et~al.}(2025)\citenamefont {Zawi\ifmmode~\acute{s}\else \'{s}\fi{}lak}, \citenamefont {\ifmmode~\check{S}\else \v{S}\fi{}indik}, \citenamefont {Stringari},\ and\ \citenamefont {Recati}}]{zawislak2024anomalous}%
  \BibitemOpen
  \bibfield  {author} {\bibinfo {author} {\bibfnamefont {T.}~\bibnamefont {Zawi\ifmmode~\acute{s}\else \'{s}\fi{}lak}}, \bibinfo {author} {\bibfnamefont {M.}~\bibnamefont {\ifmmode~\check{S}\else \v{S}\fi{}indik}}, \bibinfo {author} {\bibfnamefont {S.}~\bibnamefont {Stringari}}, \ and\ \bibinfo {author} {\bibfnamefont {A.}~\bibnamefont {Recati}},\ }\href {\doibase 10.1103/PhysRevLett.134.226001} {\bibfield  {journal} {\bibinfo  {journal} {Phys. Rev. Lett.}\ }\textbf {\bibinfo {volume} {134}},\ \bibinfo {pages} {226001} (\bibinfo {year} {2025})}\BibitemShut {NoStop}%
\bibitem [{\citenamefont {Andrews}\ \emph {et~al.}(1997)\citenamefont {Andrews}, \citenamefont {Townsend}, \citenamefont {Miesner}, \citenamefont {Durfee}, \citenamefont {Kurn},\ and\ \citenamefont {Ketterle}}]{andrews_observation_1997}%
  \BibitemOpen
  \bibfield  {author} {\bibinfo {author} {\bibfnamefont {M.}~\bibnamefont {Andrews}}, \bibinfo {author} {\bibfnamefont {C.}~\bibnamefont {Townsend}}, \bibinfo {author} {\bibfnamefont {H.-J.}\ \bibnamefont {Miesner}}, \bibinfo {author} {\bibfnamefont {D.}~\bibnamefont {Durfee}}, \bibinfo {author} {\bibfnamefont {D.}~\bibnamefont {Kurn}}, \ and\ \bibinfo {author} {\bibfnamefont {W.}~\bibnamefont {Ketterle}},\ }\href {\doibase 10.1126/science.275.5300.637} {\bibfield  {journal} {\bibinfo  {journal} {Science}\ }\textbf {\bibinfo {volume} {275}},\ \bibinfo {pages} {637} (\bibinfo {year} {1997})}\BibitemShut {NoStop}%
\bibitem [{\citenamefont {Shin}\ \emph {et~al.}(2004)\citenamefont {Shin}, \citenamefont {Saba}, \citenamefont {Pasquini}, \citenamefont {Ketterle}, \citenamefont {Pritchard},\ and\ \citenamefont {Leanhardt}}]{Shin_atom_2004}%
  \BibitemOpen
  \bibfield  {author} {\bibinfo {author} {\bibfnamefont {Y.}~\bibnamefont {Shin}}, \bibinfo {author} {\bibfnamefont {M.}~\bibnamefont {Saba}}, \bibinfo {author} {\bibfnamefont {T.~A.}\ \bibnamefont {Pasquini}}, \bibinfo {author} {\bibfnamefont {W.}~\bibnamefont {Ketterle}}, \bibinfo {author} {\bibfnamefont {D.~E.}\ \bibnamefont {Pritchard}}, \ and\ \bibinfo {author} {\bibfnamefont {A.~E.}\ \bibnamefont {Leanhardt}},\ }\href {\doibase 10.1103/PhysRevLett.92.050405} {\bibfield  {journal} {\bibinfo  {journal} {Phys. Rev. Lett.}\ }\textbf {\bibinfo {volume} {92}},\ \bibinfo {pages} {050405} (\bibinfo {year} {2004})}\BibitemShut {NoStop}%
\bibitem [{\citenamefont {Kohstall}\ \emph {et~al.}(2011)\citenamefont {Kohstall}, \citenamefont {Riedl}, \citenamefont {Guajardo}, \citenamefont {Sidorenkov}, \citenamefont {Denschlag},\ and\ \citenamefont {Grimm}}]{kohstall_observation_2011}%
  \BibitemOpen
  \bibfield  {author} {\bibinfo {author} {\bibfnamefont {C.}~\bibnamefont {Kohstall}}, \bibinfo {author} {\bibfnamefont {S.}~\bibnamefont {Riedl}}, \bibinfo {author} {\bibfnamefont {E.~S.}\ \bibnamefont {Guajardo}}, \bibinfo {author} {\bibfnamefont {L.}~\bibnamefont {Sidorenkov}}, \bibinfo {author} {\bibfnamefont {J.~H.}\ \bibnamefont {Denschlag}}, \ and\ \bibinfo {author} {\bibfnamefont {R.}~\bibnamefont {Grimm}},\ }\href {\doibase 10.1088/1367-2630/13/6/065027} {\bibfield  {journal} {\bibinfo  {journal} {New J. Phys.}\ }\textbf {\bibinfo {volume} {13}},\ \bibinfo {pages} {065027} (\bibinfo {year} {2011})}\BibitemShut {NoStop}%
\bibitem [{\citenamefont {Chang}\ \emph {et~al.}(2008)\citenamefont {Chang}, \citenamefont {Engels},\ and\ \citenamefont {Hoefer}}]{Engels_DS_shock}%
  \BibitemOpen
  \bibfield  {author} {\bibinfo {author} {\bibfnamefont {J.~J.}\ \bibnamefont {Chang}}, \bibinfo {author} {\bibfnamefont {P.}~\bibnamefont {Engels}}, \ and\ \bibinfo {author} {\bibfnamefont {M.~A.}\ \bibnamefont {Hoefer}},\ }\href {\doibase 10.1103/PhysRevLett.101.170404} {\bibfield  {journal} {\bibinfo  {journal} {Phys. Rev. Lett.}\ }\textbf {\bibinfo {volume} {101}},\ \bibinfo {pages} {170404} (\bibinfo {year} {2008})}\BibitemShut {NoStop}%
\bibitem [{\citenamefont {Hoefer}\ \emph {et~al.}(2006)\citenamefont {Hoefer}, \citenamefont {Ablowitz}, \citenamefont {Coddington}, \citenamefont {Cornell}, \citenamefont {Engels},\ and\ \citenamefont {Schweikhard}}]{Engels_DSW}%
  \BibitemOpen
  \bibfield  {author} {\bibinfo {author} {\bibfnamefont {M.~A.}\ \bibnamefont {Hoefer}}, \bibinfo {author} {\bibfnamefont {M.~J.}\ \bibnamefont {Ablowitz}}, \bibinfo {author} {\bibfnamefont {I.}~\bibnamefont {Coddington}}, \bibinfo {author} {\bibfnamefont {E.~A.}\ \bibnamefont {Cornell}}, \bibinfo {author} {\bibfnamefont {P.}~\bibnamefont {Engels}}, \ and\ \bibinfo {author} {\bibfnamefont {V.}~\bibnamefont {Schweikhard}},\ }\href {\doibase 10.1103/PhysRevA.74.023623} {\bibfield  {journal} {\bibinfo  {journal} {Phys. Rev. A}\ }\textbf {\bibinfo {volume} {74}},\ \bibinfo {pages} {023623} (\bibinfo {year} {2006})}\BibitemShut {NoStop}%
\bibitem [{\citenamefont {Burger}\ \emph {et~al.}(1999)\citenamefont {Burger}, \citenamefont {Bongs}, \citenamefont {Dettmer}, \citenamefont {Ertmer}, \citenamefont {Sengstock}, \citenamefont {Sanpera}, \citenamefont {Shlyapnikov},\ and\ \citenamefont {Lewenstein}}]{Burger_dark_1999}%
  \BibitemOpen
  \bibfield  {author} {\bibinfo {author} {\bibfnamefont {S.}~\bibnamefont {Burger}}, \bibinfo {author} {\bibfnamefont {K.}~\bibnamefont {Bongs}}, \bibinfo {author} {\bibfnamefont {S.}~\bibnamefont {Dettmer}}, \bibinfo {author} {\bibfnamefont {W.}~\bibnamefont {Ertmer}}, \bibinfo {author} {\bibfnamefont {K.}~\bibnamefont {Sengstock}}, \bibinfo {author} {\bibfnamefont {A.}~\bibnamefont {Sanpera}}, \bibinfo {author} {\bibfnamefont {G.~V.}\ \bibnamefont {Shlyapnikov}}, \ and\ \bibinfo {author} {\bibfnamefont {M.}~\bibnamefont {Lewenstein}},\ }\href {\doibase 10.1103/PhysRevLett.83.5198} {\bibfield  {journal} {\bibinfo  {journal} {Phys. Rev. Lett.}\ }\textbf {\bibinfo {volume} {83}},\ \bibinfo {pages} {5198} (\bibinfo {year} {1999})}\BibitemShut {NoStop}%
\bibitem [{\citenamefont {Matthews}\ \emph {et~al.}(1999)\citenamefont {Matthews}, \citenamefont {Anderson}, \citenamefont {Haljan}, \citenamefont {Hall}, \citenamefont {Wieman},\ and\ \citenamefont {Cornell}}]{Matthews_vortices_1999}%
  \BibitemOpen
  \bibfield  {author} {\bibinfo {author} {\bibfnamefont {M.~R.}\ \bibnamefont {Matthews}}, \bibinfo {author} {\bibfnamefont {B.~P.}\ \bibnamefont {Anderson}}, \bibinfo {author} {\bibfnamefont {P.~C.}\ \bibnamefont {Haljan}}, \bibinfo {author} {\bibfnamefont {D.~S.}\ \bibnamefont {Hall}}, \bibinfo {author} {\bibfnamefont {C.~E.}\ \bibnamefont {Wieman}}, \ and\ \bibinfo {author} {\bibfnamefont {E.~A.}\ \bibnamefont {Cornell}},\ }\href {\doibase 10.1103/PhysRevLett.83.2498} {\bibfield  {journal} {\bibinfo  {journal} {Phys. Rev. Lett.}\ }\textbf {\bibinfo {volume} {83}},\ \bibinfo {pages} {2498} (\bibinfo {year} {1999})}\BibitemShut {NoStop}%
\bibitem [{\citenamefont {Anglin}(2008)}]{anglin2008quantum}%
  \BibitemOpen
  \bibfield  {author} {\bibinfo {author} {\bibfnamefont {J.}~\bibnamefont {Anglin}},\ }\href {https://www.nature.com/articles/nphys980} {\bibfield  {journal} {\bibinfo  {journal} {Nat. Phys.}\ }\textbf {\bibinfo {volume} {4}},\ \bibinfo {pages} {437} (\bibinfo {year} {2008})}\BibitemShut {NoStop}%
\bibitem [{\citenamefont {Paw{\l}owski}\ and\ \citenamefont {Rz{\k{a}}{\.z}ewski}(2015)}]{pawlowski2015dipolar}%
  \BibitemOpen
  \bibfield  {author} {\bibinfo {author} {\bibfnamefont {K.}~\bibnamefont {Paw{\l}owski}}\ and\ \bibinfo {author} {\bibfnamefont {K.}~\bibnamefont {Rz{\k{a}}{\.z}ewski}},\ }\href {\doibase 10.1088/1367-2630/17/10/105006} {\bibfield  {journal} {\bibinfo  {journal} {New J. Phys.}\ }\textbf {\bibinfo {volume} {17}},\ \bibinfo {pages} {105006} (\bibinfo {year} {2015})}\BibitemShut {NoStop}%
\bibitem [{\citenamefont {Bland}\ \emph {et~al.}(2015)\citenamefont {Bland}, \citenamefont {Edmonds}, \citenamefont {Proukakis}, \citenamefont {Martin}, \citenamefont {O'Dell},\ and\ \citenamefont {Parker}}]{bland2015controllable}%
  \BibitemOpen
  \bibfield  {author} {\bibinfo {author} {\bibfnamefont {T.}~\bibnamefont {Bland}}, \bibinfo {author} {\bibfnamefont {M.}~\bibnamefont {Edmonds}}, \bibinfo {author} {\bibfnamefont {N.}~\bibnamefont {Proukakis}}, \bibinfo {author} {\bibfnamefont {A.}~\bibnamefont {Martin}}, \bibinfo {author} {\bibfnamefont {D.}~\bibnamefont {O'Dell}}, \ and\ \bibinfo {author} {\bibfnamefont {N.}~\bibnamefont {Parker}},\ }\href {https://doi.org/10.1103/PhysRevA.92.063601} {\bibfield  {journal} {\bibinfo  {journal} {Phys. Rev. A}\ }\textbf {\bibinfo {volume} {92}},\ \bibinfo {pages} {063601} (\bibinfo {year} {2015})}\BibitemShut {NoStop}%
\bibitem [{\citenamefont {Edmonds}\ \emph {et~al.}(2016)\citenamefont {Edmonds}, \citenamefont {Bland}, \citenamefont {O'Dell},\ and\ \citenamefont {Parker}}]{edmonds2016exploring}%
  \BibitemOpen
  \bibfield  {author} {\bibinfo {author} {\bibfnamefont {M.}~\bibnamefont {Edmonds}}, \bibinfo {author} {\bibfnamefont {T.}~\bibnamefont {Bland}}, \bibinfo {author} {\bibfnamefont {D.}~\bibnamefont {O'Dell}}, \ and\ \bibinfo {author} {\bibfnamefont {N.}~\bibnamefont {Parker}},\ }\href {https://doi.org/10.1103/PhysRevA.93.063617} {\bibfield  {journal} {\bibinfo  {journal} {Phys. Rev. A}\ }\textbf {\bibinfo {volume} {93}},\ \bibinfo {pages} {063617} (\bibinfo {year} {2016})}\BibitemShut {NoStop}%
\bibitem [{\citenamefont {Bland}\ \emph {et~al.}(2017)\citenamefont {Bland}, \citenamefont {Paw\l{}owski}, \citenamefont {Edmonds}, \citenamefont {Rz\k{a}\ifmmode~\dot{z}\else \.{z}\fi{}ewski},\ and\ \citenamefont {Parker}}]{Bland_interaction_2017}%
  \BibitemOpen
  \bibfield  {author} {\bibinfo {author} {\bibfnamefont {T.}~\bibnamefont {Bland}}, \bibinfo {author} {\bibfnamefont {K.}~\bibnamefont {Paw\l{}owski}}, \bibinfo {author} {\bibfnamefont {M.~J.}\ \bibnamefont {Edmonds}}, \bibinfo {author} {\bibfnamefont {K.}~\bibnamefont {Rz\k{a}\ifmmode~\dot{z}\else \.{z}\fi{}ewski}}, \ and\ \bibinfo {author} {\bibfnamefont {N.~G.}\ \bibnamefont {Parker}},\ }\href {\doibase 10.1103/PhysRevA.95.063622} {\bibfield  {journal} {\bibinfo  {journal} {Phys. Rev. A}\ }\textbf {\bibinfo {volume} {95}},\ \bibinfo {pages} {063622} (\bibinfo {year} {2017})}\BibitemShut {NoStop}%
\bibitem [{\citenamefont {Kopyci{\'n}ski}\ \emph {et~al.}(2023)\citenamefont {Kopyci{\'n}ski}, \citenamefont {{\L}ebek}, \citenamefont {G{\'o}recki},\ and\ \citenamefont {Paw{\l}owski}}]{kopycinski2023ultrawide}%
  \BibitemOpen
  \bibfield  {author} {\bibinfo {author} {\bibfnamefont {J.}~\bibnamefont {Kopyci{\'n}ski}}, \bibinfo {author} {\bibfnamefont {M.}~\bibnamefont {{\L}ebek}}, \bibinfo {author} {\bibfnamefont {W.}~\bibnamefont {G{\'o}recki}}, \ and\ \bibinfo {author} {\bibfnamefont {K.}~\bibnamefont {Paw{\l}owski}},\ }\href {https://doi.org/10.1103/PhysRevLett.130.043401} {\bibfield  {journal} {\bibinfo  {journal} {Physi. Rev. Lett.}\ }\textbf {\bibinfo {volume} {130}},\ \bibinfo {pages} {043401} (\bibinfo {year} {2023})}\BibitemShut {NoStop}%
\bibitem [{\citenamefont {W\"achtler}\ and\ \citenamefont {Santos}(2016)}]{Santos2016filemanets}%
  \BibitemOpen
  \bibfield  {author} {\bibinfo {author} {\bibfnamefont {F.}~\bibnamefont {W\"achtler}}\ and\ \bibinfo {author} {\bibfnamefont {L.}~\bibnamefont {Santos}},\ }\href {\doibase 10.1103/PhysRevA.93.061603} {\bibfield  {journal} {\bibinfo  {journal} {Phys. Rev. A}\ }\textbf {\bibinfo {volume} {93}},\ \bibinfo {pages} {061603} (\bibinfo {year} {2016})}\BibitemShut {NoStop}%
\bibitem [{\citenamefont {Bisset}\ \emph {et~al.}(2016)\citenamefont {Bisset}, \citenamefont {Wilson}, \citenamefont {Baillie},\ and\ \citenamefont {Blakie}}]{bisset_ground-state_2016}%
  \BibitemOpen
  \bibfield  {author} {\bibinfo {author} {\bibfnamefont {R.~N.}\ \bibnamefont {Bisset}}, \bibinfo {author} {\bibfnamefont {R.~M.}\ \bibnamefont {Wilson}}, \bibinfo {author} {\bibfnamefont {D.}~\bibnamefont {Baillie}}, \ and\ \bibinfo {author} {\bibfnamefont {P.~B.}\ \bibnamefont {Blakie}},\ }\href {\doibase 10.1103/PhysRevA.94.033619} {\bibfield  {journal} {\bibinfo  {journal} {Phys. Rev. A}\ }\textbf {\bibinfo {volume} {94}},\ \bibinfo {pages} {033619} (\bibinfo {year} {2016})}\BibitemShut {NoStop}%
\bibitem [{\citenamefont {Hertkorn}\ \emph {et~al.}(2024)\citenamefont {Hertkorn}, \citenamefont {St\"urmer}, \citenamefont {Mukherjee}, \citenamefont {Ng}, \citenamefont {Uerlings}, \citenamefont {Hellstern}, \citenamefont {Lavoine}, \citenamefont {Reimann}, \citenamefont {Pfau},\ and\ \citenamefont {Klemt}}]{hertkorn_decoupled_2024}%
  \BibitemOpen
  \bibfield  {author} {\bibinfo {author} {\bibfnamefont {J.}~\bibnamefont {Hertkorn}}, \bibinfo {author} {\bibfnamefont {P.}~\bibnamefont {St\"urmer}}, \bibinfo {author} {\bibfnamefont {K.}~\bibnamefont {Mukherjee}}, \bibinfo {author} {\bibfnamefont {K.~S.~H.}\ \bibnamefont {Ng}}, \bibinfo {author} {\bibfnamefont {P.}~\bibnamefont {Uerlings}}, \bibinfo {author} {\bibfnamefont {F.}~\bibnamefont {Hellstern}}, \bibinfo {author} {\bibfnamefont {L.}~\bibnamefont {Lavoine}}, \bibinfo {author} {\bibfnamefont {S.~M.}\ \bibnamefont {Reimann}}, \bibinfo {author} {\bibfnamefont {T.}~\bibnamefont {Pfau}}, \ and\ \bibinfo {author} {\bibfnamefont {R.}~\bibnamefont {Klemt}},\ }\href {\doibase 10.1103/PhysRevResearch.6.L042056} {\bibfield  {journal} {\bibinfo  {journal} {Phys. Rev. Res.}\ }\textbf {\bibinfo {volume} {6}},\ \bibinfo {pages} {L042056} (\bibinfo {year} {2024})}\BibitemShut {NoStop}%
\bibitem [{\citenamefont {Guo}\ \emph {et~al.}(2019)\citenamefont {Guo}, \citenamefont {Böttcher}, \citenamefont {Hertkorn}, \citenamefont {Schmidt}, \citenamefont {Wenzel}, \citenamefont {Büchler}, \citenamefont {Langen},\ and\ \citenamefont {Pfau}}]{Guo2019}%
  \BibitemOpen
  \bibfield  {author} {\bibinfo {author} {\bibfnamefont {M.}~\bibnamefont {Guo}}, \bibinfo {author} {\bibfnamefont {F.}~\bibnamefont {Böttcher}}, \bibinfo {author} {\bibfnamefont {J.}~\bibnamefont {Hertkorn}}, \bibinfo {author} {\bibfnamefont {J.-N.}\ \bibnamefont {Schmidt}}, \bibinfo {author} {\bibfnamefont {M.}~\bibnamefont {Wenzel}}, \bibinfo {author} {\bibfnamefont {H.~P.}\ \bibnamefont {Büchler}}, \bibinfo {author} {\bibfnamefont {T.}~\bibnamefont {Langen}}, \ and\ \bibinfo {author} {\bibfnamefont {T.}~\bibnamefont {Pfau}},\ }\href {\doibase 10.1038/s41586-019-1569-5} {\bibfield  {journal} {\bibinfo  {journal} {Nature}\ }\textbf {\bibinfo {volume} {574}},\ \bibinfo {pages} {386–389} (\bibinfo {year} {2019})}\BibitemShut {NoStop}%
\bibitem [{\citenamefont {Tanzi}\ \emph {et~al.}(2019{\natexlab{b}})\citenamefont {Tanzi}, \citenamefont {Roccuzzo}, \citenamefont {Lucioni}, \citenamefont {Fama}, \citenamefont {Fioretti}, \citenamefont {Gabbanini}, \citenamefont {Modugno}, \citenamefont {Recati},\ and\ \citenamefont {Stringari}}]{tanzi2019supersolid}%
  \BibitemOpen
  \bibfield  {author} {\bibinfo {author} {\bibfnamefont {L.}~\bibnamefont {Tanzi}}, \bibinfo {author} {\bibfnamefont {S.~M.}\ \bibnamefont {Roccuzzo}}, \bibinfo {author} {\bibfnamefont {E.}~\bibnamefont {Lucioni}}, \bibinfo {author} {\bibfnamefont {F.}~\bibnamefont {Fama}}, \bibinfo {author} {\bibfnamefont {A.}~\bibnamefont {Fioretti}}, \bibinfo {author} {\bibfnamefont {C.}~\bibnamefont {Gabbanini}}, \bibinfo {author} {\bibfnamefont {G.}~\bibnamefont {Modugno}}, \bibinfo {author} {\bibfnamefont {A.}~\bibnamefont {Recati}}, \ and\ \bibinfo {author} {\bibfnamefont {S.}~\bibnamefont {Stringari}},\ }\href {\doibase 10.1038/s41586-019-1568-6} {\bibfield  {journal} {\bibinfo  {journal} {Nature}\ }\textbf {\bibinfo {volume} {574}},\ \bibinfo {pages} {382–385} (\bibinfo {year} {2019}{\natexlab{b}})}\BibitemShut {NoStop}%
\bibitem [{sup()}]{supp}%
  \BibitemOpen
  \href@noop {} {}\bibinfo {note} {In the Supplementary Material [URL] more information can be found regarding the eGPE framework, soliton creation in superfluids, transverse excitations, the rigid dynamics for larger number of droplet peaks, a coupled oscillators model with next-to-nearest neighbor couplings, tunneling rates, and details of the simulations. Moreover, the Supplementary Material includes the following references~\cite{Parker_structure_2009,Bohn_collapse_2009,schutzhold_mean_2006,Kivshar_dark_1998,Frantzeskakis_dark_2010,Ros_demand_2021,Weller_experimental_2008,Becker_oscillations_2008,Busch_motion_2000,Mistakidis2024,crank_practical_1947,antoine_computational_2013,arfken_mathematical_1972,Goral_ground_2002}.}\BibitemShut {Stop}%
\bibitem [{\citenamefont {Chin}\ \emph {et~al.}(2010)\citenamefont {Chin}, \citenamefont {Grimm}, \citenamefont {Julienne},\ and\ \citenamefont {Tiesinga}}]{Chin_Feshbach_2010}%
  \BibitemOpen
  \bibfield  {author} {\bibinfo {author} {\bibfnamefont {C.}~\bibnamefont {Chin}}, \bibinfo {author} {\bibfnamefont {R.}~\bibnamefont {Grimm}}, \bibinfo {author} {\bibfnamefont {P.}~\bibnamefont {Julienne}}, \ and\ \bibinfo {author} {\bibfnamefont {E.}~\bibnamefont {Tiesinga}},\ }\href {\doibase 10.1103/RevModPhys.82.1225} {\bibfield  {journal} {\bibinfo  {journal} {Rev. Mod. Phys.}\ }\textbf {\bibinfo {volume} {82}},\ \bibinfo {pages} {1225} (\bibinfo {year} {2010})}\BibitemShut {NoStop}%
\bibitem [{\citenamefont {Mukherjee}\ and\ \citenamefont {Reimann}(2023)}]{Mukherjee_classical_2023}%
  \BibitemOpen
  \bibfield  {author} {\bibinfo {author} {\bibfnamefont {K.}~\bibnamefont {Mukherjee}}\ and\ \bibinfo {author} {\bibfnamefont {S.~M.}\ \bibnamefont {Reimann}},\ }\href {\doibase 10.1103/PhysRevA.107.043319} {\bibfield  {journal} {\bibinfo  {journal} {Phys. Rev. A}\ }\textbf {\bibinfo {volume} {107}},\ \bibinfo {pages} {043319} (\bibinfo {year} {2023})}\BibitemShut {NoStop}%
\bibitem [{\citenamefont {Turmanov}\ \emph {et~al.}(2021)\citenamefont {Turmanov}, \citenamefont {Baizakov}, \citenamefont {Abdullaev},\ and\ \citenamefont {Salerno}}]{Turmanov_oscillations_2021}%
  \BibitemOpen
  \bibfield  {author} {\bibinfo {author} {\bibfnamefont {B.~K.}\ \bibnamefont {Turmanov}}, \bibinfo {author} {\bibfnamefont {B.~B.}\ \bibnamefont {Baizakov}}, \bibinfo {author} {\bibfnamefont {F.~K.}\ \bibnamefont {Abdullaev}}, \ and\ \bibinfo {author} {\bibfnamefont {M.}~\bibnamefont {Salerno}},\ }\href {\doibase 10.1088/1361-6455/ac15a4} {\bibfield  {journal} {\bibinfo  {journal} {J. Phys. B: At., Mol. and Opt. Phys.}\ }\textbf {\bibinfo {volume} {54}},\ \bibinfo {pages} {145302} (\bibinfo {year} {2021})}\BibitemShut {NoStop}%
\bibitem [{\citenamefont {Liebster}\ \emph {et~al.}(2025)\citenamefont {Liebster}, \citenamefont {Sparn}, \citenamefont {Kath}, \citenamefont {Duchene}, \citenamefont {Strobel},\ and\ \citenamefont {Oberthaler}}]{liebster2025observation}%
  \BibitemOpen
  \bibfield  {author} {\bibinfo {author} {\bibfnamefont {N.}~\bibnamefont {Liebster}}, \bibinfo {author} {\bibfnamefont {M.}~\bibnamefont {Sparn}}, \bibinfo {author} {\bibfnamefont {E.}~\bibnamefont {Kath}}, \bibinfo {author} {\bibfnamefont {J.}~\bibnamefont {Duchene}}, \bibinfo {author} {\bibfnamefont {H.}~\bibnamefont {Strobel}}, \ and\ \bibinfo {author} {\bibfnamefont {M.~K.}\ \bibnamefont {Oberthaler}},\ }\href {\doibase 10.1038/s41567-025-02927-4} {\bibfield  {journal} {\bibinfo  {journal} {Nat. Phys.}\ }\textbf {\bibinfo {volume} {21}},\ \bibinfo {pages} {1064} (\bibinfo {year} {2025})}\BibitemShut {NoStop}%
\bibitem [{\citenamefont {Dove}(1993)}]{dove_introduction_1993}%
  \BibitemOpen
  \bibfield  {author} {\bibinfo {author} {\bibfnamefont {M.~T.}\ \bibnamefont {Dove}},\ }\href@noop {} {\emph {\bibinfo {title} {Introduction to lattice dynamics}}},\ \bibinfo {number} {4}\ (\bibinfo  {publisher} {Cambridge university press},\ \bibinfo {year} {1993})\BibitemShut {NoStop}%
\bibitem [{\citenamefont {Bland}(1960)}]{bland_linear_1960}%
  \BibitemOpen
  \bibfield  {author} {\bibinfo {author} {\bibfnamefont {D.}~\bibnamefont {Bland}},\ }\href@noop {} {\emph {\bibinfo {title} {The Theory of Linear Viscoelasticity}}}\ (\bibinfo  {publisher} {Pergamon Press},\ \bibinfo {address} {Oxford},\ \bibinfo {year} {1960})\BibitemShut {NoStop}%
\bibitem [{\citenamefont {Bland}\ \emph {et~al.}(2022)\citenamefont {Bland}, \citenamefont {Poli}, \citenamefont {Politi}, \citenamefont {Klaus}, \citenamefont {Norcia}, \citenamefont {Ferlaino}, \citenamefont {Santos},\ and\ \citenamefont {Bisset}}]{bland_two-dimensional_2022}%
  \BibitemOpen
  \bibfield  {author} {\bibinfo {author} {\bibfnamefont {T.}~\bibnamefont {Bland}}, \bibinfo {author} {\bibfnamefont {E.}~\bibnamefont {Poli}}, \bibinfo {author} {\bibfnamefont {C.}~\bibnamefont {Politi}}, \bibinfo {author} {\bibfnamefont {L.}~\bibnamefont {Klaus}}, \bibinfo {author} {\bibfnamefont {M.~A.}\ \bibnamefont {Norcia}}, \bibinfo {author} {\bibfnamefont {F.}~\bibnamefont {Ferlaino}}, \bibinfo {author} {\bibfnamefont {L.}~\bibnamefont {Santos}}, \ and\ \bibinfo {author} {\bibfnamefont {R.}~\bibnamefont {Bisset}},\ }\href {\doibase 10.1103/PhysRevLett.128.195302} {\bibfield  {journal} {\bibinfo  {journal} {Phys. Rev. Lett.}\ }\textbf {\bibinfo {volume} {128}},\ \bibinfo {pages} {195302} (\bibinfo {year} {2022})}\BibitemShut {NoStop}%
\bibitem [{\citenamefont {Hadzibabic}\ \emph {et~al.}(2004)\citenamefont {Hadzibabic}, \citenamefont {Stock}, \citenamefont {Battelier}, \citenamefont {Bretin},\ and\ \citenamefont {Dalibard}}]{Hadzibabic_interference_2004}%
  \BibitemOpen
  \bibfield  {author} {\bibinfo {author} {\bibfnamefont {Z.}~\bibnamefont {Hadzibabic}}, \bibinfo {author} {\bibfnamefont {S.}~\bibnamefont {Stock}}, \bibinfo {author} {\bibfnamefont {B.}~\bibnamefont {Battelier}}, \bibinfo {author} {\bibfnamefont {V.}~\bibnamefont {Bretin}}, \ and\ \bibinfo {author} {\bibfnamefont {J.}~\bibnamefont {Dalibard}},\ }\href {\doibase 10.1103/PhysRevLett.93.180403} {\bibfield  {journal} {\bibinfo  {journal} {Phys. Rev. Lett.}\ }\textbf {\bibinfo {volume} {93}},\ \bibinfo {pages} {180403} (\bibinfo {year} {2004})}\BibitemShut {NoStop}%
\bibitem [{\citenamefont {Denschlag}\ \emph {et~al.}(2000)\citenamefont {Denschlag}, \citenamefont {Simsarian}, \citenamefont {Feder}, \citenamefont {Clark}, \citenamefont {Collins}, \citenamefont {Cubizolles}, \citenamefont {Deng}, \citenamefont {Hagley}, \citenamefont {Helmerson}, \citenamefont {Reinhardt} \emph {et~al.}}]{denschlag_generating_2000}%
  \BibitemOpen
  \bibfield  {author} {\bibinfo {author} {\bibfnamefont {J.}~\bibnamefont {Denschlag}}, \bibinfo {author} {\bibfnamefont {J.~E.}\ \bibnamefont {Simsarian}}, \bibinfo {author} {\bibfnamefont {D.~L.}\ \bibnamefont {Feder}}, \bibinfo {author} {\bibfnamefont {C.~W.}\ \bibnamefont {Clark}}, \bibinfo {author} {\bibfnamefont {L.~A.}\ \bibnamefont {Collins}}, \bibinfo {author} {\bibfnamefont {J.}~\bibnamefont {Cubizolles}}, \bibinfo {author} {\bibfnamefont {L.}~\bibnamefont {Deng}}, \bibinfo {author} {\bibfnamefont {E.~W.}\ \bibnamefont {Hagley}}, \bibinfo {author} {\bibfnamefont {K.}~\bibnamefont {Helmerson}}, \bibinfo {author} {\bibfnamefont {W.~P.}\ \bibnamefont {Reinhardt}},  \emph {et~al.},\ }\href {\doibase 10.1126/science.287.5450.97} {\bibfield  {journal} {\bibinfo  {journal} {Science}\ }\textbf {\bibinfo {volume} {287}},\ \bibinfo {pages} {97} (\bibinfo {year} {2000})}\BibitemShut {NoStop}%
\bibitem [{\citenamefont {Meyer}\ \emph {et~al.}(2017)\citenamefont {Meyer}, \citenamefont {Proud}, \citenamefont {Perea-Ortiz}, \citenamefont {O’Neale}, \citenamefont {Baumert}, \citenamefont {Holynski}, \citenamefont {Kronj{\"a}ger}, \citenamefont {Barontini},\ and\ \citenamefont {Bongs}}]{meyer2017observation}%
  \BibitemOpen
  \bibfield  {author} {\bibinfo {author} {\bibfnamefont {N.}~\bibnamefont {Meyer}}, \bibinfo {author} {\bibfnamefont {H.}~\bibnamefont {Proud}}, \bibinfo {author} {\bibfnamefont {M.}~\bibnamefont {Perea-Ortiz}}, \bibinfo {author} {\bibfnamefont {C.}~\bibnamefont {O’Neale}}, \bibinfo {author} {\bibfnamefont {M.}~\bibnamefont {Baumert}}, \bibinfo {author} {\bibfnamefont {M.}~\bibnamefont {Holynski}}, \bibinfo {author} {\bibfnamefont {J.}~\bibnamefont {Kronj{\"a}ger}}, \bibinfo {author} {\bibfnamefont {G.}~\bibnamefont {Barontini}}, \ and\ \bibinfo {author} {\bibfnamefont {K.}~\bibnamefont {Bongs}},\ }\href {https://doi.org/10.1103/PhysRevLett.119.150403} {\bibfield  {journal} {\bibinfo  {journal} {Phys. Rev. Lett.}\ }\textbf {\bibinfo {volume} {119}},\ \bibinfo {pages} {150403} (\bibinfo {year} {2017})}\BibitemShut {NoStop}%
\bibitem [{\citenamefont {Leanhardt}\ \emph {et~al.}(2002)\citenamefont {Leanhardt}, \citenamefont {G\"orlitz}, \citenamefont {Chikkatur}, \citenamefont {Kielpinski}, \citenamefont {Shin}, \citenamefont {Pritchard},\ and\ \citenamefont {Ketterle}}]{Leanhardt_imprinting_2002}%
  \BibitemOpen
  \bibfield  {author} {\bibinfo {author} {\bibfnamefont {A.~E.}\ \bibnamefont {Leanhardt}}, \bibinfo {author} {\bibfnamefont {A.}~\bibnamefont {G\"orlitz}}, \bibinfo {author} {\bibfnamefont {A.~P.}\ \bibnamefont {Chikkatur}}, \bibinfo {author} {\bibfnamefont {D.}~\bibnamefont {Kielpinski}}, \bibinfo {author} {\bibfnamefont {Y.}~\bibnamefont {Shin}}, \bibinfo {author} {\bibfnamefont {D.~E.}\ \bibnamefont {Pritchard}}, \ and\ \bibinfo {author} {\bibfnamefont {W.}~\bibnamefont {Ketterle}},\ }\href {\doibase 10.1103/PhysRevLett.89.190403} {\bibfield  {journal} {\bibinfo  {journal} {Phys. Rev. Lett.}\ }\textbf {\bibinfo {volume} {89}},\ \bibinfo {pages} {190403} (\bibinfo {year} {2002})}\BibitemShut {NoStop}%
\bibitem [{\citenamefont {Gallem\'{\i}}\ \emph {et~al.}(2020)\citenamefont {Gallem\'{\i}}, \citenamefont {Roccuzzo}, \citenamefont {Stringari},\ and\ \citenamefont {Recati}}]{Gallemi_quantized_2020}%
  \BibitemOpen
  \bibfield  {author} {\bibinfo {author} {\bibfnamefont {A.}~\bibnamefont {Gallem\'{\i}}}, \bibinfo {author} {\bibfnamefont {S.~M.}\ \bibnamefont {Roccuzzo}}, \bibinfo {author} {\bibfnamefont {S.}~\bibnamefont {Stringari}}, \ and\ \bibinfo {author} {\bibfnamefont {A.}~\bibnamefont {Recati}},\ }\href {\doibase 10.1103/PhysRevA.102.023322} {\bibfield  {journal} {\bibinfo  {journal} {Phys. Rev. A}\ }\textbf {\bibinfo {volume} {102}},\ \bibinfo {pages} {023322} (\bibinfo {year} {2020})}\BibitemShut {NoStop}%
\bibitem [{\citenamefont {Ancilotto}\ \emph {et~al.}(2021)\citenamefont {Ancilotto}, \citenamefont {Barranco}, \citenamefont {Pi},\ and\ \citenamefont {Reatto}}]{Ancilotto_vortex_2021}%
  \BibitemOpen
  \bibfield  {author} {\bibinfo {author} {\bibfnamefont {F.}~\bibnamefont {Ancilotto}}, \bibinfo {author} {\bibfnamefont {M.}~\bibnamefont {Barranco}}, \bibinfo {author} {\bibfnamefont {M.}~\bibnamefont {Pi}}, \ and\ \bibinfo {author} {\bibfnamefont {L.}~\bibnamefont {Reatto}},\ }\href {\doibase 10.1103/PhysRevA.103.033314} {\bibfield  {journal} {\bibinfo  {journal} {Phys. Rev. A}\ }\textbf {\bibinfo {volume} {103}},\ \bibinfo {pages} {033314} (\bibinfo {year} {2021})}\BibitemShut {NoStop}%
\bibitem [{\citenamefont {Nilsson~Tengstrand}\ \emph {et~al.}(2023)\citenamefont {Nilsson~Tengstrand}, \citenamefont {St{\"u}rmer}, \citenamefont {Ribbing},\ and\ \citenamefont {Reimann}}]{tengstrand2023toroidal}%
  \BibitemOpen
  \bibfield  {author} {\bibinfo {author} {\bibfnamefont {M.}~\bibnamefont {Nilsson~Tengstrand}}, \bibinfo {author} {\bibfnamefont {P.}~\bibnamefont {St{\"u}rmer}}, \bibinfo {author} {\bibfnamefont {J.}~\bibnamefont {Ribbing}}, \ and\ \bibinfo {author} {\bibfnamefont {S.~M.}\ \bibnamefont {Reimann}},\ }\href {https://doi.org/10.1103/PhysRevA.107.063316} {\bibfield  {journal} {\bibinfo  {journal} {Phys. Rev. A}\ }\textbf {\bibinfo {volume} {107}},\ \bibinfo {pages} {063316} (\bibinfo {year} {2023})}\BibitemShut {NoStop}%
\bibitem [{\citenamefont {S{\'a}nchez-Baena}\ \emph {et~al.}(2023)\citenamefont {S{\'a}nchez-Baena}, \citenamefont {Politi}, \citenamefont {Maucher}, \citenamefont {Ferlaino},\ and\ \citenamefont {Pohl}}]{sanchez_heating_2023}%
  \BibitemOpen
  \bibfield  {author} {\bibinfo {author} {\bibfnamefont {J.}~\bibnamefont {S{\'a}nchez-Baena}}, \bibinfo {author} {\bibfnamefont {C.}~\bibnamefont {Politi}}, \bibinfo {author} {\bibfnamefont {F.}~\bibnamefont {Maucher}}, \bibinfo {author} {\bibfnamefont {F.}~\bibnamefont {Ferlaino}}, \ and\ \bibinfo {author} {\bibfnamefont {T.}~\bibnamefont {Pohl}},\ }\href {\doibase 10.1038/s41467-023-37207-3} {\bibfield  {journal} {\bibinfo  {journal} {Nat. Commun.}\ }\textbf {\bibinfo {volume} {14}},\ \bibinfo {pages} {1868} (\bibinfo {year} {2023})}\BibitemShut {NoStop}%
\bibitem [{\citenamefont {Prasad}\ \emph {et~al.}(2019)\citenamefont {Prasad}, \citenamefont {Bland}, \citenamefont {Mulkerin}, \citenamefont {Parker},\ and\ \citenamefont {Martin}}]{prasad_instability_2019}%
  \BibitemOpen
  \bibfield  {author} {\bibinfo {author} {\bibfnamefont {S.~B.}\ \bibnamefont {Prasad}}, \bibinfo {author} {\bibfnamefont {T.}~\bibnamefont {Bland}}, \bibinfo {author} {\bibfnamefont {B.~C.}\ \bibnamefont {Mulkerin}}, \bibinfo {author} {\bibfnamefont {N.~G.}\ \bibnamefont {Parker}}, \ and\ \bibinfo {author} {\bibfnamefont {A.~M.}\ \bibnamefont {Martin}},\ }\href {\doibase 10.1103/PhysRevLett.122.050401} {\bibfield  {journal} {\bibinfo  {journal} {Phys. Rev. Lett.}\ }\textbf {\bibinfo {volume} {122}},\ \bibinfo {pages} {050401} (\bibinfo {year} {2019})},\ \bibinfo {note} {publisher: American Physical Society}\BibitemShut {NoStop}%
\bibitem [{\citenamefont {Klaus}\ \emph {et~al.}(2022)\citenamefont {Klaus}, \citenamefont {Bland}, \citenamefont {Poli}, \citenamefont {Politi}, \citenamefont {Lamporesi}, \citenamefont {Casotti}, \citenamefont {Bisset}, \citenamefont {Mark},\ and\ \citenamefont {Ferlaino}}]{Klaus2022}%
  \BibitemOpen
  \bibfield  {author} {\bibinfo {author} {\bibfnamefont {L.}~\bibnamefont {Klaus}}, \bibinfo {author} {\bibfnamefont {T.}~\bibnamefont {Bland}}, \bibinfo {author} {\bibfnamefont {E.}~\bibnamefont {Poli}}, \bibinfo {author} {\bibfnamefont {C.}~\bibnamefont {Politi}}, \bibinfo {author} {\bibfnamefont {G.}~\bibnamefont {Lamporesi}}, \bibinfo {author} {\bibfnamefont {E.}~\bibnamefont {Casotti}}, \bibinfo {author} {\bibfnamefont {R.~N.}\ \bibnamefont {Bisset}}, \bibinfo {author} {\bibfnamefont {M.~J.}\ \bibnamefont {Mark}}, \ and\ \bibinfo {author} {\bibfnamefont {F.}~\bibnamefont {Ferlaino}},\ }\href {\doibase 10.1038/s41567-022-01793-8} {\bibfield  {journal} {\bibinfo  {journal} {Nat. Phys.}\ }\textbf {\bibinfo {volume} {18}},\ \bibinfo {pages} {1453} (\bibinfo {year} {2022})}\BibitemShut {NoStop}%
\bibitem [{\citenamefont {Bland}\ \emph {et~al.}(2023)\citenamefont {Bland}, \citenamefont {Lamporesi}, \citenamefont {Mark},\ and\ \citenamefont {Ferlaino}}]{bland_vortices_2023}%
  \BibitemOpen
  \bibfield  {author} {\bibinfo {author} {\bibfnamefont {T.}~\bibnamefont {Bland}}, \bibinfo {author} {\bibfnamefont {G.}~\bibnamefont {Lamporesi}}, \bibinfo {author} {\bibfnamefont {M.~J.}\ \bibnamefont {Mark}}, \ and\ \bibinfo {author} {\bibfnamefont {F.}~\bibnamefont {Ferlaino}},\ }\href {\doibase 10.5802/crphys.160} {\bibfield  {journal} {\bibinfo  {journal} {Comptes Rendus. Physique}\ }\textbf {\bibinfo {volume} {24}},\ \bibinfo {pages} {133} (\bibinfo {year} {2023})}\BibitemShut {NoStop}%
\bibitem [{\citenamefont {Bougas}\ \emph {et~al.}(2026)\citenamefont {Bougas}, \citenamefont {Bland}, \citenamefont {Sadeghpour},\ and\ \citenamefont {Mistakidis}}]{prep}%
  \BibitemOpen
  \bibfield  {author} {\bibinfo {author} {\bibfnamefont {G.~A.}\ \bibnamefont {Bougas}}, \bibinfo {author} {\bibfnamefont {T.}~\bibnamefont {Bland}}, \bibinfo {author} {\bibfnamefont {H.~R.}\ \bibnamefont {Sadeghpour}}, \ and\ \bibinfo {author} {\bibfnamefont {S.~I.}\ \bibnamefont {Mistakidis}},\ }\href@noop {} {\emph {\bibinfo {title} {{\rm unpublished}}}}\ (\bibinfo {year} {2026})\BibitemShut {NoStop}%
\bibitem [{\citenamefont {Theocharis}\ \emph {et~al.}(2003)\citenamefont {Theocharis}, \citenamefont {Frantzeskakis}, \citenamefont {Kevrekidis}, \citenamefont {Malomed},\ and\ \citenamefont {Kivshar}}]{Theocharis_ring}%
  \BibitemOpen
  \bibfield  {author} {\bibinfo {author} {\bibfnamefont {G.}~\bibnamefont {Theocharis}}, \bibinfo {author} {\bibfnamefont {D.~J.}\ \bibnamefont {Frantzeskakis}}, \bibinfo {author} {\bibfnamefont {P.~G.}\ \bibnamefont {Kevrekidis}}, \bibinfo {author} {\bibfnamefont {B.~A.}\ \bibnamefont {Malomed}}, \ and\ \bibinfo {author} {\bibfnamefont {Y.~S.}\ \bibnamefont {Kivshar}},\ }\href {\doibase 10.1103/PhysRevLett.90.120403} {\bibfield  {journal} {\bibinfo  {journal} {Phys. Rev. Lett.}\ }\textbf {\bibinfo {volume} {90}},\ \bibinfo {pages} {120403} (\bibinfo {year} {2003})}\BibitemShut {NoStop}%
\bibitem [{\citenamefont {Parker}\ \emph {et~al.}(2009)\citenamefont {Parker}, \citenamefont {Ticknor}, \citenamefont {Martin},\ and\ \citenamefont {O'Dell}}]{Parker_structure_2009}%
  \BibitemOpen
  \bibfield  {author} {\bibinfo {author} {\bibfnamefont {N.~G.}\ \bibnamefont {Parker}}, \bibinfo {author} {\bibfnamefont {C.}~\bibnamefont {Ticknor}}, \bibinfo {author} {\bibfnamefont {A.~M.}\ \bibnamefont {Martin}}, \ and\ \bibinfo {author} {\bibfnamefont {D.~H.~J.}\ \bibnamefont {O'Dell}},\ }\href {\doibase 10.1103/PhysRevA.79.013617} {\bibfield  {journal} {\bibinfo  {journal} {Phys. Rev. A}\ }\textbf {\bibinfo {volume} {79}},\ \bibinfo {pages} {013617} (\bibinfo {year} {2009})}\BibitemShut {NoStop}%
\bibitem [{\citenamefont {Bohn}\ \emph {et~al.}(2009)\citenamefont {Bohn}, \citenamefont {Wilson},\ and\ \citenamefont {Ronen}}]{Bohn_collapse_2009}%
  \BibitemOpen
  \bibfield  {author} {\bibinfo {author} {\bibfnamefont {J.~L.}\ \bibnamefont {Bohn}}, \bibinfo {author} {\bibfnamefont {R.~M.}\ \bibnamefont {Wilson}}, \ and\ \bibinfo {author} {\bibfnamefont {S.}~\bibnamefont {Ronen}},\ }\href {\doibase 10.1134/S1054660X09040021} {\bibfield  {journal} {\bibinfo  {journal} {Laser Phys.}\ }\textbf {\bibinfo {volume} {19}},\ \bibinfo {pages} {547} (\bibinfo {year} {2009})}\BibitemShut {NoStop}%
\bibitem [{\citenamefont {Sch{\"u}tzhold}\ \emph {et~al.}(2006)\citenamefont {Sch{\"u}tzhold}, \citenamefont {Uhlmann}, \citenamefont {Xu},\ and\ \citenamefont {Fischer}}]{schutzhold_mean_2006}%
  \BibitemOpen
  \bibfield  {author} {\bibinfo {author} {\bibfnamefont {R.}~\bibnamefont {Sch{\"u}tzhold}}, \bibinfo {author} {\bibfnamefont {M.}~\bibnamefont {Uhlmann}}, \bibinfo {author} {\bibfnamefont {Y.}~\bibnamefont {Xu}}, \ and\ \bibinfo {author} {\bibfnamefont {U.~R.}\ \bibnamefont {Fischer}},\ }\href {\doibase 10.1142/S0217979206035631} {\bibfield  {journal} {\bibinfo  {journal} {Int. J. Mod. Phys. B}\ }\textbf {\bibinfo {volume} {20}},\ \bibinfo {pages} {3555} (\bibinfo {year} {2006})}\BibitemShut {NoStop}%
\bibitem [{\citenamefont {Kivshar}\ and\ \citenamefont {Luther-Davies}(1998)}]{Kivshar_dark_1998}%
  \BibitemOpen
  \bibfield  {author} {\bibinfo {author} {\bibfnamefont {Y.~S.}\ \bibnamefont {Kivshar}}\ and\ \bibinfo {author} {\bibfnamefont {B.}~\bibnamefont {Luther-Davies}},\ }\href {\doibase https://doi.org/10.1016/S0370-1573(97)00073-2} {\bibfield  {journal} {\bibinfo  {journal} {Phys. Rep.}\ }\textbf {\bibinfo {volume} {298}},\ \bibinfo {pages} {81} (\bibinfo {year} {1998})}\BibitemShut {NoStop}%
\bibitem [{\citenamefont {Frantzeskakis}(2010)}]{Frantzeskakis_dark_2010}%
  \BibitemOpen
  \bibfield  {author} {\bibinfo {author} {\bibfnamefont {D.~J.}\ \bibnamefont {Frantzeskakis}},\ }\href {\doibase 10.1088/1751-8113/43/21/213001} {\bibfield  {journal} {\bibinfo  {journal} {J. Phys. A: Math. Theor.}\ }\textbf {\bibinfo {volume} {43}},\ \bibinfo {pages} {213001} (\bibinfo {year} {2010})}\BibitemShut {NoStop}%
\bibitem [{\citenamefont {Romero-Ros}\ \emph {et~al.}(2021)\citenamefont {Romero-Ros}, \citenamefont {Katsimiga}, \citenamefont {Kevrekidis}, \citenamefont {Prinari}, \citenamefont {Biondini},\ and\ \citenamefont {Schmelcher}}]{Ros_demand_2021}%
  \BibitemOpen
  \bibfield  {author} {\bibinfo {author} {\bibfnamefont {A.}~\bibnamefont {Romero-Ros}}, \bibinfo {author} {\bibfnamefont {G.~C.}\ \bibnamefont {Katsimiga}}, \bibinfo {author} {\bibfnamefont {P.~G.}\ \bibnamefont {Kevrekidis}}, \bibinfo {author} {\bibfnamefont {B.}~\bibnamefont {Prinari}}, \bibinfo {author} {\bibfnamefont {G.}~\bibnamefont {Biondini}}, \ and\ \bibinfo {author} {\bibfnamefont {P.}~\bibnamefont {Schmelcher}},\ }\href {\doibase 10.1103/PhysRevA.103.023329} {\bibfield  {journal} {\bibinfo  {journal} {Phys. Rev. A}\ }\textbf {\bibinfo {volume} {103}},\ \bibinfo {pages} {023329} (\bibinfo {year} {2021})}\BibitemShut {NoStop}%
\bibitem [{\citenamefont {Weller}\ \emph {et~al.}(2008)\citenamefont {Weller}, \citenamefont {Ronzheimer}, \citenamefont {Gross}, \citenamefont {Esteve}, \citenamefont {Oberthaler}, \citenamefont {Frantzeskakis}, \citenamefont {Theocharis},\ and\ \citenamefont {Kevrekidis}}]{Weller_experimental_2008}%
  \BibitemOpen
  \bibfield  {author} {\bibinfo {author} {\bibfnamefont {A.}~\bibnamefont {Weller}}, \bibinfo {author} {\bibfnamefont {J.~P.}\ \bibnamefont {Ronzheimer}}, \bibinfo {author} {\bibfnamefont {C.}~\bibnamefont {Gross}}, \bibinfo {author} {\bibfnamefont {J.}~\bibnamefont {Esteve}}, \bibinfo {author} {\bibfnamefont {M.~K.}\ \bibnamefont {Oberthaler}}, \bibinfo {author} {\bibfnamefont {D.~J.}\ \bibnamefont {Frantzeskakis}}, \bibinfo {author} {\bibfnamefont {G.}~\bibnamefont {Theocharis}}, \ and\ \bibinfo {author} {\bibfnamefont {P.~G.}\ \bibnamefont {Kevrekidis}},\ }\href {\doibase 10.1103/PhysRevLett.101.130401} {\bibfield  {journal} {\bibinfo  {journal} {Phys. Rev. Lett.}\ }\textbf {\bibinfo {volume} {101}},\ \bibinfo {pages} {130401} (\bibinfo {year} {2008})}\BibitemShut {NoStop}%
\bibitem [{\citenamefont {Becker}\ \emph {et~al.}(2008)\citenamefont {Becker}, \citenamefont {Stellmer}, \citenamefont {Soltan-Panahi}, \citenamefont {D{\"o}rscher}, \citenamefont {Baumert}, \citenamefont {Richter}, \citenamefont {Kronj{\"a}ger}, \citenamefont {Bongs},\ and\ \citenamefont {Sengstock}}]{Becker_oscillations_2008}%
  \BibitemOpen
  \bibfield  {author} {\bibinfo {author} {\bibfnamefont {C.}~\bibnamefont {Becker}}, \bibinfo {author} {\bibfnamefont {S.}~\bibnamefont {Stellmer}}, \bibinfo {author} {\bibfnamefont {P.}~\bibnamefont {Soltan-Panahi}}, \bibinfo {author} {\bibfnamefont {S.}~\bibnamefont {D{\"o}rscher}}, \bibinfo {author} {\bibfnamefont {M.}~\bibnamefont {Baumert}}, \bibinfo {author} {\bibfnamefont {E.-M.}\ \bibnamefont {Richter}}, \bibinfo {author} {\bibfnamefont {J.}~\bibnamefont {Kronj{\"a}ger}}, \bibinfo {author} {\bibfnamefont {K.}~\bibnamefont {Bongs}}, \ and\ \bibinfo {author} {\bibfnamefont {K.}~\bibnamefont {Sengstock}},\ }\href {\doibase 10.1038/nphys962} {\bibfield  {journal} {\bibinfo  {journal} {Nat. Phys.}\ }\textbf {\bibinfo {volume} {4}},\ \bibinfo {pages} {496} (\bibinfo {year} {2008})}\BibitemShut {NoStop}%
\bibitem [{\citenamefont {Busch}\ and\ \citenamefont {Anglin}(2000)}]{Busch_motion_2000}%
  \BibitemOpen
  \bibfield  {author} {\bibinfo {author} {\bibfnamefont {T.}~\bibnamefont {Busch}}\ and\ \bibinfo {author} {\bibfnamefont {J.~R.}\ \bibnamefont {Anglin}},\ }\href {\doibase 10.1103/PhysRevLett.84.2298} {\bibfield  {journal} {\bibinfo  {journal} {Phys. Rev. Lett.}\ }\textbf {\bibinfo {volume} {84}},\ \bibinfo {pages} {2298} (\bibinfo {year} {2000})}\BibitemShut {NoStop}%
\bibitem [{\citenamefont {Mistakidis}\ \emph {et~al.}(2024)\citenamefont {Mistakidis}, \citenamefont {Mukherjee}, \citenamefont {Reimann},\ and\ \citenamefont {Sadeghpour}}]{Mistakidis2024}%
  \BibitemOpen
  \bibfield  {author} {\bibinfo {author} {\bibfnamefont {S.~I.}\ \bibnamefont {Mistakidis}}, \bibinfo {author} {\bibfnamefont {K.}~\bibnamefont {Mukherjee}}, \bibinfo {author} {\bibfnamefont {S.~M.}\ \bibnamefont {Reimann}}, \ and\ \bibinfo {author} {\bibfnamefont {H.~R.}\ \bibnamefont {Sadeghpour}},\ }\href {\doibase 10.1103/PhysRevA.110.013323} {\bibfield  {journal} {\bibinfo  {journal} {Phys. Rev. A}\ }\textbf {\bibinfo {volume} {110}},\ \bibinfo {pages} {013323} (\bibinfo {year} {2024})}\BibitemShut {NoStop}%
\bibitem [{\citenamefont {Crank}\ and\ \citenamefont {Nicolson}(1947)}]{crank_practical_1947}%
  \BibitemOpen
  \bibfield  {author} {\bibinfo {author} {\bibfnamefont {J.}~\bibnamefont {Crank}}\ and\ \bibinfo {author} {\bibfnamefont {P.}~\bibnamefont {Nicolson}},\ }in\ \href@noop {} {\emph {\bibinfo {booktitle} {Math. Proc. Camb. Philos. Soc.}}},\ Vol.~\bibinfo {volume} {43}\ (\bibinfo {organization} {Cambridge University Press},\ \bibinfo {year} {1947})\ pp.\ \bibinfo {pages} {50--67}\BibitemShut {NoStop}%
\bibitem [{\citenamefont {Antoine}\ \emph {et~al.}(2013)\citenamefont {Antoine}, \citenamefont {Bao},\ and\ \citenamefont {Besse}}]{antoine_computational_2013}%
  \BibitemOpen
  \bibfield  {author} {\bibinfo {author} {\bibfnamefont {X.}~\bibnamefont {Antoine}}, \bibinfo {author} {\bibfnamefont {W.}~\bibnamefont {Bao}}, \ and\ \bibinfo {author} {\bibfnamefont {C.}~\bibnamefont {Besse}},\ }\href {\doibase https://doi.org/10.1016/j.cpc.2013.07.012} {\bibfield  {journal} {\bibinfo  {journal} {Comput. Phys. Commun.}\ }\textbf {\bibinfo {volume} {184}},\ \bibinfo {pages} {2621} (\bibinfo {year} {2013})}\BibitemShut {NoStop}%
\bibitem [{\citenamefont {Arfken}\ and\ \citenamefont {Weber}(1972)}]{arfken_mathematical_1972}%
  \BibitemOpen
  \bibfield  {author} {\bibinfo {author} {\bibfnamefont {G.~B.}\ \bibnamefont {Arfken}}\ and\ \bibinfo {author} {\bibfnamefont {H.-J.}\ \bibnamefont {Weber}},\ }\href@noop {} {\emph {\bibinfo {title} {Mathematical methods for physicists}}}\ (\bibinfo  {publisher} {Academic Press Orlando, FL},\ \bibinfo {year} {1972})\BibitemShut {NoStop}%
\bibitem [{\citenamefont {G\'oral}\ and\ \citenamefont {Santos}(2002)}]{Goral_ground_2002}%
  \BibitemOpen
  \bibfield  {author} {\bibinfo {author} {\bibfnamefont {K.}~\bibnamefont {G\'oral}}\ and\ \bibinfo {author} {\bibfnamefont {L.}~\bibnamefont {Santos}},\ }\href {\doibase 10.1103/PhysRevA.66.023613} {\bibfield  {journal} {\bibinfo  {journal} {Phys. Rev. A}\ }\textbf {\bibinfo {volume} {66}},\ \bibinfo {pages} {023613} (\bibinfo {year} {2002})}\BibitemShut {NoStop}%
\end{thebibliography}%


\begin{thebibliography}{27}%
\makeatletter
\providecommand \@ifxundefined [1]{%
 \@ifx{#1\undefined}
}%
\providecommand \@ifnum [1]{%
 \ifnum #1\expandafter \@firstoftwo
 \else \expandafter \@secondoftwo
 \fi
}%
\providecommand \@ifx [1]{%
 \ifx #1\expandafter \@firstoftwo
 \else \expandafter \@secondoftwo
 \fi
}%
\providecommand \natexlab [1]{#1}%
\providecommand \enquote  [1]{``#1''}%
\providecommand \bibnamefont  [1]{#1}%
\providecommand \bibfnamefont [1]{#1}%
\providecommand \citenamefont [1]{#1}%
\providecommand \href@noop [0]{\@secondoftwo}%
\providecommand \href [0]{\begingroup \@sanitize@url \@href}%
\providecommand \@href[1]{\@@startlink{#1}\@@href}%
\providecommand \@@href[1]{\endgroup#1\@@endlink}%
\providecommand \@sanitize@url [0]{\catcode `\\12\catcode `\$12\catcode `\&12\catcode `\#12\catcode `\^12\catcode `\_12\catcode `\%12\relax}%
\providecommand \@@startlink[1]{}%
\providecommand \@@endlink[0]{}%
\providecommand \url  [0]{\begingroup\@sanitize@url \@url }%
\providecommand \@url [1]{\endgroup\@href {#1}{\urlprefix }}%
\providecommand \urlprefix  [0]{URL }%
\providecommand \Eprint [0]{\href }%
\providecommand \doibase [0]{http://dx.doi.org/}%
\providecommand \selectlanguage [0]{\@gobble}%
\providecommand \bibinfo  [0]{\@secondoftwo}%
\providecommand \bibfield  [0]{\@secondoftwo}%
\providecommand \translation [1]{[#1]}%
\providecommand \BibitemOpen [0]{}%
\providecommand \bibitemStop [0]{}%
\providecommand \bibitemNoStop [0]{.\EOS\space}%
\providecommand \EOS [0]{\spacefactor3000\relax}%
\providecommand \BibitemShut  [1]{\csname bibitem#1\endcsname}%
\let\auto@bib@innerbib\@empty
\bibitem [{\citenamefont {W\"achtler}\ and\ \citenamefont {Santos}(2016)}]{Santos2016filemanets}%
  \BibitemOpen
  \bibfield  {author} {\bibinfo {author} {\bibfnamefont {F.}~\bibnamefont {W\"achtler}}\ and\ \bibinfo {author} {\bibfnamefont {L.}~\bibnamefont {Santos}},\ }\href {\doibase 10.1103/PhysRevA.93.061603} {\bibfield  {journal} {\bibinfo  {journal} {Phys. Rev. A}\ }\textbf {\bibinfo {volume} {93}},\ \bibinfo {pages} {061603} (\bibinfo {year} {2016})}\BibitemShut {NoStop}%
\bibitem [{\citenamefont {Bisset}\ \emph {et~al.}(2016)\citenamefont {Bisset}, \citenamefont {Wilson}, \citenamefont {Baillie},\ and\ \citenamefont {Blakie}}]{bisset_ground-state_2016}%
  \BibitemOpen
  \bibfield  {author} {\bibinfo {author} {\bibfnamefont {R.~N.}\ \bibnamefont {Bisset}}, \bibinfo {author} {\bibfnamefont {R.~M.}\ \bibnamefont {Wilson}}, \bibinfo {author} {\bibfnamefont {D.}~\bibnamefont {Baillie}}, \ and\ \bibinfo {author} {\bibfnamefont {P.~B.}\ \bibnamefont {Blakie}},\ }\href {\doibase 10.1103/PhysRevA.94.033619} {\bibfield  {journal} {\bibinfo  {journal} {Phys. Rev. A}\ }\textbf {\bibinfo {volume} {94}},\ \bibinfo {pages} {033619} (\bibinfo {year} {2016})}\BibitemShut {NoStop}%
\bibitem [{\citenamefont {Chomaz}\ \emph {et~al.}(2016)\citenamefont {Chomaz}, \citenamefont {Baier}, \citenamefont {Petter}, \citenamefont {Mark}, \citenamefont {W\"achtler}, \citenamefont {Santos},\ and\ \citenamefont {Ferlaino}}]{chomaz2016quantum}%
  \BibitemOpen
  \bibfield  {author} {\bibinfo {author} {\bibfnamefont {L.}~\bibnamefont {Chomaz}}, \bibinfo {author} {\bibfnamefont {S.}~\bibnamefont {Baier}}, \bibinfo {author} {\bibfnamefont {D.}~\bibnamefont {Petter}}, \bibinfo {author} {\bibfnamefont {M.~J.}\ \bibnamefont {Mark}}, \bibinfo {author} {\bibfnamefont {F.}~\bibnamefont {W\"achtler}}, \bibinfo {author} {\bibfnamefont {L.}~\bibnamefont {Santos}}, \ and\ \bibinfo {author} {\bibfnamefont {F.}~\bibnamefont {Ferlaino}},\ }\href {\doibase 10.1103/PhysRevX.6.041039} {\bibfield  {journal} {\bibinfo  {journal} {Phys. Rev. X}\ }\textbf {\bibinfo {volume} {6}},\ \bibinfo {pages} {041039} (\bibinfo {year} {2016})}\BibitemShut {NoStop}%
\bibitem [{\citenamefont {Ferrier-Barbut}\ \emph {et~al.}(2016)\citenamefont {Ferrier-Barbut}, \citenamefont {Kadau}, \citenamefont {Schmitt}, \citenamefont {Wenzel},\ and\ \citenamefont {Pfau}}]{ferrier-barbut_observation_2016}%
  \BibitemOpen
  \bibfield  {author} {\bibinfo {author} {\bibfnamefont {I.}~\bibnamefont {Ferrier-Barbut}}, \bibinfo {author} {\bibfnamefont {H.}~\bibnamefont {Kadau}}, \bibinfo {author} {\bibfnamefont {M.}~\bibnamefont {Schmitt}}, \bibinfo {author} {\bibfnamefont {M.}~\bibnamefont {Wenzel}}, \ and\ \bibinfo {author} {\bibfnamefont {T.}~\bibnamefont {Pfau}},\ }\href {\doibase 10.1103/PhysRevLett.116.215301} {\bibfield  {journal} {\bibinfo  {journal} {Phys. Rev. Lett.}\ }\textbf {\bibinfo {volume} {116}},\ \bibinfo {pages} {215301} (\bibinfo {year} {2016})}\BibitemShut {NoStop}%
\bibitem [{\citenamefont {Parker}\ \emph {et~al.}(2009)\citenamefont {Parker}, \citenamefont {Ticknor}, \citenamefont {Martin},\ and\ \citenamefont {O'Dell}}]{Parker_structure_2009}%
  \BibitemOpen
  \bibfield  {author} {\bibinfo {author} {\bibfnamefont {N.~G.}\ \bibnamefont {Parker}}, \bibinfo {author} {\bibfnamefont {C.}~\bibnamefont {Ticknor}}, \bibinfo {author} {\bibfnamefont {A.~M.}\ \bibnamefont {Martin}}, \ and\ \bibinfo {author} {\bibfnamefont {D.~H.~J.}\ \bibnamefont {O'Dell}},\ }\href {\doibase 10.1103/PhysRevA.79.013617} {\bibfield  {journal} {\bibinfo  {journal} {Phys. Rev. A}\ }\textbf {\bibinfo {volume} {79}},\ \bibinfo {pages} {013617} (\bibinfo {year} {2009})}\BibitemShut {NoStop}%
\bibitem [{\citenamefont {Bohn}\ \emph {et~al.}(2009)\citenamefont {Bohn}, \citenamefont {Wilson},\ and\ \citenamefont {Ronen}}]{Bohn_collapse_2009}%
  \BibitemOpen
  \bibfield  {author} {\bibinfo {author} {\bibfnamefont {J.~L.}\ \bibnamefont {Bohn}}, \bibinfo {author} {\bibfnamefont {R.~M.}\ \bibnamefont {Wilson}}, \ and\ \bibinfo {author} {\bibfnamefont {S.}~\bibnamefont {Ronen}},\ }\href {\doibase 10.1134/S1054660X09040021} {\bibfield  {journal} {\bibinfo  {journal} {Laser Phys.}\ }\textbf {\bibinfo {volume} {19}},\ \bibinfo {pages} {547} (\bibinfo {year} {2009})}\BibitemShut {NoStop}%
\bibitem [{\citenamefont {Kadau}\ \emph {et~al.}(2016)\citenamefont {Kadau}, \citenamefont {Schmitt}, \citenamefont {Wenzel}, \citenamefont {Wink}, \citenamefont {Maier}, \citenamefont {Ferrier-Barbut},\ and\ \citenamefont {Pfau}}]{kadau_observing_2016}%
  \BibitemOpen
  \bibfield  {author} {\bibinfo {author} {\bibfnamefont {H.}~\bibnamefont {Kadau}}, \bibinfo {author} {\bibfnamefont {M.}~\bibnamefont {Schmitt}}, \bibinfo {author} {\bibfnamefont {M.}~\bibnamefont {Wenzel}}, \bibinfo {author} {\bibfnamefont {C.}~\bibnamefont {Wink}}, \bibinfo {author} {\bibfnamefont {T.}~\bibnamefont {Maier}}, \bibinfo {author} {\bibfnamefont {I.}~\bibnamefont {Ferrier-Barbut}}, \ and\ \bibinfo {author} {\bibfnamefont {T.}~\bibnamefont {Pfau}},\ }\href {\doibase 10.1038/nature16485} {\bibfield  {journal} {\bibinfo  {journal} {Nature}\ }\textbf {\bibinfo {volume} {530}},\ \bibinfo {pages} {194} (\bibinfo {year} {2016})}\BibitemShut {NoStop}%
\bibitem [{\citenamefont {Chomaz}\ \emph {et~al.}(2022)\citenamefont {Chomaz}, \citenamefont {Ferrier-Barbut}, \citenamefont {Ferlaino}, \citenamefont {Laburthe-Tolra}, \citenamefont {Lev},\ and\ \citenamefont {Pfau}}]{chomaz_dipolar_2022}%
  \BibitemOpen
  \bibfield  {author} {\bibinfo {author} {\bibfnamefont {L.}~\bibnamefont {Chomaz}}, \bibinfo {author} {\bibfnamefont {I.}~\bibnamefont {Ferrier-Barbut}}, \bibinfo {author} {\bibfnamefont {F.}~\bibnamefont {Ferlaino}}, \bibinfo {author} {\bibfnamefont {B.}~\bibnamefont {Laburthe-Tolra}}, \bibinfo {author} {\bibfnamefont {B.~L.}\ \bibnamefont {Lev}}, \ and\ \bibinfo {author} {\bibfnamefont {T.}~\bibnamefont {Pfau}},\ }\href {\doibase 10.1088/1361-6633/aca814} {\bibfield  {journal} {\bibinfo  {journal} {Rep. Prog. Phys.}\ }\textbf {\bibinfo {volume} {86}},\ \bibinfo {pages} {026401} (\bibinfo {year} {2022})}\BibitemShut {NoStop}%
\bibitem [{\citenamefont {Lima}\ and\ \citenamefont {Pelster}(2011)}]{Lima_quantum_2011}%
  \BibitemOpen
  \bibfield  {author} {\bibinfo {author} {\bibfnamefont {A.~R.~P.}\ \bibnamefont {Lima}}\ and\ \bibinfo {author} {\bibfnamefont {A.}~\bibnamefont {Pelster}},\ }\href {\doibase 10.1103/PhysRevA.84.041604} {\bibfield  {journal} {\bibinfo  {journal} {Phys. Rev. A}\ }\textbf {\bibinfo {volume} {84}},\ \bibinfo {pages} {041604} (\bibinfo {year} {2011})}\BibitemShut {NoStop}%
\bibitem [{\citenamefont {Sch{\"u}tzhold}\ \emph {et~al.}(2006)\citenamefont {Sch{\"u}tzhold}, \citenamefont {Uhlmann}, \citenamefont {Xu},\ and\ \citenamefont {Fischer}}]{schutzhold_mean_2006}%
  \BibitemOpen
  \bibfield  {author} {\bibinfo {author} {\bibfnamefont {R.}~\bibnamefont {Sch{\"u}tzhold}}, \bibinfo {author} {\bibfnamefont {M.}~\bibnamefont {Uhlmann}}, \bibinfo {author} {\bibfnamefont {Y.}~\bibnamefont {Xu}}, \ and\ \bibinfo {author} {\bibfnamefont {U.~R.}\ \bibnamefont {Fischer}},\ }\href {\doibase 10.1142/S0217979206035631} {\bibfield  {journal} {\bibinfo  {journal} {Int. J. Mod. Phys. B}\ }\textbf {\bibinfo {volume} {20}},\ \bibinfo {pages} {3555} (\bibinfo {year} {2006})}\BibitemShut {NoStop}%
\bibitem [{\citenamefont {Kivshar}\ and\ \citenamefont {Luther-Davies}(1998)}]{Kivshar_dark_1998}%
  \BibitemOpen
  \bibfield  {author} {\bibinfo {author} {\bibfnamefont {Y.~S.}\ \bibnamefont {Kivshar}}\ and\ \bibinfo {author} {\bibfnamefont {B.}~\bibnamefont {Luther-Davies}},\ }\href {\doibase https://doi.org/10.1016/S0370-1573(97)00073-2} {\bibfield  {journal} {\bibinfo  {journal} {Phys. Rep.}\ }\textbf {\bibinfo {volume} {298}},\ \bibinfo {pages} {81} (\bibinfo {year} {1998})}\BibitemShut {NoStop}%
\bibitem [{\citenamefont {Frantzeskakis}(2010)}]{Frantzeskakis_dark_2010}%
  \BibitemOpen
  \bibfield  {author} {\bibinfo {author} {\bibfnamefont {D.~J.}\ \bibnamefont {Frantzeskakis}},\ }\href {\doibase 10.1088/1751-8113/43/21/213001} {\bibfield  {journal} {\bibinfo  {journal} {J. Phys. A: Math. Theor.}\ }\textbf {\bibinfo {volume} {43}},\ \bibinfo {pages} {213001} (\bibinfo {year} {2010})}\BibitemShut {NoStop}%
\bibitem [{\citenamefont {Paw{\l}owski}\ and\ \citenamefont {Rz{\k{a}}{\.z}ewski}(2015)}]{pawlowski2015dipolar}%
  \BibitemOpen
  \bibfield  {author} {\bibinfo {author} {\bibfnamefont {K.}~\bibnamefont {Paw{\l}owski}}\ and\ \bibinfo {author} {\bibfnamefont {K.}~\bibnamefont {Rz{\k{a}}{\.z}ewski}},\ }\href {\doibase 10.1088/1367-2630/17/10/105006} {\bibfield  {journal} {\bibinfo  {journal} {New J. Phys.}\ }\textbf {\bibinfo {volume} {17}},\ \bibinfo {pages} {105006} (\bibinfo {year} {2015})}\BibitemShut {NoStop}%
\bibitem [{\citenamefont {Bland}\ \emph {et~al.}(2015)\citenamefont {Bland}, \citenamefont {Edmonds}, \citenamefont {Proukakis}, \citenamefont {Martin}, \citenamefont {O'Dell},\ and\ \citenamefont {Parker}}]{bland2015controllable}%
  \BibitemOpen
  \bibfield  {author} {\bibinfo {author} {\bibfnamefont {T.}~\bibnamefont {Bland}}, \bibinfo {author} {\bibfnamefont {M.}~\bibnamefont {Edmonds}}, \bibinfo {author} {\bibfnamefont {N.}~\bibnamefont {Proukakis}}, \bibinfo {author} {\bibfnamefont {A.}~\bibnamefont {Martin}}, \bibinfo {author} {\bibfnamefont {D.}~\bibnamefont {O'Dell}}, \ and\ \bibinfo {author} {\bibfnamefont {N.}~\bibnamefont {Parker}},\ }\href {https://doi.org/10.1103/PhysRevA.92.063601} {\bibfield  {journal} {\bibinfo  {journal} {Phys. Rev. A}\ }\textbf {\bibinfo {volume} {92}},\ \bibinfo {pages} {063601} (\bibinfo {year} {2015})}\BibitemShut {NoStop}%
\bibitem [{\citenamefont {Bland}\ \emph {et~al.}(2017)\citenamefont {Bland}, \citenamefont {Paw\l{}owski}, \citenamefont {Edmonds}, \citenamefont {Rz\k{a}\ifmmode~\dot{z}\else \.{z}\fi{}ewski},\ and\ \citenamefont {Parker}}]{Bland_interaction_2017}%
  \BibitemOpen
  \bibfield  {author} {\bibinfo {author} {\bibfnamefont {T.}~\bibnamefont {Bland}}, \bibinfo {author} {\bibfnamefont {K.}~\bibnamefont {Paw\l{}owski}}, \bibinfo {author} {\bibfnamefont {M.~J.}\ \bibnamefont {Edmonds}}, \bibinfo {author} {\bibfnamefont {K.}~\bibnamefont {Rz\k{a}\ifmmode~\dot{z}\else \.{z}\fi{}ewski}}, \ and\ \bibinfo {author} {\bibfnamefont {N.~G.}\ \bibnamefont {Parker}},\ }\href {\doibase 10.1103/PhysRevA.95.063622} {\bibfield  {journal} {\bibinfo  {journal} {Phys. Rev. A}\ }\textbf {\bibinfo {volume} {95}},\ \bibinfo {pages} {063622} (\bibinfo {year} {2017})}\BibitemShut {NoStop}%
\bibitem [{\citenamefont {Edmonds}\ \emph {et~al.}(2016)\citenamefont {Edmonds}, \citenamefont {Bland}, \citenamefont {O'Dell},\ and\ \citenamefont {Parker}}]{edmonds2016exploring}%
  \BibitemOpen
  \bibfield  {author} {\bibinfo {author} {\bibfnamefont {M.}~\bibnamefont {Edmonds}}, \bibinfo {author} {\bibfnamefont {T.}~\bibnamefont {Bland}}, \bibinfo {author} {\bibfnamefont {D.}~\bibnamefont {O'Dell}}, \ and\ \bibinfo {author} {\bibfnamefont {N.}~\bibnamefont {Parker}},\ }\href {https://doi.org/10.1103/PhysRevA.93.063617} {\bibfield  {journal} {\bibinfo  {journal} {Phys. Rev. A}\ }\textbf {\bibinfo {volume} {93}},\ \bibinfo {pages} {063617} (\bibinfo {year} {2016})}\BibitemShut {NoStop}%
\bibitem [{\citenamefont {Romero-Ros}\ \emph {et~al.}(2021)\citenamefont {Romero-Ros}, \citenamefont {Katsimiga}, \citenamefont {Kevrekidis}, \citenamefont {Prinari}, \citenamefont {Biondini},\ and\ \citenamefont {Schmelcher}}]{Ros_demand_2021}%
  \BibitemOpen
  \bibfield  {author} {\bibinfo {author} {\bibfnamefont {A.}~\bibnamefont {Romero-Ros}}, \bibinfo {author} {\bibfnamefont {G.~C.}\ \bibnamefont {Katsimiga}}, \bibinfo {author} {\bibfnamefont {P.~G.}\ \bibnamefont {Kevrekidis}}, \bibinfo {author} {\bibfnamefont {B.}~\bibnamefont {Prinari}}, \bibinfo {author} {\bibfnamefont {G.}~\bibnamefont {Biondini}}, \ and\ \bibinfo {author} {\bibfnamefont {P.}~\bibnamefont {Schmelcher}},\ }\href {\doibase 10.1103/PhysRevA.103.023329} {\bibfield  {journal} {\bibinfo  {journal} {Phys. Rev. A}\ }\textbf {\bibinfo {volume} {103}},\ \bibinfo {pages} {023329} (\bibinfo {year} {2021})}\BibitemShut {NoStop}%
\bibitem [{\citenamefont {Weller}\ \emph {et~al.}(2008)\citenamefont {Weller}, \citenamefont {Ronzheimer}, \citenamefont {Gross}, \citenamefont {Esteve}, \citenamefont {Oberthaler}, \citenamefont {Frantzeskakis}, \citenamefont {Theocharis},\ and\ \citenamefont {Kevrekidis}}]{Weller_experimental_2008}%
  \BibitemOpen
  \bibfield  {author} {\bibinfo {author} {\bibfnamefont {A.}~\bibnamefont {Weller}}, \bibinfo {author} {\bibfnamefont {J.~P.}\ \bibnamefont {Ronzheimer}}, \bibinfo {author} {\bibfnamefont {C.}~\bibnamefont {Gross}}, \bibinfo {author} {\bibfnamefont {J.}~\bibnamefont {Esteve}}, \bibinfo {author} {\bibfnamefont {M.~K.}\ \bibnamefont {Oberthaler}}, \bibinfo {author} {\bibfnamefont {D.~J.}\ \bibnamefont {Frantzeskakis}}, \bibinfo {author} {\bibfnamefont {G.}~\bibnamefont {Theocharis}}, \ and\ \bibinfo {author} {\bibfnamefont {P.~G.}\ \bibnamefont {Kevrekidis}},\ }\href {\doibase 10.1103/PhysRevLett.101.130401} {\bibfield  {journal} {\bibinfo  {journal} {Phys. Rev. Lett.}\ }\textbf {\bibinfo {volume} {101}},\ \bibinfo {pages} {130401} (\bibinfo {year} {2008})}\BibitemShut {NoStop}%
\bibitem [{\citenamefont {Denschlag}\ \emph {et~al.}(2000)\citenamefont {Denschlag}, \citenamefont {Simsarian}, \citenamefont {Feder}, \citenamefont {Clark}, \citenamefont {Collins}, \citenamefont {Cubizolles}, \citenamefont {Deng}, \citenamefont {Hagley}, \citenamefont {Helmerson}, \citenamefont {Reinhardt} \emph {et~al.}}]{denschlag_generating_2000}%
  \BibitemOpen
  \bibfield  {author} {\bibinfo {author} {\bibfnamefont {J.}~\bibnamefont {Denschlag}}, \bibinfo {author} {\bibfnamefont {J.~E.}\ \bibnamefont {Simsarian}}, \bibinfo {author} {\bibfnamefont {D.~L.}\ \bibnamefont {Feder}}, \bibinfo {author} {\bibfnamefont {C.~W.}\ \bibnamefont {Clark}}, \bibinfo {author} {\bibfnamefont {L.~A.}\ \bibnamefont {Collins}}, \bibinfo {author} {\bibfnamefont {J.}~\bibnamefont {Cubizolles}}, \bibinfo {author} {\bibfnamefont {L.}~\bibnamefont {Deng}}, \bibinfo {author} {\bibfnamefont {E.~W.}\ \bibnamefont {Hagley}}, \bibinfo {author} {\bibfnamefont {K.}~\bibnamefont {Helmerson}}, \bibinfo {author} {\bibfnamefont {W.~P.}\ \bibnamefont {Reinhardt}},  \emph {et~al.},\ }\href {\doibase 10.1126/science.287.5450.97} {\bibfield  {journal} {\bibinfo  {journal} {Science}\ }\textbf {\bibinfo {volume} {287}},\ \bibinfo {pages} {97} (\bibinfo {year} {2000})}\BibitemShut {NoStop}%
\bibitem [{\citenamefont {Becker}\ \emph {et~al.}(2008)\citenamefont {Becker}, \citenamefont {Stellmer}, \citenamefont {Soltan-Panahi}, \citenamefont {D{\"o}rscher}, \citenamefont {Baumert}, \citenamefont {Richter}, \citenamefont {Kronj{\"a}ger}, \citenamefont {Bongs},\ and\ \citenamefont {Sengstock}}]{Becker_oscillations_2008}%
  \BibitemOpen
  \bibfield  {author} {\bibinfo {author} {\bibfnamefont {C.}~\bibnamefont {Becker}}, \bibinfo {author} {\bibfnamefont {S.}~\bibnamefont {Stellmer}}, \bibinfo {author} {\bibfnamefont {P.}~\bibnamefont {Soltan-Panahi}}, \bibinfo {author} {\bibfnamefont {S.}~\bibnamefont {D{\"o}rscher}}, \bibinfo {author} {\bibfnamefont {M.}~\bibnamefont {Baumert}}, \bibinfo {author} {\bibfnamefont {E.-M.}\ \bibnamefont {Richter}}, \bibinfo {author} {\bibfnamefont {J.}~\bibnamefont {Kronj{\"a}ger}}, \bibinfo {author} {\bibfnamefont {K.}~\bibnamefont {Bongs}}, \ and\ \bibinfo {author} {\bibfnamefont {K.}~\bibnamefont {Sengstock}},\ }\href {\doibase 10.1038/nphys962} {\bibfield  {journal} {\bibinfo  {journal} {Nat. Phys.}\ }\textbf {\bibinfo {volume} {4}},\ \bibinfo {pages} {496} (\bibinfo {year} {2008})}\BibitemShut {NoStop}%
\bibitem [{\citenamefont {Busch}\ and\ \citenamefont {Anglin}(2000)}]{Busch_motion_2000}%
  \BibitemOpen
  \bibfield  {author} {\bibinfo {author} {\bibfnamefont {T.}~\bibnamefont {Busch}}\ and\ \bibinfo {author} {\bibfnamefont {J.~R.}\ \bibnamefont {Anglin}},\ }\href {\doibase 10.1103/PhysRevLett.84.2298} {\bibfield  {journal} {\bibinfo  {journal} {Phys. Rev. Lett.}\ }\textbf {\bibinfo {volume} {84}},\ \bibinfo {pages} {2298} (\bibinfo {year} {2000})}\BibitemShut {NoStop}%
\bibitem [{\citenamefont {Mukherjee}\ and\ \citenamefont {Reimann}(2023)}]{Mukherjee_classical_2023}%
  \BibitemOpen
  \bibfield  {author} {\bibinfo {author} {\bibfnamefont {K.}~\bibnamefont {Mukherjee}}\ and\ \bibinfo {author} {\bibfnamefont {S.~M.}\ \bibnamefont {Reimann}},\ }\href {\doibase 10.1103/PhysRevA.107.043319} {\bibfield  {journal} {\bibinfo  {journal} {Phys. Rev. A}\ }\textbf {\bibinfo {volume} {107}},\ \bibinfo {pages} {043319} (\bibinfo {year} {2023})}\BibitemShut {NoStop}%
\bibitem [{\citenamefont {Mistakidis}\ \emph {et~al.}(2024)\citenamefont {Mistakidis}, \citenamefont {Mukherjee}, \citenamefont {Reimann},\ and\ \citenamefont {Sadeghpour}}]{Mistakidis2024}%
  \BibitemOpen
  \bibfield  {author} {\bibinfo {author} {\bibfnamefont {S.~I.}\ \bibnamefont {Mistakidis}}, \bibinfo {author} {\bibfnamefont {K.}~\bibnamefont {Mukherjee}}, \bibinfo {author} {\bibfnamefont {S.~M.}\ \bibnamefont {Reimann}}, \ and\ \bibinfo {author} {\bibfnamefont {H.~R.}\ \bibnamefont {Sadeghpour}},\ }\href {\doibase 10.1103/PhysRevA.110.013323} {\bibfield  {journal} {\bibinfo  {journal} {Phys. Rev. A}\ }\textbf {\bibinfo {volume} {110}},\ \bibinfo {pages} {013323} (\bibinfo {year} {2024})}\BibitemShut {NoStop}%
\bibitem [{\citenamefont {Crank}\ and\ \citenamefont {Nicolson}(1947)}]{crank_practical_1947}%
  \BibitemOpen
  \bibfield  {author} {\bibinfo {author} {\bibfnamefont {J.}~\bibnamefont {Crank}}\ and\ \bibinfo {author} {\bibfnamefont {P.}~\bibnamefont {Nicolson}},\ }in\ \href@noop {} {\emph {\bibinfo {booktitle} {Math. Proc. Camb. Philos. Soc.}}},\ Vol.~\bibinfo {volume} {43}\ (\bibinfo {organization} {Cambridge University Press},\ \bibinfo {year} {1947})\ pp.\ \bibinfo {pages} {50--67}\BibitemShut {NoStop}%
\bibitem [{\citenamefont {Antoine}\ \emph {et~al.}(2013)\citenamefont {Antoine}, \citenamefont {Bao},\ and\ \citenamefont {Besse}}]{antoine_computational_2013}%
  \BibitemOpen
  \bibfield  {author} {\bibinfo {author} {\bibfnamefont {X.}~\bibnamefont {Antoine}}, \bibinfo {author} {\bibfnamefont {W.}~\bibnamefont {Bao}}, \ and\ \bibinfo {author} {\bibfnamefont {C.}~\bibnamefont {Besse}},\ }\href {\doibase https://doi.org/10.1016/j.cpc.2013.07.012} {\bibfield  {journal} {\bibinfo  {journal} {Comput. Phys. Commun.}\ }\textbf {\bibinfo {volume} {184}},\ \bibinfo {pages} {2621} (\bibinfo {year} {2013})}\BibitemShut {NoStop}%
\bibitem [{\citenamefont {Arfken}\ and\ \citenamefont {Weber}(1972)}]{arfken_mathematical_1972}%
  \BibitemOpen
  \bibfield  {author} {\bibinfo {author} {\bibfnamefont {G.~B.}\ \bibnamefont {Arfken}}\ and\ \bibinfo {author} {\bibfnamefont {H.-J.}\ \bibnamefont {Weber}},\ }\href@noop {} {\emph {\bibinfo {title} {Mathematical methods for physicists}}}\ (\bibinfo  {publisher} {Academic Press Orlando, FL},\ \bibinfo {year} {1972})\BibitemShut {NoStop}%
\bibitem [{\citenamefont {G\'oral}\ and\ \citenamefont {Santos}(2002)}]{Goral_ground_2002}%
  \BibitemOpen
  \bibfield  {author} {\bibinfo {author} {\bibfnamefont {K.}~\bibnamefont {G\'oral}}\ and\ \bibinfo {author} {\bibfnamefont {L.}~\bibnamefont {Santos}},\ }\href {\doibase 10.1103/PhysRevA.66.023613} {\bibfield  {journal} {\bibinfo  {journal} {Phys. Rev. A}\ }\textbf {\bibinfo {volume} {66}},\ \bibinfo {pages} {023613} (\bibinfo {year} {2002})}\BibitemShut {NoStop}%
\end{thebibliography}%


\begin{thebibliography}{0}%
\makeatletter
\providecommand \@ifxundefined [1]{%
 \@ifx{#1\undefined}
}%
\providecommand \@ifnum [1]{%
 \ifnum #1\expandafter \@firstoftwo
 \else \expandafter \@secondoftwo
 \fi
}%
\providecommand \@ifx [1]{%
 \ifx #1\expandafter \@firstoftwo
 \else \expandafter \@secondoftwo
 \fi
}%
\providecommand \natexlab [1]{#1}%
\providecommand \enquote  [1]{``#1''}%
\providecommand \bibnamefont  [1]{#1}%
\providecommand \bibfnamefont [1]{#1}%
\providecommand \citenamefont [1]{#1}%
\providecommand \href@noop [0]{\@secondoftwo}%
\providecommand \href [0]{\begingroup \@sanitize@url \@href}%
\providecommand \@href[1]{\@@startlink{#1}\@@href}%
\providecommand \@@href[1]{\endgroup#1\@@endlink}%
\providecommand \@sanitize@url [0]{\catcode `\\12\catcode `\$12\catcode `\&12\catcode `\#12\catcode `\^12\catcode `\_12\catcode `\%12\relax}%
\providecommand \@@startlink[1]{}%
\providecommand \@@endlink[0]{}%
\providecommand \url  [0]{\begingroup\@sanitize@url \@url }%
\providecommand \@url [1]{\endgroup\@href {#1}{\urlprefix }}%
\providecommand \urlprefix  [0]{URL }%
\providecommand \Eprint [0]{\href }%
\providecommand \doibase [0]{http://dx.doi.org/}%
\providecommand \selectlanguage [0]{\@gobble}%
\providecommand \bibinfo  [0]{\@secondoftwo}%
\providecommand \bibfield  [0]{\@secondoftwo}%
\providecommand \translation [1]{[#1]}%
\providecommand \BibitemOpen [0]{}%
\providecommand \bibitemStop [0]{}%
\providecommand \bibitemNoStop [0]{.\EOS\space}%
\providecommand \EOS [0]{\spacefactor3000\relax}%
\providecommand \BibitemShut  [1]{\csname bibitem#1\endcsname}%
\let\auto@bib@innerbib\@empty
\end{thebibliography}%
\end{document}